\documentclass{article}

\usepackage{PRIMEarxiv}
\usepackage{fancyhdr}
\usepackage{microtype}

\usepackage{amsmath}
\usepackage{amssymb}
\usepackage{romannum}
\usepackage[T1]{fontenc}

\usepackage{graphicx}
\usepackage{listings}
\usepackage{multirow}
\usepackage{multicol}
\usepackage[figuresright]{rotating}

\usepackage[colorlinks]{hyperref}
\usepackage{xcolor}

\usepackage{indentfirst}
\def\api#1{\texttt{#1}}

\def\ket#1{|#1\rangle}
\def\scale#1{\tilde{#1}}
\def\change#1{\textcolor{black}{#1}}

\title{\api{TBPLaS}: a Tight-Binding Package for Large-scale Simulation}
\author{
	Yunhai~Li, Zhen~Zhan, Xueheng~Kuang, Yonggang~Li \\
	Key Laboratory of Artificial Micro- and Nano-structures of Ministry of Education and \\School of Physical Science and Technology \\
	Wuhan University \\
	Wuhan 430072, China\\
    \And
	Shengjun~Yuan$^*$ \\
	Key Laboratory of Artificial Micro- and Nano-structures of Ministry of Education and \\School of Physical Science and Technology \\
	Wuhan University \\
	Wuhan 430072, China\\
    E-mail:~\texttt{s.yuan@whu.edu.cn} \\
}

\begin{document}
\maketitle

\begin{abstract}
\api{TBPLaS} is an open-source software package for the accurate simulation of physical systems with arbitrary geometry and dimensionality utilizing the tight-binding (TB) theory. It has an intuitive object-oriented Python application interface (API) and Cython/Fortran extensions for the performance critical parts, ensuring both flexibility and efficiency. Under the hood, numerical calculations are mainly performed by both exact diagonalization and the tight-binding propagation method (TBPM) without diagonalization.  Especially, the TBPM is based on the numerical solution of time-dependent Schr\"{o}dinger equation, achieving linear scaling with system size in both memory and CPU costs.  Consequently, \api{TBPLaS} provides a numerically cheap approach to calculate the electronic, transport, plasmon and optical properties of large tight-binding models with billions of atomic orbitals. Current capabilities of \api{TBPLaS} include the calculation of band structure,  density of states, local density of states, quasi-eigenstates, optical conductivity, electrical conductivity, Hall conductivity, polarization function, dielectric function, plasmon dispersion, carrier mobility and velocity, localization length and free path, $\mathbb{Z}_2$ topological invariant, wave-packet propagation, etc. All the properties can be obtained with only a few lines of code. Other algorithms involving tight-binding Hamiltonians can be implemented easily thanks to its extensible and modular nature. In this paper, we discuss the theoretical framework, implementation details and common workflow of \api{TBPLaS}, and give a few demonstrations of its applications.  
\end{abstract}


\section{Introduction}
\label{intro}

Computational modelling is an essential tool for both fundamental and applied researches in the condensed matter community.
Among the widely used modelling tools, the tight-binding (TB) method is popular in both quantum chemistry and solid state physics \cite{slater1954simplified,goringe1997tight}, which can provide a fast and accurate understanding of the electronic structures of crystals with small unit cells, or large complex systems with/without translational symmetry. The TB method investigates electronic structure via both exact diagonalization and non-diagonalization techniques. With exact diagonalization, the TB method can tackle crystalline structures containing up to tens of thousands of orbitals in the unit cell. With non-diagonalization techniques, for instance the tight-binding propagation method (TBPM) \cite{yuan2010modeling,hams2000TBPM,yuan2011excitation,logemann2015modeling,slotman2015twisted} and the recursion technique \cite{haydock1972electronic}, large systems with up to billions of orbitals can be easily handled.

Recently, a plethora of exotic properties, such as superconductivity \cite{cao2018unconventional,park2021tunable,zhou2022isospin}, correlated insulator \cite{cao2018correlated,xie2019spectroscopic,lu2019superconductors}, charge-ordered states \cite{jiang2019charge}, ferromagnetism \cite{sharpe2019emergent}, quantum anomalous Hall effect \cite{serlin2020intrinsic} and unconventional ferroelectricity \cite{zheng2020unconventional}, are constantly observed in moir\'{e} superlattices, which are formed by stacking single layers of two-dimensional (2D) materials on top of each other with a small misalignment \cite{geim2013van}. To facilitate the exploration of the physical phenomena in the moir\'{e} superlattices, theoretical calculations are utilized to provide accurate and robust predictions.
In the moir\'{e} patterns, the loss of angstrom-scale periodicity poses an obviously computing challenge. For instance, in twisted bilayer graphene (TBG) with rotation angle of 1.05$^\circ$--the so-called magic angle, the number of atoms in a supercell is 11908, which is too large for \textit{state-of-the-art} first-principles methods. 
On the contrary, the TB method has been proved to be a simple and effective approach to investigate the electronic structure of moir\'{e} pattern \cite{carr2020electronic,Andrews2020}. More importantly, with the real-space TB method, the substrate effects, strains, disorders, defects, electric and magnetic fields and many other external perturbations can be naturally implemented via the modifications of the tight-binding parameters \cite{yuan2010modeling,garcia2015real}. Therefore, the TB method provides a more powerful framework to tackle realistic materials fabricated in the laboratory. 

There are some open source software packages implementing the TB method and covering different aspects of the modelling of quantum transport and electronic structure. For example, \api{Kwant} is a Python package for numerical calculations of quantum transport of nanodevices from the transmission probabilities, which is based on the Landauer-Buttiker formalism and the wave function-matching technique \cite{groth2014kwant}. \api{PythTB} is a Python package for the construction and solution of simple TB models \cite{pythtb}. It includes the tools for calculating quantities that are related to Berry phases or curvatures. \api{Pybinding} is a package with a Python interface and a C++ core, which is based on both the exact diagonalization and the kernel polynomial method (KPM) \cite{dean_moldovan_2020_4010216}. Technically, KPM utilizes convolutions with a kernel to attenuate the Gibbs oscillations caused by discontinuities or singularities, and is a general tool to study large matrix problems \cite{weisse2006kernel}. In \api{Pybinding}, the KPM is adopted to model complex systems with disorder, strains or external fields. The software supports numerical calculations of band structures, density of states (DOS), local density of states (LDOS) and conductivity. \api{TBTK} is a C++ software development kit for numerical calculations of quantum mechanical properties \cite{bjornson2019tbtk}. Particularly, it is also based on the KPM and designed for accurate real-space simulations of electronic structures and quantum transport properties of large-scale molecular and condensed systems with tens of billions of atomic orbitals \cite{joao2020kite}. \api{KITE} is an open-source software with a Python interface and a C++ core, which is based on the spectral expansions methods with an exact Chebyshev polynomial expansion of Green's function \cite{ferreira2015critical}. Several functionalities are demonstrated, ranging from calculations of DOS, LDOS, spectral function, electrical (DC) conductivity, optical (AC) conductivity and wave-packet propagation. \api{MathemaTB} is a Mathematica package for TB calculations, which provides 62 functionalities to carry out matrix manipulation, data analysis and visualizations on molecules, wave functions, Hamiltonians, coefficient matrices, and energy spectra \cite{jacobse2019mathematb}.

Previous implementations of the TB method have so far been limited to simple models or have limited functionalities. Therefore, we have developed the TBPM method,
which is based on the numerical solution of time-dependent Schr\"{o}dinger equation (TDSE) without any diagonalization \cite{hams2000TBPM}. The core concepts of TBPM are the correlation functions, which are obtained directly from the time-dependent wave function and contain part of the features of the Hamiltonian. With enough small time step and long propagation time, the whole characteristics of the Hamiltonian can be accurately captured. The correlation functions are then analyzed to yield the desired physical quantities. Compared to exact diagonalization whose costs of memory and CPU time scale as \textit{O(N$^2$)} and \textit{O(N$^3$)}, TBPM has linear scaling in both resources, allowing us to deal with models containing tens of billions of orbitals. Moreover, the calculations of electronic, optical, plasmon and transport properties can be easily implemented in TBPM without the requirement of any symmetries. Other calculations involving the TB Hamiltonian can also be implemented easily.

We implement TBPM in the open source software package named \textbf{T}ight-\textbf{B}inding \textbf{P}ackage for \textbf{La}rge-scale \textbf{S}imulation, or \api{TBPLaS} in shot. In \api{TBPLaS}, TB models can be constructed from scratch using the the application interface (API), or imported from Wannier90 output files directly. Physical quantities can be obtained via four methods: (\romannum{1}) exact diagonalization to calculate the band structure, DOS, eigenfunction, polarization function \cite{Kuang2022} and AC conductivity ; (\romannum{2}) recursive Green's function to get LDOS \cite{shi2020large,liu2021realizing}; (\romannum{3}) KPM to obtain DC and Hall conductivity \cite{yuan2015transport,yuan2016quantum}; (\romannum{4}) TBPM to calculate DOS, LDOS, carrier density, AC conductivity, absorption spectrum, DC conductivity, time-dependent diffusion coefficient, carrier velocity and mobility, elastic mean free path, Anderson localization length, polarization function, response function, dielectric function, energy loss function, plasmon dispersion, plasmon lifetime, damping rate, quasi-eigenstate, real-space charge density, and wave packet propagation \cite{yuan2010modeling,yuan2015electronic,yuan2010electronic,yuan2011landau,yuan2011optical,yuan2012enhanced,logemann2015modeling,yuan2016quantum,van2019tuning}. At the core of \api{TBPLaS}, we use TBPM to achieve nearly linear scaling performance. Furthermore, crystalline defects, vacancies, adsorbates, charged impurities, strains and external electric and/or magnetic \change{fields} can be easily set up with \api{TBPLaS}'s API. These features make it possible for the simulation of systems with low concentrations of disorder \cite{yuan2010modeling, jin2015BPplasmons} or large unit cells, such as twisted bilayer and multilayer systems \cite{slotman2015twisted}. What is more, the computations are performed in real space, so it also allows us to consider systems that lack translation symmetry, such as fractals \cite{van2016quantum, van2017optical} and quasicrystals \cite{yu2019dodecagonal,yu2022interlayer}.

The numerical calculations in \api{TBPLaS} are separated into two stages. In the first stage, the TB model can be constructed in Python using the API in an intuitive object-oriented manner. Many of the concepts of the API are natural in solid state physics, such as lattices, orbitals, hopping terms, vacancies, external electric and magnetic fields, etc. \change{Moreover, the TB model can also be imported from Wannier90 output files directly.} In the second stage, the Hamiltonian matrix is set up from the TB model and passed to backends written in Cython and Fortran, where the quantities are calculated by using either exact diagonalization, recursion method, KPM or post-processing of the correlation functions obtained from the TBPM. The advantage of the two-state paradigm is that it provides both excellent flexibility and high efficiency. Up to now, \api{TBPLaS} has been utilized to investigate the electronic structures of a plenty of 2D materials, such as graphene \cite{slotman2015twisted,shi2020large}, transition metal dichalcogenides \cite{Kuang2022,zhan2020tunability}, tin disulfide \cite{yu2018effective}, arsenene \cite{yu2018tunable}, antimonene \cite{slotman2018plasmon}, black phosphorus \cite{yuan2015transport,jin2015BPplasmons}, tin diselenide \cite{zhong2020electronic}, MoSi$_2$N$_4$ \cite{wang2021electronic}. Moreover, \api{TBPLaS} is a powerful tool to tackle complex systems, for example, graphene with vacancies \cite{yuan2015electronic,yuan2010electronic,yuan2011optical}, twisted multilayer graphene \cite{shi2020large,kuang2021collective,wu2021lattice,wu2021magic}, twisted multilayer transition metal dichalcogenides \cite{zhang2020tuning,zhan2020tunability,Kuang2022}, graphene-boron nitride heterostructures \cite{slotman2015twisted,long2022atomistic}, dodecagonal bilayer graphene quasicrystals \cite{yu2019dodecagonal,yu2020electronic,yu2020pressure,yu2022interlayer,wang2022polarization} and fractals \cite{van2016quantum, van2017optical,westerhout2018plasmon,iliasov2019power,iliasov2020hall,yang2020confined}.   

The paper is organized as following. In Sec. \ref{method} we discuss the concepts and theories of TBPM and other methods. Then the implementation details of \api{TBPLaS} are described in Sec. \ref{impl}, followed by the usages in Sec. \ref{usage}. In Sec. \ref{examples}, we give some examples of calculations that can be done with \api{TBPLaS}. Finally, in Sec. \ref{summary} we give the conclusions, outlooks and possible future developments.

\section{Methodology}
\label{method}
In this section, we discuss briefly the underlying concepts and theories of \api{TBPLaS} with which to calculate the electronic, optical, plasmon and transport properties. Note that if not explicitly given, we will take $\hbar=1$ and omit it from the formula.

\subsection{Tight-binding models}
The Hamiltonian of any non-periodic system containing $n$ orbitals follows
\begin{equation}
	\hat{H} = \sum_i \epsilon_i c_i^{\dagger} c_i^{\phantom{\dagger}} - \sum_{i \neq j} t_{ij} c_i^{\dagger} c_j^{\phantom{\dagger}}
    \label{hal}
\end{equation}
which can be rewritten in a compact matrix form
\begin{equation}
    \hat{H} = \mathbf{c}^{\dagger} H \mathbf{c}
\end{equation}
with
\begin{align}
    \mathbf{c}^{\dagger} &= \left[ c_1^\dagger, c_2^\dagger, \cdots, c_n^\dagger \right] \\
    \mathbf{c} &= \left[ c_1^{\phantom{\dagger}}, c_2^{\phantom{\dagger}}, \cdots, c_n^{\phantom{\dagger}} \right]^\mathrm{T} \\
    \label{eq:ham_mat_finite} H_{ij} &= \epsilon_i\delta_{ij} - t_{ij}(1-\delta_{ij})
\end{align}
Here $\epsilon_i$ denotes the on-site energy of orbital $i$, $t_{ij}$ denotes the hopping integral between orbitals $i$ and $j$, $c^{\dagger}$ and $c$ are the creation and annihilation operators, respectively. The on-site energy $\epsilon_i$ is defined as
\begin{equation}
    \epsilon_i = \int{\phi_{i}^{*}(\mathbf{r}) \hat{h}_0(\mathbf{r}) \phi_{i}^{\phantom{*}}(\mathbf{r})} \mathrm{d}\mathbf{r}
    \label{eq:on_site}
\end{equation}
and the hopping integral $t_{ij}$ is defined as
\begin{equation}
    t_{ij} = -\int{\phi_{i}^{*}(\mathbf{r}) \hat{h}_0(\mathbf{r}) \phi_j^{\phantom{*}}(\mathbf{r})} \mathrm{d}\mathbf{r}
    \label{eq:hop}
\end{equation}
with $\hat{h}_0$ being the single-particle Hamiltonian
\begin{equation}
    \hat{h}_0(\mathbf{r}) = -\frac{\hbar^2}{2m}\nabla^2 + V(\mathbf{r})
    \label{eq:h0}
\end{equation}
and $\phi_i$ being the reference single particle state. In actual calculations, the reference states are typically chosen to be localized states centered at $\tau_i$, e.g., atomic wave functions or maximally localized generalized Wannier functions (MLWF). The on-site energies and hopping integrals can be determined by either direct evaluation following Eqs. (\ref{eq:on_site})-(\ref{eq:h0}), the Slater-Koster formula  \cite{slater1954simplified,cappelluti2013tight}, numerical fitting to experimental or \textit{ab initio} data. Once the parameters are determined, the eigenvalues and eigenstates can be obtained by diagonalizing the Hamiltonian matrix defined in Eq. (\ref{eq:ham_mat_finite}).

For periodic systems, the reference state gets an additional cell index $\mathbf{R}$
\begin{equation}
    \phi_{i\mathbf{R}}(\mathbf{r}) = \phi_{i}(\mathbf{r}-\mathbf{R})
\end{equation}
We define the Bloch basis functions and creation (annihilation) operators by Fourier transform
\begin{align}
    \label{eq:bloch_k1}
    \chi_{i\mathbf{k}}(\mathbf{r}) &= \frac{1}{\sqrt{N}} \sum_{\mathbf{R}} \mathrm{e}^{\mathrm{i}\mathbf{k}\cdot(\mathbf{R}+\tau_i)}\phi_{i\mathbf{R}}(\mathbf{r}) \\
    c_i^{\dagger}(\mathbf{k}) &= \frac{1}{\sqrt{N}} \sum_{\mathbf{R}} \mathrm{e}^{\mathrm{i}\mathbf{k}\cdot(\mathbf{R}+\tau_i)} c_i^{\dagger}(\mathbf{R}) \\
    c_i^{\phantom{\dagger}}(\mathbf{k}) &= \frac{1}{\sqrt{N}} \sum_{\mathbf{R}} \mathrm{e}^{-\mathrm{i}\mathbf{k}\cdot(\mathbf{R}+\tau_i)} c_i^{\phantom{\dagger}}(\mathbf{R})
\end{align}
where $N$ is the number of unit cells. Then the Hamiltonian in Bloch basis can be written as
\begin{equation}
    \hat{H} = N \sum_{\mathbf{k}} \left[ \sum_{i\in\mathrm{uc}} \epsilon_i c_i^{\dagger}(\mathbf{k})c_i^{\phantom{\dagger}}(\mathbf{k})
    - \sum_{\mathbf{R} \neq \mathbf{0} \lor i \neq j} t_{ij}(\mathbf{R}) \mathrm{e}^{\mathrm{i}\mathbf{k}\cdot(\mathbf{R}+\tau_j-\tau_i)} c_i^{\dagger}(\mathbf{k}) c_j^{\phantom{\dagger}}(\mathbf{k})
    \right]
\end{equation}
Here the third summation is performed for all cell indices $\mathbf{R}$ and orbital pairs $(i, j)$, except the diagonal terms with $\mathbf{R}=\mathbf{0}$ and $i=j$. The Hamiltonian can also be rewritten in matrix form as
\begin{equation}
    \hat{H} = N \sum_{\mathbf{k}} \mathbf{c}^{\dagger}(\mathbf{k}) H(\mathbf{k}) \mathbf{c}(\mathbf{k})
\end{equation}
with
\begin{align}
    \mathbf{c}^{\dagger}(\mathbf{k}) &= \left[ c_1^\dagger(\mathbf{k}), c_2^\dagger(\mathbf{k}), \cdots, c_n^\dagger(\mathbf{k}) \right] \\
    \mathbf{c}(\mathbf{k}) &= \left[ c_1^{\phantom{\dagger}}(\mathbf{k}), c_2^{\phantom{\dagger}}(\mathbf{k}), \cdots, c_n^{\phantom{\dagger}}(\mathbf{k}) \right]^\mathrm{T} \\
    H_{ij}(\mathbf{k}) &= \epsilon_i\delta_{ij} - \sum_{\mathbf{R} \neq \mathbf{0} \lor i \neq j} t_{ij}(\mathbf{R}) \mathrm{e}^{\mathrm{i}\mathbf{k}\cdot(\mathbf{R}+\tau_j-\tau_i)}
    \label{eq:ham_k1}
\end{align}
Here $t_{ij} (\mathbf{R})$ is the hopping integral between $\phi_{i\mathbf{0}}$ and $\phi_{j\mathbf{R}}$.

There is another convention to construct the Bloch basis functions and creation (annihilation) operators, which excludes the orbital position $\mathbf{\tau}_i$ in the Fourier transform
\begin{align}
    \label{eq:bloch_k2}
    \chi_{i\mathbf{k}}(\mathbf{r}) &= \frac{1}{\sqrt{N}} \sum_{\mathbf{R}} \mathrm{e}^{\mathrm{i}\mathbf{k}\cdot\mathbf{R}}\phi_{i\mathbf{R}}(\mathbf{r}) \\
    c_i^{\dagger}(\mathbf{k}) &= \frac{1}{\sqrt{N}} \sum_{\mathbf{R}} \mathrm{e}^{\mathrm{i}\mathbf{k}\cdot\mathbf{R}} c_i^{\dagger}(\mathbf{R}) \\
    c_i^{\phantom{\dagger}}(\mathbf{k}) &= \frac{1}{\sqrt{N}} \sum_{\mathbf{R}} \mathrm{e}^{-\mathrm{i}\mathbf{k}\cdot\mathbf{R}} c_i^{\phantom{\dagger}}(\mathbf{R})
\end{align}
Then Eq. (\ref{eq:ham_k1}) becomes
\begin{equation}
    H_{ij}(\mathbf{k}) = \epsilon_i \delta_{ij} - \sum_{\mathbf{R} \neq \mathbf{0} \lor i \neq j} t_{ij}(\mathbf{R}) \mathrm{e}^{\mathrm{i}\mathbf{k}\cdot\mathbf{R}}
    \label{eq:ham_k2}
\end{equation}
Both conventions have been implemented in \api{TBPLaS}, while the first convention is enabled by default.

External electric and magnetic fields can be introduced into the tight-binding model by modifying the on-site energies and hopping integrals. For example, homogeneous electric fields towards $-z$ direction can be described by
\begin{equation}
    \epsilon_i \to \epsilon_i + E \cdot (z_i-z_0)
\end{equation}
where $E$ is the intensity of electric field, $z_i$ is the position of orbital $i$ along $z$-axis, and $z_0$ is the position of zero-potential plane. Magnetic fields, on the other hand, can be described by the vector potential $\mathbf{A}$ and Peierls substitution \cite{vonsovsky1989quantum}
\begin{equation}
    t_{ij} \to t_{ij} \cdot \exp\left({\mathrm{i}}\frac{e}{\hbar c}\int_i^j \mathbf{A}\cdot \mathrm{d} \mathbf{l} \right) = t_{ij} \cdot \exp\left({\mathrm{i}}\frac{2 \pi}{\Phi_0}\int_i^j \mathbf{A}\cdot \mathrm{d} \mathbf{l} \right)
\end{equation}
where $\int_i^j \mathbf{A}\cdot\mathrm{d}\mathbf{l}$ is the line integral of the vector potential from orbital $i$ to orbital $j$, and $\Phi_0=ch/e$ is the flux quantum. For homogeneous magnetic field towards $-z$, we follow the Landau gauge $\mathbf{A}=(By, 0, 0)$. Note that for numerical stability, the size of the system should be larger than the magnetic length.

Finally, we mention that we have omitted the spin notations in above formulation for clarity. However, spin-related terms such as spin-orbital coupling (SOC), can be easily incorporated into the Hamiltonian and treated in the same approach in TBPM and \api{TBPLaS}.

\subsection{Tight-binding propagation method}

Exact diagonalization of the Hamiltonian matrix in Eq. (\ref{eq:ham_mat_finite}), (\ref{eq:ham_k1}) and (\ref{eq:ham_k2}) yields the eigenvalues and eigenstates  of the model, eventually all the physical quantities. However, the memory and CPU time costs of exact diagonalization scale as \textit{O(N$^2$)} and \textit{O(N$^3$)} with  the model size $N$, making it infeasible for large models. The TBPM, on the contrary, tackles the eigenvalue problem with a totally different philosophy. The memory and CPU time costs of TBPM scale linearly with the model size, so models with tens of billions of orbitals can be easily handled.

In TBPM, a set of randomly generated states are prepared as the initial wave functions. Then the wave functions are propagated following
\begin{equation}
    \ket{\psi(t)} = \mathrm{e}^{-\mathrm{i}\hat{H}t}\ket{\psi(0)}
    \label{state_t}
\end{equation}
and correlation functions are evaluated at each time step. The correlation functions contain a fraction of the features of the Hamiltonian. With enough small time step and long propagation time, the whole characteristics of the Hamiltonian will be accurately captured. Finally, the correlation functions are averaged and analyzed to yield the physical quantities. Taking the correlation function of DOS for example, which is defined as
\begin{equation}
	C_{\text{DOS}}(t) = \langle \psi(0) | \psi(t) \rangle
	\label{eq:corr_dos}
\end{equation}
It can be proved that the inner product is related to the eigenvalues via
\begin{align}
   \langle \psi(0) | \psi(t) \rangle = \sum_{ijk} U^{\phantom{*}}_{kj}U^{*}_{ij} a^{\phantom{*}}_ia^{*}_k \mathrm{e}^{-\mathrm{i}\epsilon_jt} 
\end{align}
with $\epsilon_j$ being the $j$-th eigenvalue, $U_{kj}$ being the $k$-th component of $j$-th eigenstate, respectively. The initial wave function $\psi(0)$ is a random superposition of all basis states
\begin{equation}
	|\psi(0)\rangle = \sum_i a_i \left| \phi_i \right\rangle
	\label{ini_state}
\end{equation}
where $a_i$ are random complex numbers with $\sum_i |a_i|^2 = 1$, and $\phi_i$ are the basis states. It is clear that the correlation function can be viewed as a linear combination of oscillations with frequencies of $\epsilon_j$. With inverse Fourier transform, the eigenvalues and DOS can be determined.

To propagate the wave function, one needs to numerically decompose the time evolution operator. As the TB Hamiltonian matrix is sparse, it is convenient to use the Chebyshev polynomial method for the decomposition, which is proved to be unconditionally stable for solving TDSE \cite{jin2021random}.
Suppose $x \in [-1,1]$, then
\begin{equation}
    \mathrm{e}^{-\mathrm{i}zx}=J_0(z)+2\displaystyle\sum_{m=1}^{\infty}(-\mathrm{i})^mJ_m(z)T_m(x)
\end{equation}
where $J_m(z)$ is the Bessel function of integer order $m$, $T_m(x)=\cos{[m\arccos{x}]}$ is the Chebyshev polynomial of the first kind. $T_m(x)$ follows a recurrence relation as
\begin{equation}
    T_{m+1}(x)+T_{m-1}(x)=2xT_m(x)
\end{equation}
To utilize the Chebyshev polynomial method, we need to rescale the Hamiltonian as $\scale{H} = \hat H/||\hat H||$ such that $\scale{H}$ has eigenvalues in the range $[-1,1]$. Then, the time evolution of the states can be represented as
\begin{equation}
	|\psi(t) \rangle = \left[ J_0(\scale{t}) \hat T_0(\scale{H}) + 2\sum_{m=1}^{\infty}
	J_m(\scale{t}) \hat T_m(\scale{H}) \right] |\psi(0) \rangle
	\label{eq:psi_t}
\end{equation}
where $\scale{t}=t\cdot||\hat H||$, $J_m(\scale{t})$ is the Bessel function of integer order $m$, $\hat T(\scale{H})$ is the modified Chebyshev polynomials, which can be calculated up to machine precision with the recurrence relation
\begin{equation}
    \hat T_{m+1}(\scale{H}) |\psi \rangle = -2\mathrm{i}\scale{H}\hat T_m(\scale{H})|\psi \rangle + \hat T_{m-1}(\scale{H})|\psi \rangle
\end{equation}
with
\begin{align}
	\hat T_0(\scale{H}) |\psi \rangle = |\psi \rangle, \qquad
	\hat T_1(\scale{H})|\psi \rangle = -\mathrm{i}\scale{H}|\psi \rangle
\end{align}

The other operators appear in TBPM can also be decomposed numerically using the Chebyshev polynomial method. A function $f(x)$ whose values are in the range [-1, 1] can be expressed as
\begin{equation}
    f(x)=\frac{1}{2}c_0T_0(x)+\displaystyle\sum_{k=1}^{\infty}c_kT_k(x)
\end{equation}
where $T_k(x)=\cos{(k\arccos{x})}$ and the coefficients $c_k$ are
\begin{equation}
    c_k=\frac{2}{\pi}\int_{-1}^{1}\frac{\mathrm d x}{\sqrt{1-x^2}}f(x)T_k(x)
\label{ck_coef}
\end{equation}
Assume $x=\cos{\theta}$ and substitute it into Eq. (\ref{ck_coef}), we have
\begin{equation}
    c_k=\frac{2}{\pi}\int_{0}^{\pi}f(\cos{\theta})\cos{k\theta}\mathrm d \theta =\mathrm{Re}\left[\frac{2}{\pi}\displaystyle\sum_{n=0}^{N-1}f\left(\cos{\frac{2\pi n}{N}} \right) \exp\left(\mathrm{i} \frac{2\pi n}{N}k\right) \right]
\end{equation}
which can be calculated by fast Fourier transform. For the Fermi-Dirac operator as frequently used in TBPM, it is more convinced to expressed it as $f=z\mathrm{e}^{-\beta H}/(1+z\mathrm{e}^{-\beta H})$ \cite{yuan2010modeling}, where $z=\mathrm{e}^{\beta \mu}$ is the fugacity, $\beta=1/k_B T$, $k_B$ is the Boltzmann constant, $T$ is the temperature and $\mu$ is the chemical potential. We define $\tilde{\beta}=\beta\cdot||H||$, then
\begin{equation}
    f(\tilde{H})=\frac{z\mathrm{e}^{-\tilde{\beta}\tilde{H}}}{1+z\mathrm{e}^{-\tilde{\beta}\tilde{H}}}=\displaystyle\sum_{k=0}^{\infty}c_kT_k(\tilde{H})
\end{equation}
where $c_k$ are the Chebyshev expansion coefficients of the function $f(x)=z\mathrm{e}^{-\tilde{\beta}x}/(1+z\mathrm{e}^{-\tilde{\beta}x)}$. The Chebyshev polynomials $T_k(\tilde{H})$ have the following recursion relation
\begin{equation}
    T_{k+1}(\tilde{H})-2\tilde{H}T_k(\tilde{H})+T_{k-1}(\tilde{H})=0
\end{equation}
with
\begin{equation}
  T_0(\tilde{H})=1, \qquad T_1(\tilde{H})=\tilde{H}
\end{equation}
For more details we refer to Ref. \cite{yuan2010modeling}.

\subsection{Band structure}
The band structure of a periodic system can be determined by diagonalizing the Hamiltonian matrix in Eq. (\ref{eq:ham_k1}) or (\ref{eq:ham_k2}) for a list of $\mathbf{k}$-points. Both conventions yield the same band structure. Typically, the $\mathbf{k}$-points are sampled on a $\mathbf{k}$-path connecting highly symmetric $\mathbf{k}$-points in the first Brillouin zone. A recommended set of highly symmetric $\mathbf{k}$-points can be found in Ref. \cite{SETYAWAN2010299}.

\subsection{Density of states}
\label{tbpm}
In \api{TBPLaS}, we have two approaches to calculate DOS. The first approach is based on exact diagonalization, which consists of getting the eigenvalues of the Hamiltonian matrix on a dense $\mathbf{k}$-grid, and a summation over the eigenvalues to collect the contributions
\begin{equation}
    D(E) = \sum_{i\mathbf{k}}\delta(E-\epsilon_{i\mathbf{k}})
    \label{eq:dos_diag}
\end{equation}
where $\epsilon_{i\mathbf{k}}$ is the $i$-th eigenvalue at point $\mathbf{k}$. In actual calculations the delta function is approximated with a Gaussian function
\begin{equation}
    G(E-\epsilon_{i\mathbf{k}}) = \frac{1}{\sqrt{2\pi}\sigma}\exp \left[-\frac{(E-\epsilon_{i\mathbf{k}})^2}{2\sigma^2}\right]
\end{equation}
or a Lorentzian function
\begin{equation}
    L(E-\epsilon_{i\mathbf{k}}) = \frac{1}{\pi\sigma}\frac{\sigma^2}{(E-\epsilon_{i\mathbf{k}})^2+\sigma^2}
\end{equation}
Here $\sigma$ is the broadening parameter.

The other approach is the TBPM method, which evaluates the correlation function according to Eq. (\ref{eq:corr_dos}). The DOS is then calculated by inverse Fourier transform of the averaged correlation function
\begin{equation}
	D(E) = \frac{1}{S} \sum_{p=1}^{S} \frac{1}{2\pi} \int_{-\infty}^{\infty} \mathrm{e}^{\mathrm{i}Et} C_{\text{DOS}}(t) \mathrm{d}t
	\label{dos_tbpm}
\end{equation}
Here $S$ is the number of random samples for the average. The inverse Fourier transform in Eq. (\ref{dos_tbpm}) can be performed by fast Fourier transform, or integrated numerically if higher energy resolution is desired. We use a window function to alleviate the effects of the finite time used in the numerical time integration of TDSE. Currently, three types of window functions have been implemented, namely Hanning window\cite{press2007numerical}, Gaussian window and exponential window. 

The statistical error in the calculation of DOS follows $1/\sqrt{SN}$, where $N$ is the model size. Thus the accuracy can be improved by either using large models or averaging over many initial states. For a large enough model ($>10^8$ orbitals), one random initial state is generally enough to ensure convergence. The same conclusion holds for other quantities obtained from TBPM. The energy resolution of DOS is determined by the number of propagation steps. Distinct eigenvalues that differ more than the resolution appear as separate peaks in DOS. If the eigenvalue is isolated from the rest of the spectrum, then the number of propagation steps determines the width of the peak. More details about the methodology of calculating DOS can be found in Ref. \cite{yuan2010modeling,hams2000TBPM}. We emphasize that the $1/\sqrt{SN}$ dependence of the statistical error is a general conclusion which is also valid for other quantities calculated with TBPM, and the above discussions for improving accuracy and energy resolution work for these quantities as well.

\subsection{Local density of states}

\api{TBPLaS} provides three approaches to calculate the LDOS. The first approach is based on exact diagonalization, which is similar to the evaluation of DOS
\begin{equation}
    d_i(E) = \sum_{j\mathbf{k}}\delta(E-\epsilon_{j\mathbf{k}})|U_{ij\mathbf{k}}|^2
    \label{eq:ldos_diag}
\end{equation}
where $U_{ij\mathbf{k}}$ is the $i$-th component of $j$-th eigenstate at point $\mathbf{k}$. The second approach is the TBPM method, which also has much in common with DOS. The only difference is that the initial wave function $|\psi(0) \rangle$ in Eq. (\ref{eq:corr_dos}) is redefined. For instance, to calculate the LDOS on a particular orbital $i$, we set only the component $a_i$ in Eq. (\ref{ini_state}) as nonzero. Then the correlation function can be evaluated and analyzed in the same approach as DOS, following Eq. (\ref{eq:corr_dos}) and (\ref{dos_tbpm}). It can be proved that in this case the correlation function becomes
\begin{align}
   \langle \psi(0) | \psi(t) \rangle = \sum_{j} |U_{ij}a_i|^2 \mathrm{e}^{-\mathrm{i}\epsilon_jt}
\end{align}
which contains the contributions from the $i$-th components of all the eigenstates.

The third approach evaluates LDOS utilizing the recursion method in real space based on Lanczos algorithm\cite{haydock1972electronic,bose1984local}. The LDOS on a particular orbital $i$ is
\begin{equation}
	d_i(E)=-\lim\limits_{\epsilon \to 0^+}\frac{1}{\pi}\mathrm{Im}\langle \phi_i|G(E+\mathrm{i}\epsilon)|\phi_i\rangle
\end{equation}
Then, we use the recursion method to obtain the diagonal matrix elements of the Green's function $G(E)$
\begin{align}
	G_0(E) &= \langle l_0|G(E)|l_0\rangle \nonumber\\
	\quad&=1/(E-a_0-b_1^2/(E-a_1-b_2^2/(E-a_2-b_3^2/\dots)))
\end{align}
where $l_0$ is a unit vector with non-zero component at orbital $i$ only. The elements $a_n$ and $b_n$ are determined with the following recursion relation
\begin{align}
    a_i &= \langle l_i \left| H \right| l_i \rangle \\
    |m_{i+1}\rangle &= (H - a_i)|l_i\rangle - b_i|l_{i-1}\rangle \\
    b_{i+1} &= \sqrt{\langle m_{i+1} | m_{i+1}\rangle} \\
    |l_{i+1}\rangle &= \frac{1}{b_{i+1}} |m_{i+1}\rangle
\end{align}
with $|l_{-1}\rangle=|0\rangle$.

\subsection{Quasieigenstates}
For a general Hamiltonian in Eq. (\ref{hal}) and for samples containing millions of orbitals, it is computationally expensive to get the eigenstates by exact diagonalization. An approximation of the eigenstates at a certain energy $E$ can be calculated without diagonalization following the method in  Ref. \cite{yuan2010modeling}, which has been introduced for the calculation of electric transport properties of large complex models. With an inverse Fourier transform of the time-dependent wave function $|\psi(t) \rangle$, one gets the following expression
\begin{align}
	|\Psi(E) \rangle &=\frac{1}{2\pi} \int_{-\infty}^{\infty} \mathrm{e}^{\mathrm{i}Et} |\psi(t) \rangle \mathrm{d}t \nonumber \\
	&=\frac{1}{2\pi}\sum_i a_i \int_{-\infty}^{\infty} \mathrm{e}^{\mathrm{i}(E-E_i)t} |\phi_i \rangle \mathrm{d}t \nonumber \\
	&=\sum_i a_i \delta(E-E_i)|\phi_i \rangle
\label{quasi}
\end{align}
which can be normalized as
\begin{equation}
	|\tilde\Psi(E) \rangle=\frac{1}{\sqrt{\sum_i|a_i|^2\delta(E-E_i)}}\sum_i a_i\delta(E-E_i)|\phi_i \rangle
\end{equation}
Here, $E_i$ is the $i$-th eigenvalue of the scaled Hamiltonian $\scale{H}$. Note that $|\tilde\Psi(E) \rangle$ is an eigenstate if it is a single (non-degenerate) state\cite{kosloff1983fourier}, or a superposition of the degenerate eigenstates with the energy $E$. That is why it is called the  \textit{quasieigenstate}. Although $|\tilde\Psi(E) \rangle$ is written in the energy basis, the time-dependent wave function $|\psi(t) \rangle$ can be expanded in any orthogonal and complete basis sets. Two methods can be adopted to improve the accuracy of quasieigenstates. The first one is to perform inverse Fourier transform on the states from both positive and negative time, which keeps the original form of the integral in Eq. (\ref{quasi}). The other method is to multiply the wave function $|\psi(t) \rangle$ by a window function, which improves the approximation to the integrals. Theoretically, the spatial distribution of the quasieigenstates reveals directly the electronic structure of the eigenstates with certain eigenvalue. It has been proved that the LDOS mapping from the quasieigenstates is highly consistent with the experimentally scanning tunneling microscopy (STM) $\mathrm{d}I/\mathrm{d}V$ mapping \cite{shi2020large}.

\subsection{Optical conductivity}
In \api{TBPLaS}, we use both TBPM and exact diagonalization-based methods to compute the optical conductivity \cite{kubo1957statistical}. In the TBPM method, we combine the Kubo formula with the random state technology. For a non-interacting electronic system, the real part of the optical conductivity in direction $\alpha$ due to a field in direction $\beta$ is (omitting the Drude contribution at $\omega=0$)\cite{yuan2010modeling}
\begin{equation}
	\mathrm{Re}\;\sigma_{\alpha \beta}(\hbar\omega)
	= \lim_{E \rightarrow 0^+} \frac{\mathrm{e}^{-\beta \hbar \omega}-1}{\hbar \omega A}
	\int_{0}^{\infty} \mathrm{e}^{-Et} \sin (\omega t)\times 2\mathrm{Im}\langle \psi| f(H)J_\alpha (t)[1-f(H)]J_\beta |\psi \rangle \mathrm{d}t
	\label{AC}
\end{equation}
Here, $A$ is the area or volume of the model in two or three dimensional cases, respectively. For a generic tight-binding Hamiltonian, the current density operator is defined as
\begin{equation}
	J=-\frac{\mathrm{i}e}{\hbar}\sum_{i,j}t_{ij}(\hat r_j - \hat r_i)c^{\dagger}_i c^{\phantom{\dagger}}_j
\end{equation}
where $\hat r$ is the position operator. The Fermi-Dirac distribution defined as
\begin{equation}
    f(H)=\frac{1}{e^{\beta (H-\mu)}+1}
\end{equation}
In actual calculations, the accuracy of the optical conductivity is ensured by performing the Eq. (\ref{AC}) over a random superposition of all the basis states in the real space, similar to the calculation of the DOS. Moreover, the Fermi distribution operator $f(\scale{H})$ and $1-f(\scale{H})$ can be obtained by the standard Chebyshev polynomial decomposition in section \ref{tbpm}. We introduce two wave functions 
\begin{align}
	| \psi_1 (t) \rangle_{\alpha} &=  \mathrm{e}^{-\mathrm{i}\scale{H}t} [1 - f(\scale{H})] J_{\alpha} | \psi(0) \rangle \\
	| \psi_2 (t) \rangle &=  \mathrm{e}^{-\mathrm{i}\scale{H}t} f(\scale{H}) | \psi(0) \rangle
\end{align}
Then the real part of $\sigma_{\alpha \beta}(\omega)$ is
\begin{equation}
    \mathrm{Re}\;\sigma_{\alpha \beta}(\hbar\omega)= \lim_{E \rightarrow 0^+} \frac{\mathrm{e}^{-\beta \hbar \omega}-1}{\hbar \omega A}
	\int_{0}^{\infty} \mathrm{e}^{-Et} \sin (\omega t)\times 2 \mathrm{Im}\langle \psi_2(t)|J_\alpha|\psi_1(t)\rangle_{\beta} \mathrm{d} t
\end{equation}
while the imaginary part can be extracted with the Kramers-Kronig relation
\begin{equation}
    \mathrm{Im}\;\sigma_{\alpha \beta}(\hbar\omega) = -\frac{1}{\pi}\mathcal{P}\int_{-\infty}^{\infty}
    \frac{\mathrm{Re}\;\sigma_{\alpha \beta}(\hbar\omega^\prime)}{\omega^\prime-\omega} \mathrm{d}\omega^\prime
\end{equation}

In the diagonalization-based method, the optical conductivity is evaluated as
\begin{equation}
\sigma_{\alpha\beta}(\hbar\omega)=\frac{\mathrm{i} e^2 \hbar}{N_\mathbf{k} \Omega_c}\sum_{\mathbf{k}}\sum_{m,n} \frac{f_{m\mathbf{k}} - f_{n\mathbf{k}}}{\epsilon_{m\mathbf{k}} - \epsilon_{n\mathbf{k}}} \frac{\langle\psi_{n\mathbf{k} }|v_\alpha|\psi_{m\mathbf{k}}\rangle \langle\psi_{m\mathbf{k}}|v_\beta|\psi_{n\mathbf{k}}\rangle}{\epsilon_{m\mathbf{k}} - \epsilon_{n\mathbf{k}}-(\hbar\omega+\mathrm i\eta^+)}
\label{eq:ac_diag}
\end{equation}
where $N_\mathbf{k}$ is the number of $\mathbf{k}$-points in the first Brillouin zone, and $\Omega_c$ is the volume of unit cell, respectively. $\psi_{m\mathbf{k}}$ and $\psi_{n\mathbf{k}}$ are the eigenstates of Hamiltonian defined in Eq. (\ref{eq:ham_k1}), with $\epsilon_{m\mathbf{k}}$ and $\epsilon_{n\mathbf{k}}$ being the corresponding eigenvalues, and $f_{m\mathbf{k}}$ and $f_{n\mathbf{k}}$ being the occupation numbers. $v_\alpha$ and $v_\beta$ are components of velocity operator defined as $v=-J/e$, and $\eta^+$ is the positive infestimal.

\subsection{DC conductivity}

The DC conductivity can be calculated by taking the limit $\omega \to 0$ in the Kubo formula\cite{kubo1957statistical}. Based on the DOS and quasieigenstates obtained in Eqs. (\ref{dos_tbpm}) and (\ref{quasi}), we can calculate the diagonal term of DC conductivity $\sigma_{\alpha \alpha}$ in direction $\alpha$ at temperature $T=0$ with
\begin{align}
	\sigma_{\alpha \alpha}(E) &=\lim\limits_{\tau \to \infty}\sigma_{\alpha \alpha}(E,\tau)\nonumber\\
	&= \lim\limits_{\tau \to \infty}\frac{D(E)}{A} \int_0^\tau \text{Re}\left[ \mathrm{e}^{-\mathrm{i}Et} C_{\text{DC}}(t) \right] \mathrm{d}t
	\label{dc_kubo}
\end{align}
where the DC correlation function is defined as
\begin{equation}
	C_{\text{DC}}(t) = \frac{\langle \psi(0) |
		J_\alpha \mathrm{e}^{\mathrm{i}\scale{H}t} J_\alpha |\tilde \Psi(E) \rangle}{| \langle \psi(0) |\tilde \Psi(E) \rangle |}
\end{equation}
and $A$ is the area of volume of the unit cell depending on system dimension. It is important to note that $| \psi(0) \rangle$ must be the same random initial state used in the calculation of $| \tilde \Psi(E) \rangle$. The semiclassic DC conductivity $\sigma^{sc}(E)$ without considering the effect of Anderson localization is defined as
\begin{equation}
    \sigma^{sc}(E) = \sigma_{\alpha \alpha}^{max}(E,\tau)
\end{equation}
The measured field-effect carrier mobility is related to the semiclassic DC conductivity as 
\begin{equation}
u(E)=\frac{\sigma^{sc}(E)}{en_e(E)}
\end{equation}
where the carrier density $n_e(E)$ is obtained from the integral of density of states via $n_e(E)=\int_0^ED(\varepsilon)d\varepsilon$.

In \api{TBPLaS}, there is another efficient approach to evaluate DC conductivity, which is based on a real-space implementation of the Kubo formalism, where both the diagonal and off-diagonal terms of conductivity are treated on the same footing\cite{garcia2015real}. The DC conductivity tensor for non-interacting electronic system is given by the Kubo-Bastin formula\cite{garcia2015real,bastin1971quantum}
\begin{align}
	\sigma_{\alpha\beta}(\mu,T)&=\frac{\mathrm{i}\hbar e^2}{A}\int_{-\infty}^{\infty}\mathrm{d}E f(E)\mathrm{Tr}\Big\langle v_\alpha \delta(E-H)v_\beta\frac{\mathrm{d}G^+(E)}{\mathrm{d}E}\nonumber\\
	&\qquad\qquad-v_\alpha\frac{G^-(E)}{\mathrm{d}E}v_\beta\delta(E-H)\Big\rangle
\end{align}
where $v_\alpha$ is the $\alpha$ component of the velocity operator, $G^\pm(E)=1/(E-H\pm \mathrm{i}\eta)$ are the Green's functions. Firstly, we rescale the Hamiltonian and energy, and denote them as $\tilde H$ and $\tilde E$, respectively. Then the delta $\delta$ and the Green's function $G^\pm(E)$ can be expanded in terms of Chebyshev polynomials using the kernel polynomial method (KPM)
\begin{align}
	\delta(\tilde E-\tilde H)&=\frac{2}{\pi\sqrt{1-\tilde E^2}}\displaystyle\sum_{m=0}^{M}g_m\frac{T_m(\tilde E)}{\delta_{m,0}+1}T_m(\tilde H) \\
	G^\pm(\tilde E,\tilde H)&=\mp\frac{2\mathrm{i}}{\sqrt{1-\tilde E^2}}\displaystyle\sum_{m=0}^{M}g_m\frac{\mathrm{e}^{\pm \mathrm{i}m\arccos{(\tilde E)}}}{\delta_{m,0}+1}T_m(\tilde H)
\end{align}
Truncation of the above expansions gives rise to Gibbs oscillations, which can be smoothed with a Jackson kernel $g_m$\cite{weisse2006kernel}. Then the conductivity tensor can be written as\cite{garcia2015real}
\begin{equation}
	\sigma_{\alpha\beta}(\mu,T)=\frac{4e^2\hbar}{\pi A}\frac{4}{\Delta E^2}\int_{-1}^1 \mathrm{d}\tilde{E}\frac{f(\tilde E)}{(1- \tilde E^2)^2}\displaystyle\sum_{m,n}\Gamma_{nm}(\tilde E)\mu_{nm}^{\alpha\beta}(\tilde H)
\end{equation}
where $\Delta E=E^+_{max}-E^-_{min}$ is the energy range of the spectrum, $\tilde E$ is the rescaled energy within [-1,1], $\Gamma_{nm}(\tilde E)$ and $\mu_{nm}^{\alpha\beta}(\tilde H)$ are functions of the energy and the Hamiltonian, respectively
\begin{align}
	\Gamma_{nm}(\tilde E)&=T_m(\tilde E)(\tilde E-\mathrm{i}n\sqrt{1-\tilde E^2})\mathrm{e}^{\mathrm{i}n\arccos{(\tilde E)}} \nonumber\\
	&\;+T_n(\tilde E)(\tilde E+\mathrm{i}m\sqrt{1-\tilde E^2})\mathrm{e}^{-\mathrm{i}m\arccos{(\tilde E)}}\\
	\mu_{nm}^{\alpha\beta}(\tilde H)&=\frac{g_mg_n}{(1+\delta_{n0})(1+\delta_{m0})}\mathrm{Tr}[v_\alpha T_m(\tilde H)v_\beta T_n(\tilde H)]
\end{align}

\subsection{Diffusion coefficient}
In the Kubo formalism, the DC conductivity in Eq. (\ref{dc_kubo}) can also be written as a function of diffusion coefficient 
\begin{equation}
	\sigma_{\alpha\alpha}(E)=\frac{e^2}{A}D(E)\lim\limits_{\tau \to \infty}\mathcal D_{diff}(E,\tau)
\end{equation}
Therefore, the time-dependent diffusion coefficient is obtained as
\begin{equation}
	\mathcal D_{diff}(E,\tau)=\frac{1}{e^2}\int_0^\tau \text{Re}\left[ \mathrm{e}^{-\mathrm{i}Et} C_{\text{DC}}(t) \right] \mathrm{d}t
\end{equation}
Once we know the $\mathcal D_{diff}(E,\tau)$, we can extract the carrier velocity from a short time behavior of the diffusivity as
\begin{equation}
v(E) = \sqrt{\mathcal D_{diff}(E,\tau)/\tau}
\end{equation}
and the elastic mean free path $\ell(E)$ from the maximum of the diffusion coefficient as
\begin{equation}
\ell(E)=\frac{\mathcal D_{diff}^{max}(E)}{2v(E)}
\end{equation}
This also allows us to estimate the Anderson localization lengths \cite{yuan2012enhanced,lherbier2011two} by
\begin{equation}
\xi(E)=\ell(E)\exp\left[\frac{\pi h}{2e^2}\sigma^{sc}(E)\right]
\end{equation}

\subsection{Dielectric function}
In TBPM, the dynamic polarization can be obtained by combining Kubo formula \cite{kubo1957statistical} and random state technology as
\begin{equation}
	\Pi_K(\mathbf{q},\hbar\omega) = - \frac{2}{A} \int_0^\infty \mathrm{e}^{\mathrm{i} \omega t}C_{\text{DP}}(t) \mathrm{d}t
	\label{kubo_dp}
\end{equation}
where the correlation function is defined as
\begin{equation}
	C_{\text{DP}}(t) = \text{Im} \langle \psi_2 (t) | \rho(\mathbf{q}) |
	\tilde{\psi}_1(\mathbf{q},t) \rangle
\end{equation}
Here, the density operator is
\begin{equation}
	\rho(\mathbf{q}) = \sum_i \mathrm{e}^{\mathrm{i} \mathbf{q} \cdot \mathbf{r}_i}
	c^{\dagger}_i c^{\phantom{\dagger}}_i
\end{equation}
and the introduced two functions are
\begin{align}
	| \tilde{\psi}_1 (\mathbf{q},t) \rangle_{\beta} = & \mathrm{e}^{-\mathrm{i}\tilde Ht} [1 - f(\tilde H)] \rho (-\mathbf{q}) | \psi(0) \rangle \\
	| \psi_2 (t) \rangle = & \mathrm{e}^{-\mathrm{i}\tilde Ht} f(\tilde H) | \psi(0) \rangle
\end{align}

The dynamical polarization function can also be obtained via diagonalization from the Lindhard function as\cite{giuliani2005quantum}
\begin{align}
	\Pi_L(\textbf{q}, \hbar\omega) &= -\frac{g_s}{(2\pi)^p}\int_\mathrm{BZ}\mathrm{d}^p\textbf{k}\sum_{m,n}
	\frac{f_{m\mathbf{k}}-f_{n\mathbf{k+q}}}{\epsilon_{m\mathbf{k}}-\epsilon_{n\mathbf{k+q}}+\hbar\omega+\mathrm{i}\eta^+} \nonumber\\
	& \quad\times |\langle \psi_{n\mathbf{k+q}} |\mathrm e^{\mathrm{i}\mathbf{q\cdot r}}| \psi_{m\mathbf{k}} \rangle |^2
	\label{lin_dp}
\end{align}
where $\psi_{m\mathbf{k}}$ and $\epsilon_{m\mathbf{k}}$ are the eigenstates and eigenvalues of the TB Hamiltonian defined in Eq. (\ref{eq:ham_k2}), respectively. $g_s$ is the spin degeneracy, and $p$ is the system dimension. With the polarization function obtained from the Kubo formula in Eq. (\ref{kubo_dp}) or the Lindhard function in Eq. (\ref{lin_dp}), the dielectric function can be written within the random phase approximation (RPA) as
\begin{equation}
	\epsilon(\mathbf{q},\omega) = \mathbf{1} - V(\mathbf{q}) \Pi(\mathbf{q},\omega)
	\label{eps_rpa}
\end{equation}
in which $V(\mathbf{q})$ is the Fourier transform of Coulomb interaction. For two-dimensional systems $V(\mathbf{q}) = 2\pi e^2/\kappa |\mathbf{q}|$, while for three-dimensional systems $V(\mathbf{q}) = 4\pi e^2/\kappa |\mathbf{q}|^2$, with $\kappa$ being the  background dielectric constant. The energy loss function can be obtained as
\begin{equation}
	S(\mathbf{q},\omega)= -\text{Im} \frac{1}{\epsilon(\mathbf{q},\omega)}
	\label{EEL}
\end{equation}
The energy loss function can be measured by means of electron energy loss spectroscopy (EELS). A plasmon mode with frequency $\omega_p$ and wave vector $\mathbf{q}$ is well defined when a peak exists in the $S(\mathbf{q},\omega)$ or $\epsilon(\mathbf{q},\omega)=0$ at $\omega_p$. The damping rate $\gamma$ of the mode is
\begin{equation}
    \gamma=\frac{\mathrm{Im}\; \Pi(\mathbf{q},\omega_p)}{\frac{\partial}{\partial \omega}\mathrm{Re}\;\Pi(\mathbf{q},\omega)|_{\omega=\omega_p}}
\end{equation}
and the dimensionless damping rate is
\begin{equation}
    \tilde{\gamma}=\frac{1}{\omega_p}\frac{\mathrm{Im}\; \Pi(\mathbf{q},\omega_p)}{\frac{\partial}{\partial \omega}\mathrm{Re}\;\Pi(\mathbf{q},\omega)|_{\omega=\omega_p}}
\end{equation}
The life time is defined as
\begin{equation}
    \tau=\frac{1}{\tilde{\gamma}\omega_p}
\end{equation}
All the plasmon related quantities can be calculated numerically from the functions obtained with TBPM.

\subsection{$\mathbb{Z}_2$ topological invariant}
The $\mathbb{Z}_2$ invariant characterizes whether a system is topologically trivial or nontrivial. All the two-dimensional band insulators with time-reversal invariance can be divided into two classes, i.e., the normal insulators with even $\mathbb{Z}_2$ numbers and topological insulators with odd $\mathbb{Z}_2$ numbers. In \api{TBPLaS}, we adopt the method proposed by Yu \textit{et al.} to calculate the $\mathbb{Z}_2$ numbers of a band insulator \cite{yu2011equivalent}. The main idea of the method is to calculate the evolution of the Wannier function center directly during a \textit{time-reversal pumping} process, which is a $\mathbb{Z}_2$ analog to the charge polarization. \change{The $\mathbb{Z}_2$ topological numbers can be determined as the remainder of the number of phase switching during a complete period of the time-reversal pumping process divided by 2,} which is equivalent to the $\mathbb{Z}_2$ number proposed by Fu and Kane \cite{fu2006time}. This method requires no gauge-fixing condition, thereby greatly simplifying the calculation. It can be easily applied to general systems that lack spacial inversion symmetry. 

The eigenstate of a TB Hamiltonian defined by Eq. (\ref{eq:ham_k1}) can be expressed as
\begin{equation}
    |\psi_{n\mathbf{k}}\rangle = \sum_{\alpha}g_{n\alpha}(\mathbf{k})|\chi_{\alpha\mathbf{k}}\rangle
\end{equation}
where the Bloch basis functions $|\chi_{\alpha\mathbf{k}}\rangle$ are defined in Eq. (\ref{eq:bloch_k1}). Let us take the 2D system as an example. In this case, each wave vector $\mathbf{k}_b$ defines a one-dimensional subsystem. \change{The $\mathbb{Z}_2$ topological invariant can be determined by looking at the evolution of Wannier function centers for such effective 1D system as the function of $\mathbf{k}_b$ in the subspace of occupied states.} The eigenvalue of the position operator $\hat X$ can be viewed as the center of the maximally localized Wannier functions, which is defined as
\begin{equation}
\hat{X}_P(k_b) = \begin{bmatrix}
       0           & F_{0,1} & 0        & 0       & 0     & 0 \\[0.3em]
       0           & 0       & F_{1,2}  & 0       & 0     & 0\\[0.3em]
       0           & 0       & 0        & F_{2,3} & 0     & 0\\[0.3em]
       0           & 0       & 0        & 0       & \dots & 0\\[0.3em]
       0           & 0       & 0        & 0       & 0     & F_{N_a-2,N_a-1}\\[0.3em]
       F_{N_a-1,0} & 0       & 0        & 0       & 0     & 0
     \end{bmatrix}
 \end{equation}
 where 
 \begin{equation}
     F_{i,i+1}^{nm}(\mathbf{k}_b)=\sum_{\alpha}g_{n\alpha}^{\ast}(\mathbf{k}_{a,i},\mathbf{k}_b)g_{m\alpha}(\mathbf{k}_{a,i+1},\mathbf{k}_b)
 \end{equation}
 are the $2N \times 2N$ matrices spanned in $2N$-occupied states and \change{$\mathbf{k}_{a,i}$ are the discrete $\mathbf{k}$ points sampled on the range of $\left[-\frac{1}{2}\mathbf{G}_a, \frac{1}{2}\mathbf{G}_a\right]$, with $\mathbf{G}_a$ being the reciprocal lattice vector along the $a$ axis.} We define a product of $F_{i,i+1}$ as
 \begin{equation}
     D(\mathbf{k}_b)=F_{0,1}F_{1,2}F_{2,3}\dots F_{N_a-2,N_a-1}F_{N_a-1,0}
 \label{d_matrix}
 \end{equation}
 $D(\mathbf{k}_b)$ is a $2N \times 2N$ matrix that has $2N$ eigenvalues
 \begin{equation}
     \lambda_m^D(\mathbf{k}_b) = |\lambda_m^D|\mathrm{e}^{\mathrm{i}\theta_m^D(\mathbf{k}_b)}, \qquad m=1,2,\dots,2N
 \end{equation}
 where $\theta_{m}^D(\mathbf{k}_b)$ is the phase of the eigenvalues
 \begin{equation}
     \theta_m^D(\mathbf{k}_b)=\mathrm{Im}\left[\mathrm{log}\lambda_m^D(\mathbf{k}_b) \right]
 \label{theta_D}
 \end{equation}
 The evolution of the Wannier function center for the effective 1D system with $\mathbf{k}_b$ can be obtained by looking at the phase factor $\theta_m^D$. Equation (\ref{d_matrix}) can be viewed as the discrete expression of the Wilson loop for the U(2$N$) non-Abelian Berry connection. It is invariant under the $U(2N)$ gauge transformation, and can be calculated directly from the wave functions obtained by first-principles method without choosing any gauge-fixing condition. In the $\mathbb{Z}_2$ invariant number calculations, for a particular system, \change{we calculate the evolution of the $\theta_m^D$ defined in Eq. (\ref{theta_D}) as the function of $\mathbf{k}_b$ from $\mathbf{0}$ to $\frac{1}{2}\mathbf{G}_b$, with $\mathbf{G}_b$ being the reciprocal lattice vector along the $b$ axis.} Then, we draw an arbitrary reference line parallel to the $\mathbf{k}_b$ axis, and compute the $\mathbb{Z}_2$ number by counting how many times the evolution lines of the Wannier centers cross the reference line. Note that the choice of reference line is arbitrary, but the the crossing numbers between the reference and evolution lines and the even/odd properties will not change. \change{The topological properties of three dimensional bulk materials can be determined by checking two planes in $\mathbf{k}$ space, with $\mathbf{k}_c = \mathbf{0}$ and $\mathbf{k}_c = \frac{1}{2}\mathbf{G}_c$, where $\mathbf{G}_c$ is the reciprocal lattice vector along the $c$ axis.} For more details we refer to  Ref. \cite{yu2011equivalent}

\section{Implementation}
\label{impl}

In this section, we introduce the implementation of \api{TBPLaS}, including the layout, main components, and parallelism. \api{TBPLaS} has been designed with emphasis on efficiency and user-friendliness. The performance-critical parts are written in Fortran and Cython. Sparse matrices are utilized to reduce the memory cost, which can be linked to vendor-provided math libraries like Intel$^\circledR$ MKL. A hybrid MPI+OpenMP parallelism has been implemented to exploit the modern architecture of high-performance computers. On top of the Fortran/Cython core, there is the API written in Python following an intuitive object-oriented manner, ensuring excellent user-friendliness and flexibility. Tight-binding models with arbitrary shape and boundary conditions can be easily created with the API. Advanced modeling tools for constructing hetero-structures, quasi crystals and fractals are also provided. The API also features a dedicated error handling system, which checks for illegal input and yields precise error message on the first occasion. Owing to all these features, \api{TBPLaS} can serve as not only an \textit{out-of-the-box} tight-binding package, but also a common platform for the development of advanced models and algorithms.

\subsection{Layout}
\label{layout}

\begin{figure}[h]
	\centering
	\includegraphics[width=1.0\linewidth]{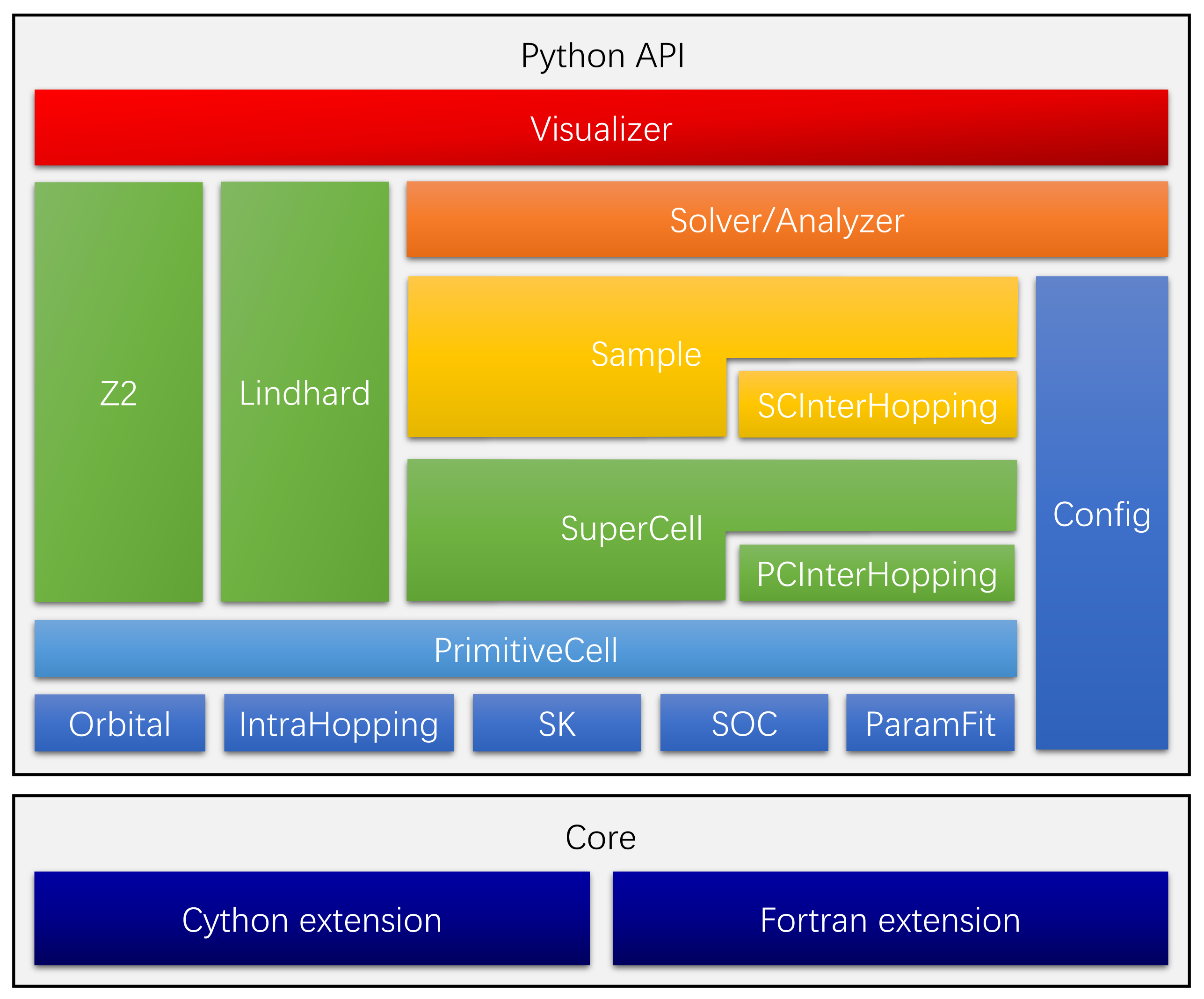}
	\caption{\change{Program layout of TBPLaS. Components of the same level in the hierarchy share the same color.}}
	\label{fig:layout}
\end{figure}

The layout of \api{TBPLaS} is shown in Fig. \ref{fig:layout}. At the root of hierarchy there are the Cython and Fortran extensions, which contain the core subroutines for building the model, constructing the Hamiltonian and performing actual calculations. The Python API consists of a comprehensive set of classes directly related to the concepts of tight-binding theory. For example, orbitals and hopping terms in a tight-binding model are represented by the \api{Orbital} and \api{IntraHopping} classes, respectively. \change{There are also auxiliary classes for setting up the orbitals and hopping terms, namely \api{SK}, \api{SOC} and \api{ParamFit}.} From the orbitals and hopping terms, as well as lattice vectors, a primitive cell can be created as an instance of the \api{PrimitiveCell} class. The goal of \api{PrimitiveCell} is to represent and solve tight-binding models of small and moderate size. Modeling tools for constructing complex primitive cells, e.g., with arbitrary shape and boundary conditions, vacancies, impurities, hetero-structures, are also available. Many properties, including band structure, DOS, dynamic polarization, dielectric function, optical conductivity \change{and $\mathbb{Z}_2$ topological invariant number} can be obtained at primitive cell level, either by calling proper functions of \api{PrimitiveCell} class, or with the help of \api{Lindhard} \change{and \api{Z2} classes.}

\api{SuperCell}, \api{SCInterHopping} and \api{Sample} are a set of classes specially designed for constructing large models from the primitive cell, especially for TBPM calculations. The computational expensive parts of these classes are written in Cython, making them extremely fast. For example, it takes less than 1 second to construct a graphene model with 1,000,000 orbitals from the \api{Sample} class on a single core of Intel$^\circledR$ Xeon$^\circledR$ E5-2690 v3 CPU. At \api{SuperCell} level the user can specify the number of replicated primitive cells, boundary conditions, vacancies, and modifier to orbital positions. Heterogenous systems, e.g., slabs with adatoms or hetero-structures with multiple layers, are modeled as separate supercells and containers (instances of the \api{SCInterHopping} class) for inter-supercell hopping terms . The \api{Sample} class is a unified interface to both homogenous and heterogenous systems, from which the band structure and DOS can be obtained via exact-diagonalization. Different kinds of perturbations, e.g., electric and magnetic fields, strain, can be specified at \api{Sample} level. Also, it is the starting point for TBPM calculations.

The parameters of TBPM calculation are stored in the \api{Config} class. Based on the sample and configuration, a solver and an analyzer can be created from \api{Solver} and \api{Analyzer} classes, respectively. The main purpose of solver is to obtain the time-dependent correlation functions, which are then analyzed by the analyzer to yield DOS, LDOS, optical conductivity, electric conductivity, Hall conductance, polarization function and quasi-eigenstates, etc. The results from TBPM calculation and exact-diagonalization at either \api{PrimitiveCell} or \api{Sample} level, can be visualized using matplotlib directly, or alternatively with the \api{Visualizer} class, which is a wrapper over matplotlib functions.

\subsection{PrimitiveCell}
\label{prim_cell}

As aforementioned in section \ref{layout}, the main purpose of \api{PrimitiveCell} class is to represent and solve tight-binding models of small and moderate size. It is also the building block for large and complex models. All calculations utilizing \api{TBPLaS} begin with creating the primitive cell. The user APIs of \api{PrimitiveCell} as well as many miscellaneous tools are summarized in Table \ref{tab:api_prim_cell}. To create the primitive cell, one needs to provide the lattice vectors, which can be generated with the \api{gen\_lattice\_vectors} function or manually specifying their Cartesian coordinates. Then the orbitals and hopping terms are added to the primitive cell with the \api{add\_orbital} and \api{add\_hopping} functions, respectively. \api{TBPLaS} utilizes the conjugate relation to reduce the hopping terms, so only half of them are needed. There are also functions to extract, modify and remove existing orbitals and hopping terms in the cell, e.g., \api{set\_orbital/get\_orbital/remove\_orbitals} and \api{get\_hopping/remove\_hopping}. Removing orbitals and hopping terms may leave dangling items in the cell. In that case, the \api{trim} function becomes useful. By default, the primitive cell is assumed to be periodic along all 3 directions. However, it can be made non-periodic along specific directions by removing hopping terms along that direction, as implemented in the \api{apply\_pbc} function. As \api{TBPLaS} utilizes the lazy evaluation technique, the \api{sync\_array} function is provided for synchronizing the array attributes after modifying the model. Once the primitive cell has been created, it can be visualized by the \api{plot} function and dumped by the \api{print} function. Geometric properties such as lattice area, volume and reciprocal lattice vectors, and electronic properties like band structure and DOS can be obtained with proper functions as listed in Table \ref{tab:api_prim_cell}. The $\mathbf k$-points required for the evaluation of band structure and DOS can be generated with the \api{gen\_kpath} and \api{gen\_kmesh} functions, respectively.

\change{\api{TBPLaS} ships with a collection of auxiliary tools for setting up the on-site energies and hopping terms. The \api{SK} class evaluates the hopping terms between atomic states up to $d$ orbitals according to the Slater-Koster formula. The \api{SOC} class evaluates the matrix element of intra-atom spin-orbital coupling term $\mathbf{L}\cdot\mathbf{S}$ in the direct product basis of $\ket{l}\otimes\ket{s}$. The \api{ParamFit} class is intended for fitting the on-site energies and hopping terms to reference data, which is typically from experiments or \textit{ab initio} calculations.}

For the user's convenience, \api{TBPLaS} provides a model repository which offers the utilities to obtain the primitive cells of popular two-dimensional materials, as summarized in Table \ref{tab:api_prim_cell}. The function \api{make\_antimonene} returns the 3-orbital or 6-orbtial primitive cell of antimonene\cite{rudenko2017electronic} depending on the inclusion of spin-orbital coupling. Diamond-shaped and rectangular primitive cells of graphene \change{based on $p_z$ orbitals} can be built with \api{make\_graphene\_diamond} and \api{make\_graphene\_rect} functions, respectively. \change{A more complicated 8-band primitive cell based on $s$, $p_x$, $p_y$ and $p_z$ orbitals can be obtained with \api{make\_graphene\_sp}.} The 4-orbital primitive cell of black phosphorus\cite{Rudenko2015} can be obtained with \api{make\_black\_phosphorus}, while the 11-orbital models of transition metal dichalcogenides\cite{Fang2015} are available with the \api{make\_tmdc} function. The primitive cell can also be created from the output of Wannier90\cite{Pizzi2020} package, namely \api{seedname.win}, \api{seedname\_centres.xyz} and \api{seedname\_hr.dat}, with the \api{wan2pc} function.

Starting from the simple primitive cell, more complex cells can be constructed through some common operations. A set of functions are provided for this purpose. \api{extend\_prim\_cell} replicates the primitive cell by given times. \api{reshape\_prim\_cell} reshapes the cell to new lattice vectors, while \api{sprical\_prim\_cell} shifts and rotates the cell with respect to c-axis, both of which are particularly useful for constructing hetero-structures. \api{make\_hetero\_layer} is a wrapper over \api{reshape\_prim\_cell} and produces one layer of the hetero-structure. \change{Inter-cell hopping terms within a hetero-structure can be searched with the \api{find\_neighbors} function and managed with the \api{PCInterHopping} class.} Finally, all the layers and intercell hopping terms can be merged into one cell by the \api{merge\_prim\_cell} function. Note all these functions work at \api{PrimitiveCell} level, i.e., they either return a new primitive cell, or modify an existing one.

\begin{table}
	\begin{center}
		\caption{\change{User APIs of \api{PrimitiveCell}, \api{SK}, \api{SOC}, \api{ParamFit}, \api{PCInterHopping} classes and and miscellaneous tools.}}
		\label{tab:api_prim_cell}
		\begin{tabular}{clp{8cm}}
			\hline\hline
			\multicolumn{1}{c}{Category} & \multicolumn{1}{c}{API} & \multicolumn{1}{c}{Purpose} \\
			\hline
			\multirow{17}{*}{PrimitiveCell} & add\_orbital & Add a new orbital \\
			                                & set\_orbital & Modify an existing orbital \\
			                                & get\_orbital & Retrieve an existing orbital \\
			                                & remove\_orbitals & Remove selected orbitals \\
			                                & add\_hopping & Add a new or modify an existing hopping term \\
			                                & get\_hopping & Retrieve an existing hopping term \\
			                                & remove\_hopping & Remove an existing hopping term \\
			                                & trim & Remove dangling orbitals and hopping terms \\
			                                & apply\_pbc & Modify the boundary conditions \\
			                                & sync\_array & Synchronize the array attributes \\
			                                & get\_lattice\_area & Calculate the area spanned by lattice vectors \\
			                                & get\_lattice\_volume & Calculate the volume spanned by lattice vectors \\
			                                & get\_reciprocal\_vectors & Calculate reciprocal lattice vectors \\
			                                & calc\_bands & Calculate band structure of the primitive cell\\
			                                & calc\_dos & Calculate DOS and LDOS of the primitive cell \\
			                                & plot & Plot the primitive cell to the screen or file \\
			                                & print & Print orbital and hopping terms \\
			\hline
			\change{SK} & \change{eval} & \change{Evaluate hopping term with Slater-Koster formula} \\
			\hline
			\change{SOC} & \change{eval} & \change{Evaluate matrix element of $\mathbf{L}\cdot\mathbf{S}$ in direct product basis} \\
			\hline
			\change{ParamFit} & \change{fit} & \change{Fit on-site energies and hopping terms to reference data} \\
			\hline
			PCInterHopping & add\_hopping & Add a new inter-cell hopping term \\
			\hline
			\multirow{6}{*}{Lattice and k-points} & gen\_lattice\_vectors & Generate lattice vectors from lattice constants \\
			                         & rotate\_coord & Rotate Cartesian coordinates \\
			                         & cart2frac & Convert coordinates from Cartesian to fractional \\
			                         & frac2cart & Convert coordinates from fractional to Cartesian \\
			                         & gen\_kpath & Generate path connecting highly-symmetric $\mathbf k$-points \\
			                         & gen\_kmesh & Generate a mesh grid in the first Brillouin zone \\
			\hline
			\multirow{7}{*}{Model repository} & make\_antimonene & Get the primitive cell of antimonene \\
			                                  & make\_graphene\_diamond & Get the diamond-shaped primitive cell of graphene \\
			                                  & make\_graphene\_rect & Get the rectangular primitive cell of graphene \\
			                                  & \change{make\_graphene\_sp} & \change{Get the 8-band primitive cell of graphene} \\
			                                  & make\_black\_phosphorus & Get the primitive cell of black phosphorus \\
			                                  & make\_tmdc & Get the primitive cells of transition metal dichalcogenides \\
			                                  & wan2pc & Create primitive cell from the output of Wannier90 \\
			\hline
			\multirow{6}{*}{Modeling tools} & extend\_prim\_cell & Replicate the primitive cell \\
			                                & reshape\_prim\_cell & Reshape primitive cell to new lattice vectors \\
			                                & spiral\_prim\_cell & Rotate and shift primitive cell \\
			                                & make\_hetero\_layer & Produce one layer of hetero-structure \\
			                                & \change{find\_neighbors} & \change{Find neighboring orbital pairs up to cutoff distance} \\
			                                & merge\_prim\_cell & Merge primitive cells and inter-cell hopping terms \\
			\hline\hline
		\end{tabular}
	\end{center}
\end{table}

\subsection{Lindhard}
\label{lindhard}

The \api{Lindhard} class evaluates response properties, i.e., dynamic polarization, dielectric function and optical conductivity of primitive cell with the help of Lindhard function. The user APIs of this class is summarized in Table \ref{tab:api_lind_z2}. To instantiate a \api{Lindhard} object, one needs to specify the primitive cell, energy range and resolution, dimension of $\mathbf k$-grid in the first Brillouin zone, system dimension, background dielectric constant and many other quantities. Since dynamic polarization and dielectric function are $\mathbf q$-dependent, three types of coordinate systems are provided to effectively represent the $\mathbf q$-points: Cartesian coordinate system in unit of $\mathrm{\AA}^{-1}$ or $\mathrm{nm}^{-1}$, fractional coordinate system in unit of reciprocal lattice vectors, and grid coordinate system in unit of dimension of $\mathbf k$-grid. Grid coordinate system is actually a variant of the fractional coordinate system. Conversion between coordinate systems can be achieved with the \api{frac2cart} and \api{cart2frac} functions.

\api{Lindhard} class offers two functions to calculate the dynamic polarization: \api{calc\_dyn\_pol\_regular} and \api{calc\_dyn\_pol\_arbitrary}. Both functions require an array of $\bold{q}$-points as input. The difference is that \api{calc\_dyn\_pol\_arbitrary} accepts arbitrary $\bold{q}$-points, while \api{calc\_dyn\_pol\_regular} requires that the $\bold{q}$-points should be on the uniform $\bold{k}$-grid in the first Brillouin zone. This is due to the term $\bold{k’=k+q}$ that appears in the Lindhard function. For regular $\bold{q}$ on $\bold{k}$-grid, $\bold{k^{\prime}}$ is still on the same grid. However, this may not be true for arbitrary $\bold{q}$-points. So, \api{calc\_dyn\_pol\_arbitrary} keeps two sets of energies and wave functions, for $\bold{k}$ and $\bold{k^{\prime}}$ grids respectively, although they may be equivalent via translational symmetry. On the contrary, \api{calc\_dyn\_pol\_regular} utilizes translational symmetry and reuses energies and wave functions when possible. So, \api{calc\_dyn\_pol\_regular} uses less computational resources, at the price that only regular $\bold{q}$-points on $\bold{k}$-grid can be taken as input. From the dynamic polarization, dielectric function can be obtained by \api{calc\_epsilon}. Unlike dynamic polarization and dielectric function, the optical conductivity considered in \api{TBPLaS} does not depend on $\bold{q}$-points. So, it can be evaluated directly by \api{calc\_ac\_cond}.

\begin{table}
    \begin{center}
    \caption{\change{User APIs of \api{Lindhard} and \api{Z2} classes.}}
    \label{tab:api_lind_z2}
    \centering
    \begin{tabular}{clp{8cm}}
		\hline\hline
		\multicolumn{1}{c}{Category} & \multicolumn{1}{c}{API} & \multicolumn{1}{c}{Purpose} \\
		\hline
		\multirow{4}{*}{Lindhard} & calc\_dyn\_pol\_regular & Calculate dynamic polarization for regular $\mathbf q$-points \\
			                      & calc\_dyn\_pol\_arbitrary & Calculate dynamic polarization for arbitrary $\mathbf q$-points \\
			                      & calc\_epsilon & Calculate dielectric function \\
			                      & calc\_ac\_cond & Calculate optical conductivity \\
		\hline
		\multirow{3}{*}{\change{Z2}} & \change{calc\_phases} & \change{Calculate phases $\theta_m^D$} \\
			                         & \change{reorder\_phases} & \change{Reorder phases improve continuity and smoothness} \\
			                         & \change{count\_crossing} & \change{Count crossing number of phases against reference line} \\
		\hline\hline
    \end{tabular}
    \end{center}
\end{table}

\subsection{\change{Z2}}
\label{z2}

\change{The \api{Z2} class evaluates and analyzes the topological phases $\theta_m^D$ to yield the $\mathbb{Z}_2$ number. The APIs of this class are summarized in Table \ref{tab:api_lind_z2}. To create a \api{Z2} calculator, the primitive cell, as well as the number of occupied bands should be provided as input. The phases $\theta_m^D$ can be obtained as the function of $\mathbf{k}_b$ with the \api{calc\_phases} function, which can then be plotted with scatter plot to count the crossing number against a reference line. If there are too many occupied states, it may be difficult to determine the crossing number with human eyes. The \api{count\_crossing} function can count the crossings automatically, provided that the phases have been correctly reordered with the \api{reorder\_phases} function. Anyway, the users are \textit{strongly} recommended to cross-validate the crossing numbers from scatter plot and \api{count\_crossing}, respectively. Finally, the $\mathbb{Z}_2$ number is determined as the remainder of crossing number divided by 2.}

\subsection{SuperCell, SCInterHopping and Sample}
\label{super_cell}

The tools discussed in section \ref{prim_cell} are sufficiently enough to build complex models of small and moderate size. However, there are occasions where large models are essential, e.g., hetero-structures with twisted layers, quasi crystals, distorted structures, etc. In particular, TBPM calculations require large models for numerical stability. To build and manipulate large models efficiently, a new set of classes, namely \api{SuperCell}, \api{SCInterHopping} and \api{Sample} are provided. The APIs of these classes are summarized in Table \ref{tab:api_super_cell}.

The purpose of \api{SuperCell} class is to represent homogenous models that are formed by replicating the primitive cell. To create a supercell, the primitive cell, supercell dimension and boundary conditions are required. Vacancies can be added to the supercell upon creation, or through the \api{add\_vacancies} and \api{set\_vacancies} functions afterwards. Modifications to the hopping terms can be added by the \api{add\_hopping} function. If the hopping terms are already included in the supercell, the original values will be overwritten. Otherwise, they will be added to the supercell as new terms. The supercell can be assgined with an orbital position modifier with the \api{set\_orb\_pos\_modifier} function, which is a Python function modifying the orbital positions \textit{in-place}. Dangling orbitals and hopping terms in the supercell can be removed by the \api{trim} function. Orbital positions, on-site energies, hopping terms and distances, as well as many properties of the supercell cell can be obtained with the \api{get\_xxx} functions, as listed in Table \ref{tab:api_super_cell}. \api{TBPLaS} utilizes the conjugate relation to reduce the hopping terms, so only half of them are returned by \api{get\_hop} and \api{get\_dr}.

Heterogenous systems, e.g., slabs with adatoms or hetero-structures with multiple layers, are modelled as separate supercells and containers for inter-supercell hopping terms. The containers are created from the \api{SCInterHopping} class, with a \textit{bra} supercell and a \textit{ket} supercell, between which the hopping terms can be added by the \api{add\_hopping} function. The \api{SCInterHopping} class also implements the \api{get\_hop} and \api{get\_dr} functions for extracting the hopping terms and distances, similar to the \api{SuperCell} class.

The \api{Sample} class is a unified interface to both homogenous and heterogenous systems. A sample may consist of single supercell, or multiple supercells and inter-supercell hopping containers. The on-site energies, orbital positions, hopping terms and distances are stored in the attributes of \api{orb\_eng}, \api{orb\_pos}, \api{hop\_i}, \api{hop\_j}, \api{hop\_v} and \api{dr}, respectively, which are all numpy arrays. To reduce the memory usage, these attributes are filled only when needed with the initialization functions. Different kinds of perturbations, e.g., electric and magnetic fields, strain, can be specified by directly calling the API, or manipulating the array attributes directly. The \api{reset\_array} function is provided to reset the array attributes of the sample, for removing the effects of perturbations. Band structure and DOS of the sample can be obtained with \api{calc\_bands} and \api{calc\_dos} respectively, similar to the \api{PrimitiveCell} class. Visualization is achieved through the \api{plot} function. Since the sample is typically large, its response properties are no longer accessible via the Lindhard function. On the contrary, TBPM is much more efficient for large samples. Since the Chebyshev polynomial decomposition of Hamiltonian requires its eigenvluates to be within [-1, 1], an API \api{rescale\_ham} is provided for this purpose. Details on TBPM will be discussed in the next section.

\begin{table}
	\begin{center}
		\caption{User APIs of \api{SuperCell}, \api{SCInterHopping} and \api{Sample} classes.}
		\label{tab:api_super_cell}
		\begin{tabular}{clp{7cm}}
			\hline\hline
			\multicolumn{1}{c}{Category} & \multicolumn{1}{c}{API} & \multicolumn{1}{c}{Purpose} \\
			\hline
			\multirow{13}{*}{SuperCell} & add\_vacancies & Add a list of vacancies to the supercell \\
			                            & set\_vacancies & Reset the list of vacancies \\
			                            & add\_hopping & Add a modification to the hopping terms \\
			                            & set\_orb\_pos\_modifier & Assign an orbital position modifier to the supercell \\
			                            & trim & Remove dangling orbitals and hopping terms \\
			                            & sync\_array & Synchronize the array attributes \\
			                            & get\_orb\_pos & Get the Cartesian coordinates of orbitals \\
			                            & get\_orb\_eng & Get the on-site energies \\
			                            & get\_hop & Get the hopping terms \\
			                            & get\_dr & Get the hopping distances \\
			                            & get\_lattice\_area & Calculate the area spanned by lattice vectors \\
			                            & get\_lattice\_volume & Calculate the volume spanned by lattice vectors \\
			                            & get\_reciprocal\_vectors & Calculate reciprocal lattice vectors \\
			\hline
			\multirow{3}{*}{SCInterHopping} & add\_hopping & Add a new inter-supercell hopping term \\
						                    & get\_hop & Get the hopping terms \\
			                                & get\_dr & Get the hopping distances \\
			\hline
			\multirow{10}{*}{Sample} & init\_orb\_eng & Initialize on-site energies on demand \\
			                         & init\_orb\_pos & Initialize orbital positions on demand \\
			                         & init\_hop & Initialize hopping terms on demand \\
			                         & init\_dr & Initialize hopping distances on demand \\
			                         & reset\_array & Reset the array atributes \\
			                         & rescale\_ham & Rescale the Hamiltonian \\
			                         & set\_magnetic\_field & Apply a perpendicular magnetic field \\
			                         & calc\_bands & Calculate band structure of the sample \\
			                         & calc\_dos & Calculate DOS and LDOS of the sample \\
			                         & plot & Plot the sample to the screen or file \\
			\hline\hline
		\end{tabular}
	\end{center}
\end{table}

\subsection{Config, Solver, Analyzer and Visualizer}
\label{config}

TBPM in \api{TBPLaS} is implemented in the classes of \api{Config}, \api{Solver} and \api{Analyzer}. \api{Config} is a simple container class holding all the parameters that controls the calculation. So, it has no API but a few Python dictionaries as attributes. The \api{Solver} class propagates the wave function and evaluates the correlation functions, which are then analyzed by \api{Analyzer} class to produce the results, including DOS, LDOS, optical conductivity, electric conductivity, etc. The user APIs of \api{Solver} and \api{Analyzer} are summarized in Table \ref{tab:api_config}. To create a solver or analyzer, one needs the sample and the configuration object. The APIs of \api{Solver} and \api{Analyzer} share a common naming convention, where \api{calc\_corr\_xxx} calculates the correlation function for property \textit{xxx} and \api{calc\_xxx} analyzes the correlation function to yield the final results. Some of the properties, such as LDOS from Green's function and time-dependent wave function, can be obtained from \api{Solver} directly without further analysis.

The \api{Visualizer} class is a thin wrapper over matplotlib for quick visualization of the results from exact-diagonalization and TBPM. Generic data, e.g., response functions, can be plotted with the \api{plot\_xy} function. There are also special functions to plot the band structure, DOS \change{and topological phases.} Quasi-eigenstates and time-dependent wave function can be plotted with the \api{plot\_wfc} function. Although \api{Visualizer} is intended for quick visualization, it can be easily extended to produce figures of publication quality, according to the user’s needs.

\begin{table}
	\begin{center}
		\caption{User APIs of \api{Solver}, \api{Analyzer}, \api{Visualizer} classes.}
		\label{tab:api_config}
		\begin{tabular}{clp{8.5cm}}
			\hline\hline
			\multicolumn{1}{c}{Category} & \multicolumn{1}{c}{API} & \multicolumn{1}{c}{Purpose} \\
			\hline
			\multirow{11}{*}{Solver} & set\_output & Prepare output directory and files \\
			                         & save\_config & Save configuration to file \\
			                         & calc\_corr\_dos & Calculate correlation function of DOS \\
			                         & calc\_corr\_ldos & Calculate correlation function of LDOS \\
			                         & calc\_corr\_dyn\_pol & Calculate correlation function of dynamical polarization \\
			                         & calc\_corr\_ac\_cond & Calculate correlation function of optical conductivity \\
			                         & calc\_corr\_dc\_cond & Calculate correlation function of electric conductivity \\
			                         & calc\_hall\_mu & Calculate $\mu_{mn}$ required for the evaluation of Hall conductivity using Kubo-Bastin formula \\
			                         & calc\_quasi\_eigenstates & Calculate quasi-eigenstates of given energies \\
			                         & calc\_ldos\_haydock & Calculate LDOS using Green's function \\
			                         & calc\_wfc\_t & Calculate propagation of wave function from given initial state \\
			\hline
			\multirow{8}{*}{Analyzer} & calc\_dos & Calculate DOS from its correlation function \\
			                          & calc\_ldos & Calculate LDOS from its correlation function \\
			                          & calc\_dyn\_pol & Calculate dynamic polarization from its correlation function \\
			                          & calc\_epsilon & Calculate dielectric function from dynamic polarization \\
			                          & calc\_ac\_cond & Calculate optical conductivity from its correlation function \\
			                          & calc\_dc\_cond & Calculate electric conductivity from its correlation function \\
			                          & calc\_diff\_coeff & Calculate diffusion coefficient from DC correlation function \\
			                          & calc\_hall\_cond & Calculate Hall conductivity from $\mu_{mn}$ \\
			\hline
			\multirow{5}{*}{Visualizer} & plot\_xy & Plot generic data of y against x \\
										& plot\_bands & Plot band structure \\
										& plot\_dos & Plot DOS \\
										& \change{plot\_phases} & \change{Plot phases $\theta_m^D$} \\
										& plot\_wfc & Plot quasi-eigenstate or time-dependent wave function in real space \\
			\hline\hline
		\end{tabular}
	\end{center}
\end{table}

\subsection{Parallelization}
\label{parallel}

Tight-binding calculations can be time-consuming when the model is large, or when ultra-fine results are desired. For example, band structure, DOS, \change{response properties from Lindhard function and topological phases from \api{Z2} require exact diagonalization for a dense $\mathbf k$-grid in the first Brillouin zone,} optionally followed by post-processing on an energy grid. TBPM calculations require large models and averaging over multiple samples to converge the results, while the time-propagation of each sample involves heavy matrix-vector multiplications. Consequently, dedicated parallelism that can exploit the modern hardware of computers are essential to promote the application of tight-binding techniques to millions or even billions of orbitals. However, the Global Interpreter Lock (GIL) of Python allows only one thread to run at one time, severely hinders the parallelization at thread level. Although the GIL can be bypassed with some tricks, thread-level parallelization is restricted to only one computational node. \api{TBPLaS} tackles these problems with a hybrid MPI+OpenMP parallelism. Tasks are firstly distributed over MPI processes that can run among multiple nodes. Since the processes are isolated mutually at operation system level and each keeps a local copy of the data, there is no need to worry about data conflicts and GIL. For the tasks assigned to each process, thread-level parallelism is implemented with OpenMP in the Cython and Fortran extensions. With a wise choice of the numbers of processes and threads, excellent scaling can be achieved with respect to the computational resources. Both MPI and OpenMP of the hybrid parallelism can be enabled or disabled separately, ensuring good flexibility.

\subsubsection{Band structure and DOS}
\label{para_bands}

For calculating the band structure, $\bold{k}$-points are firstly distributed over MPI processes, with each process dealing with some of the $\bold{k}$-points. For each $\bold{k}$-point assigned to the process, the Hamiltonian matrix has to be built in serial, while the diagonalization is further parallelized with OpenMP in the NumPy and SciPy libraries, which call OpenBLAS or MKL under the hood. The evaluation of DOS consists of getting the eigenvalues for a dense $\bold{k}$-grid, and a summation over the eigenvalues to collect the contributions following Eq. (\ref{eq:dos_diag}). Getting the eigenvalues is parallelized in the same approach as the band structure. The summation is parallelized with respect to the $\bold{k}$-points over MPI processes. Local data on each process is then collected via the \api{MPI\_Allreduce} call.

\subsubsection{Response properties from Lindhard function}
\label{para_lindhard}

Evaluation of response properties using Lindhard function is similar to that of DOS, which also consists of getting the eigenvalues and eigenvectors and subsequent post-processing. However, the post-processing is much more expensive than DOS. Taking the optical conductivity for example, whose formula follows Eq. (\ref{eq:ac_diag}). To reuse the intermediate results, we define the following arrays
\begin{equation}
    \Delta\epsilon(\bold k, m, n) = \epsilon_{m\bold k}-\epsilon_{n\bold k}
\end{equation}
and
\begin{equation}
    P(\bold k, m, n) = \frac{f_{m\bold{k}} - f_{n\bold{k}}}{\epsilon_{m\bold{k}} - \epsilon_{n\bold{k}}} \langle\psi_{n\bold{k} }|v_\alpha|\psi_{m\bold{k}}\rangle \langle\psi_{m\bold{k}}|v_\beta|\psi_{n\bold{k}}\rangle
\end{equation}
The evaluation of $\Delta\epsilon$ and $P$ are firstly parallelized with respect to $\bold k$ over MPI processes. For each process, tasks are further parallelized with respect to $m$ over OpenMP threads. Once the arrays are ready, the optical conductivity can be calculated as
\begin{equation}
    \sigma_{\alpha\beta}(\hbar\omega)=\frac{\mathrm{i} e^2 \hbar}{N_\bold{k} \Omega_c} \sum_{\bold{k}}\sum_{m,n} \frac{P(\bold k, m, n)}{\Delta\epsilon(\bold k, m, n)-(\hbar\omega+\mathrm i\eta^+)}
\end{equation}
Typically, the response properties are evaluated on a discrete frequency grid $\{\omega_i\}$. We firstly distribute $\bold k$-points over MPI processes, then distribute the frequencies over OpenMP threads. Final results are collected by MPI calls, similar to the evaluation of DOS.

\subsubsection{\change{Z2}}

\change{The evaluation of topological phases $\theta_m^D$ according to Eq. (\ref{theta_D}) can be done for each $\mathbf{k}_b$ individually. So, tasks are distributed among MPI process with respect to $\mathbf{k}_b$. For given $\mathbf{k}_b$, the $D(\mathbf{k}_b)$ matrix is evaluated in serial mode by iterative matrix multiplication according to Eq. (\ref{d_matrix}). Then it is diagonalized to yield the eigenvectors $\lambda_m^D$, from which the phases $\theta_m^D$ can be extracted. Finally the results are collected with MPI calls.}

\subsubsection{TBPM}
\label{para_tbpm}

The TBPM calculations follow a common procedure. Firstly, the time-dependent wave function is propagated from different initial conditions and correlation functions are evaluated at each time-step. Then the correlation functions are averaged and analyzed to yield the final results. The averaging and analysis are cheap and need no parallelization. The propagation of wave function, on the contrary, is much more expensive and must be parallelized. Fortunately, propagation from each initial condition is embarrassingly parallel task, i.e., it can be split into individual sub-tasks that do not exchange data mutually. So, the initial conditions are distributed among MPI processes. The propagation of wave function, according to Eq. (\ref{eq:psi_t}), involves heavy matrix-vector multiplications. In \api{TBPLaS} the matrices are stored in Compressed Sparse Row (CSR) format, significantly reducing the memory cost. The multiplication, as well as many other matrix operations, are parallelized with respect to matrix elements among OpenMP threads. Averaging of correlation functions is also done by MPI calls.

\section{Usage}
\label{usage}

In this section we demonstrate the installation and usages of \api{TBPLaS}. \api{TBPLaS} is released under the BSD license, which can be found at \url{https://opensource.org/licenses/BSD-3-Clause}. The source code is available at the home page \url{www.tbplas.net}. Detailed documentation and tutorials can also be found there.

\subsection{Installation}
\label{install}

\subsubsection{Prerequisites}
\label{prereq}

To install and run \api{TBPLaS}, a Unix-like operating system is required. You need both C and Fortran compilers, as well as vendor-provided math libraries if they are available. For Intel$^\circledR$ CPUs, it is better to use Intel compilers and Math Kernel Library (MKL). If Intel toolchain is not available, the GNU Compiler Collection (GCC) is another choice. In that case, the built-in math library will be enabled automatically.

\api{TBPLaS} requires a Python3 environment (interpreter and development files), and the packages of NumPy, SciPy, Matplotlib, Cython, Setuptools as mandatory dependencies. Optionally, the LAMMPS interface requires the ASE package. If MPI+OpenMP hybrid parallelism is to be enabled, the MPI4PY package and an MPI implementation, e.g., Open MPI or MPICH, become essential. Most of the packages can be installed via the pip command, or manually from the source code.

\subsubsection{Installation}
\label{install_actual}

The configuration of compilation is stored in \api{setup.cfg} in the top directory of the source code of \api{TBPLaS}. Examples of this file can be found in the \api{config} directory. You should adjust it according to your computer’s hardware and software settings. Here is an example utilizing Intel compilers and MKL
\begin{lstlisting}
[config_cc]
compiler = intelem

[config_fc]
fcompiler = intelem
arch = -xHost
opt = -qopenmp -O3 -ipo -heap-arrays 32
f90flags = -fpp -DMKL -mkl=parallel

[build_ext]
include_dirs = /software/intel/parallelstudio/2019/compilers_and_libraries/linux/mkl/include
library_dirs = /software/intel/parallelstudio/2019/compilers_and_libraries/linux/mkl/lib/intel64
libraries = mkl_rt iomp5 pthread m dl
\end{lstlisting}
The \api{config\_cc} and \api{config\_fc} sections contain the settings of C and Fortran compilers, while the libraries are configured in \api{build\_ext}. It is important that OpenMP should be enabled by adding proper flags to \api{config\_fc} and \api{build\_ext}, e.g., \api{-qopenmp} in \api{opt} and \api{iomp5} in \api{libraries} for Intel compilers. Here is another example utilizing GCC and the built-in math library
\begin{lstlisting}
[config_cc]
compiler = unix

[config_fc]
fcompiler = gfortran
arch = -march=native
opt = -fopenmp -O3 -mtune=native
f90flags = -fno-second-underscore -cpp

[build_ext]
libraries = gomp
\end{lstlisting}
where the OpenMP flags become \api{-fopenmp} and \api{gomp}.

Once \api{setup.cfg} has been properly configured, \api{TBPLaS} can be compiled with \api{python setup.py build}. If everything goes well, a new \api{build} directory will be created, which contains the Cython and Fortran extensions. The installation into default path is done by \api{python setup.py install}. After that, invoke the Python interpreter and try \api{import tbplas}. If no error occurs, then the installation of \api{TBPLaS} is successful.

\subsection{Overview of the workflow}
\label{workflow}

\begin{figure}[h]
	\centering
	\includegraphics[width=1.0\linewidth]{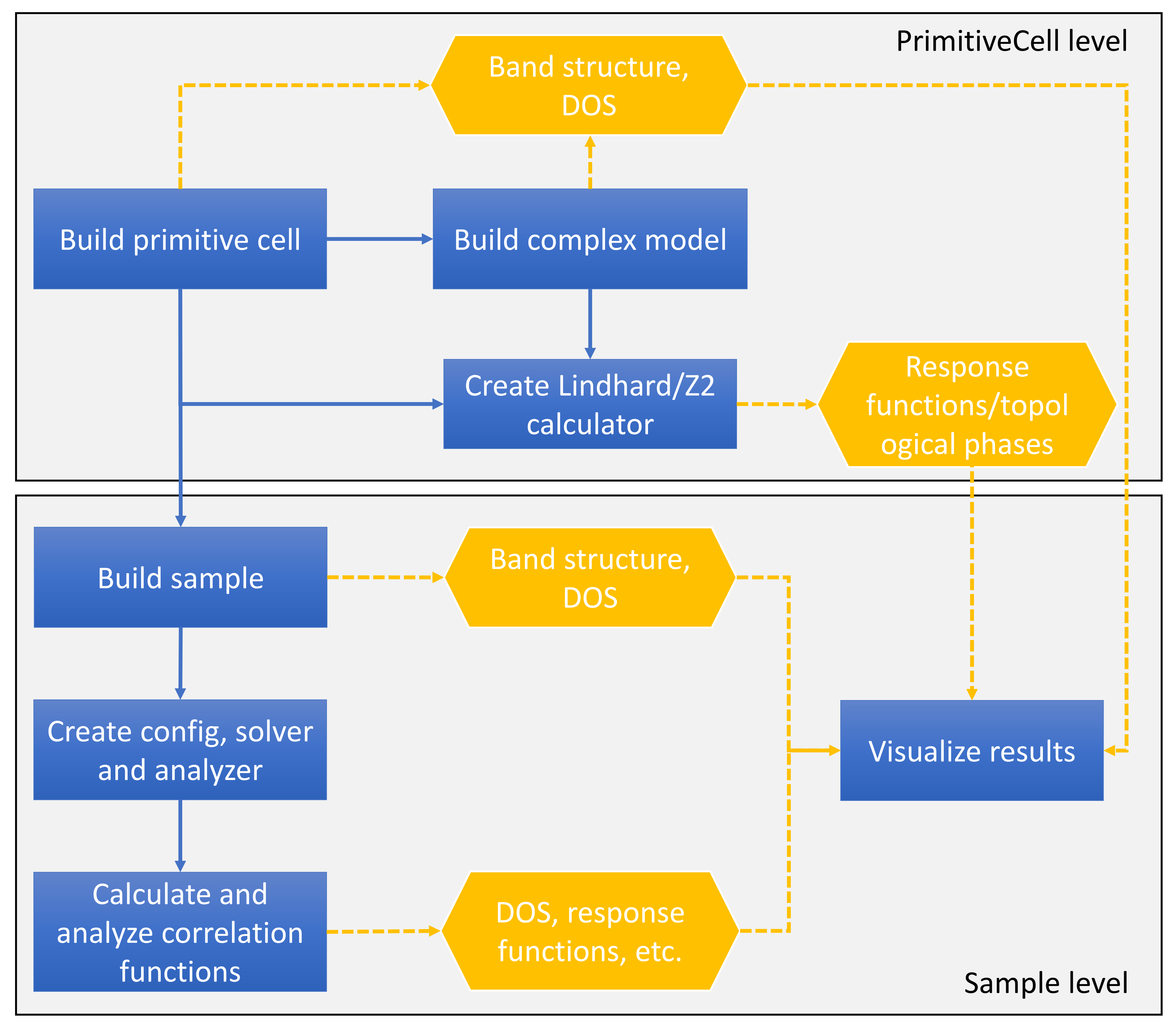}
	\caption{\change{Workflow of common usages of TBPLaS. Blue rectangles and orange hexagons denote the main steps and outputs, respectively.}}
	\label{fig:workflow}
\end{figure}

The workflow of common usages of \api{TBPLaS} is summarized in Fig. \ref{fig:workflow}. Tight-binding models can be created at either \api{PrimitiveCell} or \api{Sample} level, depending on the model size and purpose. \api{PrimitiveCell} is recommended for models of small and moderate size, and is essential for evaluating response functions utilizing the Lindhard function \change{or topological variants with the \api{Z2} class.} On the contrary, \api{Sample} is for extra-large models that may consist of millions or trillions of orbitals. Also, TBPM calculations require the model to be an instance of the \api{Sample} class. For a detailed comparison of \api{PrimitiveCell} and \api{Sample}, refer to section \ref{impl}.

Generally, all calculations utilizing \api{TBPLaS} begin with creating the primitive cell, which involves creating an empty cell from the lattice vectors, adding orbitals and adding hoping terms. Complex models, e.g., that with arbitrary shape and boundary conditions, vacancies, impurities and hetero-structures can be constructed from the simple primitive cell with the Python-based modeling tools, as discussed in section \ref{prim_cell}. Band structure and DOS of the primitive cell can be obtained via exact diagonalization with the \api{calc\_bands} and \api{calc\_dos} functions, respectively. Response functions such dynamic polarization, dielectric function and optical conductivity, need an additional step of creating a \api{Lindhard} calculator, followed by calling the corresponding functions. \change{Similar procedure applies to the topological properties, where a \api{Z2} calculator should be created and utilized.}

To build a sample, the user needs to construct a supercell with the Cython-based modeling tools. Heterogenous systems are modeled as separate supercells plus containers  for inter-supercell hopping terms. The sample is then formed by assembling the supercells and containers. Band structure and DOS of the sample can be obtained via exact diagonalization in the same approach as the primitive cell. However, these calculations may be extremely slow due to the large size of the model. In that case, TBPM is recommended. The user needs to setup the parameters using the \api{Config} class, and create a solver and an analyzer from \api{Solver} and \api{Analyzer} classes, respectively. Then evaluate and analyze the correlation functions to yield the DOS, response functions, quasieigenstates, etc. Finally, the results can be visualized using the \api{Visualizer} class, or the matplotlib library directly.

\subsection{Building the primitive cell}
\label{prim_cell_build}

In this section we show how to build the primitive cell taking monolayer graphene as the example. Monolayer graphene has lattice constants of $a=b=2.46\;\mathrm{\AA}$ and $\alpha=\beta=90^\circ$. The lattice angle $\gamma$ can be either $60^\circ$ or $120^\circ$, depending on the choice of lattice vectors. Also, we need to specify an arbitrary cell length $c$ since \api{TBPLaS} internally treats all models as three-dimensional. We will take $\gamma=60^\circ$ and $c=10\;\mathrm{\AA}$ . First of all, we need to invoke the Python interpreter and import all necessary packages
\begin{lstlisting}
import math
import numpy as np
import tbplas as tb
\end{lstlisting}
Then we generate the lattice vectors from the lattice constants with the \api{gen\_lattice\_vectors} function
\begin{lstlisting}
vectors = tb.gen_lattice_vectors(a=2.46, b=2.46, c=10.0, gamma=60)
\end{lstlisting}
The function accepts six arguments, namely \api{a}, \api{b}, \api{c}, \api{alpha}, \api{beta}, and \api{gamma}. The default value for \api{alpha} and \api{beta} is 90 degrees, if not specified. The return value \api{vectors} is a $3\times3$ array containing the Cartesian coordinates of the lattice vectors. Alternatively, we can create the lattice vectors from their Cartesian coordinates directly
\begin{lstlisting}
a = 2.46
c = 10.0
a_half = a * 0.5
sqrt3 = math.sqrt(3)

vectors = np.array([
    [a, 0, 0,],
    [a_half, sqrt3*a_half, 0],
    [0, 0, c]
])
\end{lstlisting}
From the lattice vectors, we can create an empty primitive cell by
\begin{lstlisting}
prim_cell = tb.PrimitiveCell(vectors, unit=tb.ANG)
\end{lstlisting}
where the argument \api{unit} specifies that the lattice vectors are in Angstroms.

\begin{figure}[h]
	\centering
	\includegraphics[width=1.0\linewidth]{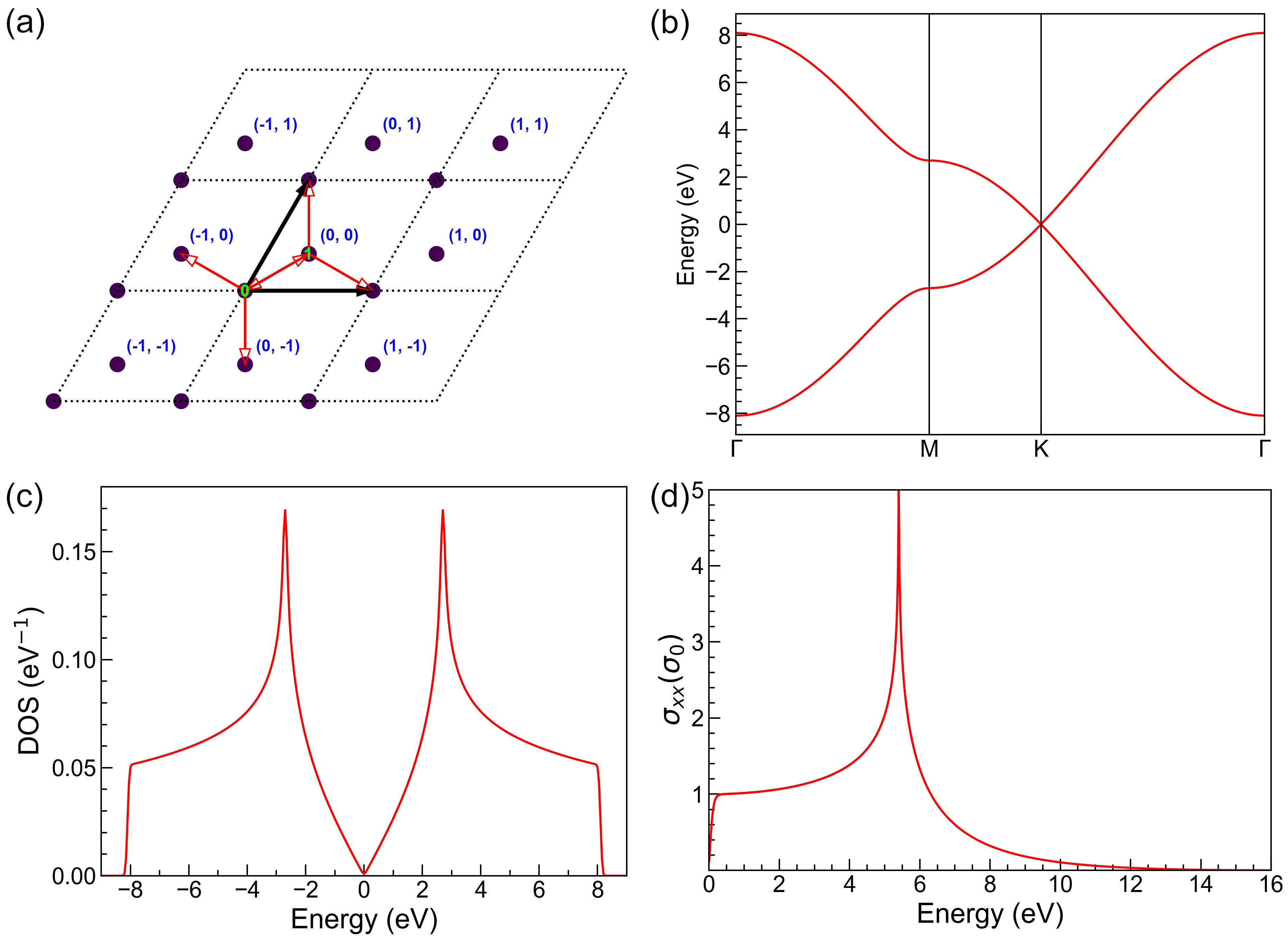}
	\caption{(a) Schematic plot of the primitive cell of monolayer graphene. Orbitals are shown as filled circles and numbered in green texts, while cells are indicated with dashed diamonds and numbered in blue texts. Thick black arrows denote the lattice vectors.  (b) Band structure, (c) DOS and (d) Optical conductivity of monolayer graphene. The optical conductivity is in the unit of $\sigma_0 = \frac{e^2}{4\hbar}$.}
	\label{fig:prim_cell}
\end{figure}

As we choose $\gamma=60^\circ$, the two carbon atoms are then located at $\tau_0 = \mathbf 0$ and $\tau_1 = \frac{1}{3}\mathbf a_1 + \frac{1}{3}\mathbf a_2$, as shown in Fig. \ref{fig:prim_cell} (a). In the simplest 2-band model of graphene, each carbon atom carries one $2p_z$ orbital. We can add the orbitals with the \api{add\_orbital} function
\begin{lstlisting}
prim_cell.add_orbital([0., 0.], energy=0.0, label="pz")
prim_cell.add_orbital([1./3, 1./3], energy=0.0, label="pz")
\end{lstlisting}
The first argument gives the position of the orbital, while \api{energy} specifies the on-site energy, which is assumed to be 0 eV if not specified. In absence of strain or external fields, the two orbitals have equal on-site energies. The argument \api{label} is a tag to denote the orbital. In addition to fractional coordinates, the orbitals can also be added using Cartesian coordinates by the \api{add\_orbital\_cart} function
\begin{lstlisting}
prim_cell.add_orbital_cart([0., 0.], unit=tb.ANG, energy=0.0, label="pz")
prim_cell.add_orbital_cart([1.23, 0.71014083], unit=tb.ANG, energy=0.0, label="pz")
\end{lstlisting}
Here we use the argument \api{unit} to specify the unit of Cartesian coordinates.

When all the orbitals have been added to the primitive cell, we can proceed with adding the hopping terms, which are defined as
\begin{equation}
    t_{ij}(\mathbf R) = \langle \phi_{i\mathbf 0} \vert \hat{h}_0 \vert \phi_{j\mathbf R} \rangle
\end{equation}
where $\mathbf R$ is the index of neighbouring cell, $i$ and $j$ are orbital indices, respectively. The hopping terms of monolayer graphene in the nearest approximation are
\begin{itemize}
	\item $\mathbf R = (0, 0), i=0, j=1$
	\item $\mathbf R = (0, 0), i=1, j=0$
	\item $\mathbf R = (1, 0), i=1, j=0$
	\item $\mathbf R = (-1, 0), i=0, j=1$
	\item $\mathbf R = (0, 1), i=1, j=0$
	\item $\mathbf R = (0, -1), i=0, j=1$
\end{itemize}
With the conjugate relation $t_{ij}^{\phantom{*}}(\mathbf R) = t_{ji}^*(-\mathbf R)$, the hopping terms can be reduced to
\begin{itemize}
	\item $\mathbf R = (0, 0), i=0, j=1$
	\item $\mathbf R = (1, 0), i=1, j=0$
	\item $\mathbf R = (0, 1), i=1, j=0$
\end{itemize}
\api{TBPLaS} takes the conjugate relation into consideration. So, we need only to add the reduced set of hopping terms. This can be done with the \api{add\_hopping} function
\begin{lstlisting}
prim_cell.add_hopping(rn=[0, 0], orb_i=0, orb_j=1, energy=-2.7)
prim_cell.add_hopping(rn=[1, 0], orb_i=1, orb_j=0, energy=-2.7)
prim_cell.add_hopping(rn=[0, 1], orb_i=1, orb_j=0, energy=-2.7)
\end{lstlisting}
The argument \api{rn} specifies the index of neighbouring cell, while \api{orb\_i} and \api{orb\_j} give the indices of orbitals of the hopping term. \api{energy} is the hopping integral, which should be a complex number in general cases.

Now we have successfully built the primitive cell. We can visualize it with the \api{plot} function:
\begin{lstlisting}
prim_cell.plot()
\end{lstlisting}
The output is shown in Fig. \ref{fig:prim_cell}(a), with orbitals shown as filled circles and hopping terms as arrows. We can also print the details of the model with the \api{print} function:
\begin{lstlisting}
prim_cell.print()
\end{lstlisting}
The output is as follows
\begin{lstlisting}
Lattice vectors (nm):
    0.24600   0.00000   0.00000
    0.12300   0.21304   0.00000
    0.00000   0.00000   1.00000
Orbitals:
    0.00000   0.00000   0.00000 0.0
    0.33333   0.33333   0.00000 0.0
Hopping terms:
    (0, 0, 0) (0, 1) -2.7
    (1, 0, 0) (1, 0) -2.7
    (0, 1, 0) (1, 0) -2.7
\end{lstlisting}

\subsection{Properties of primitive cell}
\label{prim_cell_prop}

In this section we show how to calculate the band structure, DOS and response functions of the graphene primitive cell that created in previous section. First of all, we need to generate a k-path of $\mathrm \Gamma \rightarrow \mathrm M \rightarrow \mathrm K \rightarrow \mathrm \Gamma$ with the \api{gen\_kpath} function
\begin{lstlisting}
k_points = np.array([
    [0.0, 0.0, 0.0],
    [1./2, 0.0, 0.0],
    [2./3, 1./3, 0.0],
    [0.0, 0.0, 0.0],
])
k_label = ["$\Gamma$", "M", "K", "$\Gamma$"]
k_path, k_idx = tb.gen_kpath(k_points, [40, 40, 40])
\end{lstlisting}
In this example, we interpolate with 40 intermediate $\mathbf k$-points along each segment of the $\mathbf k$-path. \api{gen\_kpath} returns two arrays, with \api{k\_path} containing the coordinates of $\mathbf k$-points and \api{k\_idx} containing the indices of highly-symmetric $\mathbf k$-points in \api{k\_path}. Then we solve the band structure with the \api{calc\_bands} function
\begin{lstlisting}
k_len, bands = prim_cell.calc_bands(k_path)
\end{lstlisting}
Here \api{k\_len} is the length of $\mathbf k$-path, while \api{bands} is a $N_k \times N_b$ matrix containing the energies. The band structure can be plotted with matplotlib
\begin{lstlisting}
num_bands = bands.shape[1]
for i in range(num_bands):
    plt.plot(k_len, bands[:, i], color="r", linewidth=1.2)
for idx in k_idx:
    plt.axvline(k_len[idx], color="k", linewidth=0.8)
plt.xlim((0, np.amax(k_len)))
plt.xticks(k_len[k_idx], k_label)
plt.ylabel("Energy (eV)")
plt.tight_layout()
plt.show()
\end{lstlisting}
Or alternatively, using the \api{Visualizer} class:
\begin{lstlisting}
vis = tb.Visualizer()
vis.plot_bands(k_len, bands, k_idx, k_label)
\end{lstlisting}
The output is shown in Fig. \ref{fig:prim_cell}(b). The Dirac cone at $\mathrm K$-point is perfectly reproduced.

To calculate the DOS, we need to sample the first Brillouin zone with a dense $\mathbf k$-grid, e.g., $240 \times 240 \times 1$
\begin{lstlisting}
k_mesh = tb.gen_kmesh((240, 240, 1))
\end{lstlisting}
where \api{k\_mesh} contains the coordinates of $\mathbf k$-points on the grid. Then we evaluate and visualize the DOS as
\begin{lstlisting}
energies, dos = prim_cell.calc_dos(k_mesh, e_min=-9, e_max=9)
vis.plot_dos(energies, dos)
\end{lstlisting}
where \api{energies} is a uniform energy grid whose lower and upper bounds are controlled by the arguments \api{e\_min} and \api{e\_max}. \api{dos} is an array containing the DOS values at the grid points in \api{energies}. The output is shown in Fig. \ref{fig:prim_cell}(c).

The evaluation of response functions requires a Lindhard calculator, which can be created by
\begin{lstlisting}
lind = tb.Lindhard(cell=prim_cell, energy_max=20, energy_step=2000, kmesh_size=(4096, 4096, 1), mu=0.0, temperature=300.0, g_s=2, back_epsilon=1.0)
\end{lstlisting}
The argument \api{cell} assigns the primitive cell to the calculator. \api{energy\_max} and \api{energy\_step} define a uniform energy grid on which response functions will be evaluated. \api{kmesh\_size} specifies the size of $\mathbf k$-grid in the first Brillouin zone. As monolayer graphene is semi-metallic, we need a very dense $\mathbf k$-grid in order to converge the response functions. \api{mu}, \api{temperature} and \api{g\_s} are the chemical potential, temperature and spin degeneracy of the system, while \api{back\_epsilon} is the background dielectic constant, respectively. The \textit{xx} component of optical conductivity, namely $\sigma_{xx}$, can be evaluated with the \api{calc\_ac\_cond} function
\begin{lstlisting}
omegas, ac_cond = lind.calc_ac_cond(component="xx")
\end{lstlisting}
where \api{omegas} is the energy grid and \api{ac\_cond} is the optical conductivity. The results can be visualized using the \api{Visualizer} class
\begin{lstlisting}
ac_cond *= 4
vis = tb.Visualizer()
vis.plot_xy(omegas, ac_cond.real, x_label="Energy (eV)", y_label="$\sigma_{xx} (\sigma_0)$")
\end{lstlisting}
The output is shown in Fig. \ref{fig:prim_cell}(d), in the unit of $\sigma_0 = \frac{e^2}{4\hbar}$.

\subsection{Building the sample}
\label{sample_build}

\begin{figure}[h]
	\centering
	\includegraphics[width=1.0\linewidth]{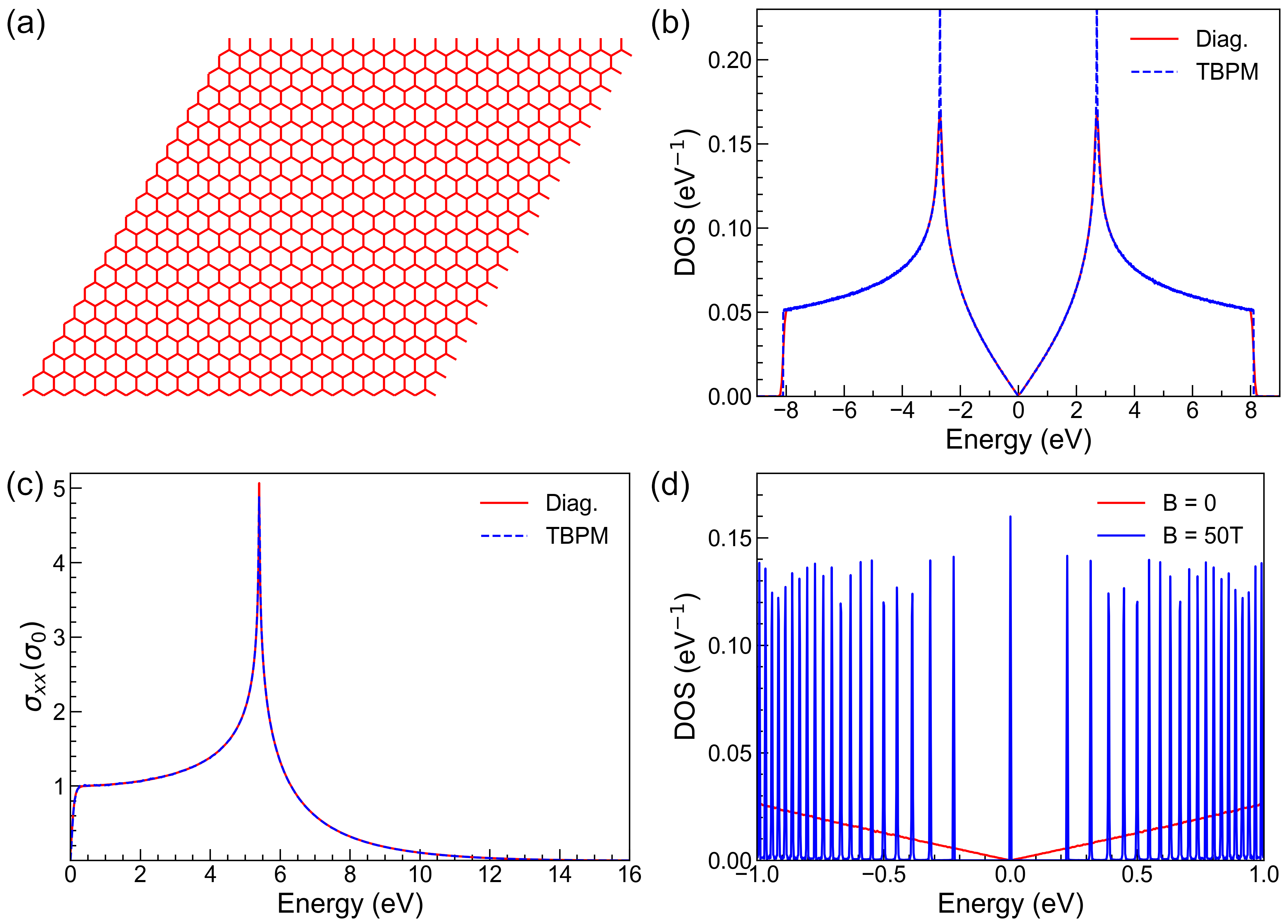}
	\caption{(a) Plot of the $20\times20\times1$ graphene sample. (b) DOS of graphene from exact-diagonalization and TBPM. (c) Optical conductivity of graphene from Lindhard function and TBPM. (d) DOS of graphene under zero and 50 Tesla magnetic fields.}
	\label{fig:sample_dos_ac}
\end{figure}

In this section we show how to construct a sample by making a graphene model with $20\times20\times1$ primitive cells. To build the sample, we need to create the supercell first
\begin{lstlisting}
super_cell = tb.SuperCell(prim_cell, dim=(20, 20, 1), pbc=(True, True, False))
\end{lstlisting}
The \api{SuperCell} class is similar to the functions of \api{extend\_prim\_cell} and \api{apply\_pbc}, where the dimension and periodic boundary conditions are set up at the same time. The sample is formed by gluing the supercells and inter-hopping terms altogether with the \api{Sample} class. In our case the sample consists of only one supercell. So it can be created and visualized by
\begin{lstlisting}
sample = tb.Sample(super_cell)
sample.plot(with_orbitals=False, with_cells=False, hop_as_arrows=False)
\end{lstlisting}
where some options are switched for boosting the plot. The output is shown in Fig. \ref{fig:sample_dos_ac}(a).

\subsection{Properties of sample}
\label{sample_prop}

The \api{Sample} class supports the evaluation of band structure and DOS via exact-diagonalization with the \api{calc\_bands} and \api{calc\_dos} functions, similar to the \api{PrimitiveCell} class. Taking the DOS as an example, in section \ref{prim_cell_prop} we have sampled the first Brillouin zone with a $\mathbf k$-grid of $240\times240\times1$. Now that we have a much larger sample, the dimension of $\mathbf k$-grid can be reduced to $12\times12\times1$ accordingly
\begin{lstlisting}
k_mesh = tb.gen_kmesh((12, 12, 1))
energies, dos = sample.calc_dos(k_mesh, e_min=-9, e_max=9)
vis.plot_dos(energies, dos)
\end{lstlisting}
The output is shown in Fig. \ref{fig:sample_dos_ac}(b), which is consistent with Fig. \ref{fig:prim_cell}(c).

Exact diagonalization-based techniques are not feasible for large models as the computational costs scale cubically with  the model size. On the contrary, TBPM involves only matrix-vector multiplication, and is less demanding on computational resources. Therefore, TBPM is particularly suitable for large models with millions of orbitals or more. Current capabilities of TBPM in TBPLaS are summarized in section \ref{config}. We demonstrate the usage of TBPM to evaluate the DOS and optical conductivity of a graphene sample with $4096\times4096\times1$ primitive cells, i.e., 33,554,432 orbitals. We begin with creating the sample
\begin{lstlisting}
super_cell = tb.SuperCell(prim_cell, dim=(4096, 4096, 1), pbc=(True, True, False))
sample = tb.Sample(super_cell)
sample.rescale_ham(9.0)
\end{lstlisting}
Since the model is extremely large, we will not visualize it as in other examples. In TBPM the time evolution and Fermi-Dirac operators are expanded in Chebyshev polynomials, which requires the eigenvalues of the Hamiltonian to be within $[-1, 1]$. So, we need to rescale the Hamiltonian with the \api{rescale\_ham} function. The scaling factor can be specified as an argument. If not provided, a reasonable default value will be estimated from the Hamiltonian. Then we set up the parameters of TBPM in an instance of the \api{Config} class
\begin{lstlisting}
config = tb.Config()
config.generic["nr_random_samples"] = 4
config.generic["nr_time_steps"] = 4096
\end{lstlisting}
Here we set two parameters: \api{nr\_random\_samples} and \api{nr\_time\_steps}. \api{nr\_random\_samples} specifies that we are going to consider 4 random initial wave functions for the propagation, while \api{nr\_time\_steps} indicates the number of steps to propagate. The time step for the propagation is $\pi / f$ (in unit of $\hbar / eV$), with $f$ being the scaling factor of Hamiltonian in eV. Now we create a pair of solver and analyzer by
\begin{lstlisting}
solver = tb.Solver(sample, config)
analyzer = tb.Analyzer(sample, config)
\end{lstlisting}
Then we calculate and analyze the correlation function to get DOS
\begin{lstlisting}
corr_dos = solver.calc_corr_dos()
energies, dos = analyzer.calc_dos(corr_dos)
vis = tb.Visualizer()
vis.plot_dos(energies, dos)
\end{lstlisting}
Here the correlation function \api{corr\_dos} is obtained with the \api{calc\_corr\_dos} function, and then analyzed by the \api{calc\_dos} function to yield the energy grid \api{energies} and DOS values \api{dos}. The result is shown in Fig. \ref{fig:sample_dos_ac}(b), consistent with the results from exact-diagonalization.

The calculation of optical conductivity is similar to DOS
\begin{lstlisting}
config.generic["correct_spin"] = True
corr_ac_cond = solver.calc_corr_ac_cond()
omegas, ac_cond = analyzer.calc_ac_cond(corr_ac_cond)
ac_cond *= 4
vis.plot_xy(omegas, ac_cond[0].real, x_label="Energy (eV)", y_label="$\sigma_{xx} (\sigma_0)$")
\end{lstlisting}
Note that we set the spin-degeneracy of the model to 2 by setting the \api{correct\_spin} argument to \api{True}, for consistency with the example in section \ref{prim_cell_prop}. The optical conductivity is shown in Fig. \ref{fig:sample_dos_ac}(c), which matches perfectly with the results from Lindhard function.

\subsection{Advanced modeling}
\label{advanced_model}

In this section, we demonstrate how to construct complex models, including hetero structure, quasicrystal and fractal. For the hetero structure, we are going to take the twisted bilayer graphene as an example, while for the fractal we will consider the Sierpi\'nski carpet.

\subsubsection{Hetero-structure}
\label{model_tbg}

\begin{figure}[h]
	\centering
	\includegraphics[width=0.8\linewidth]{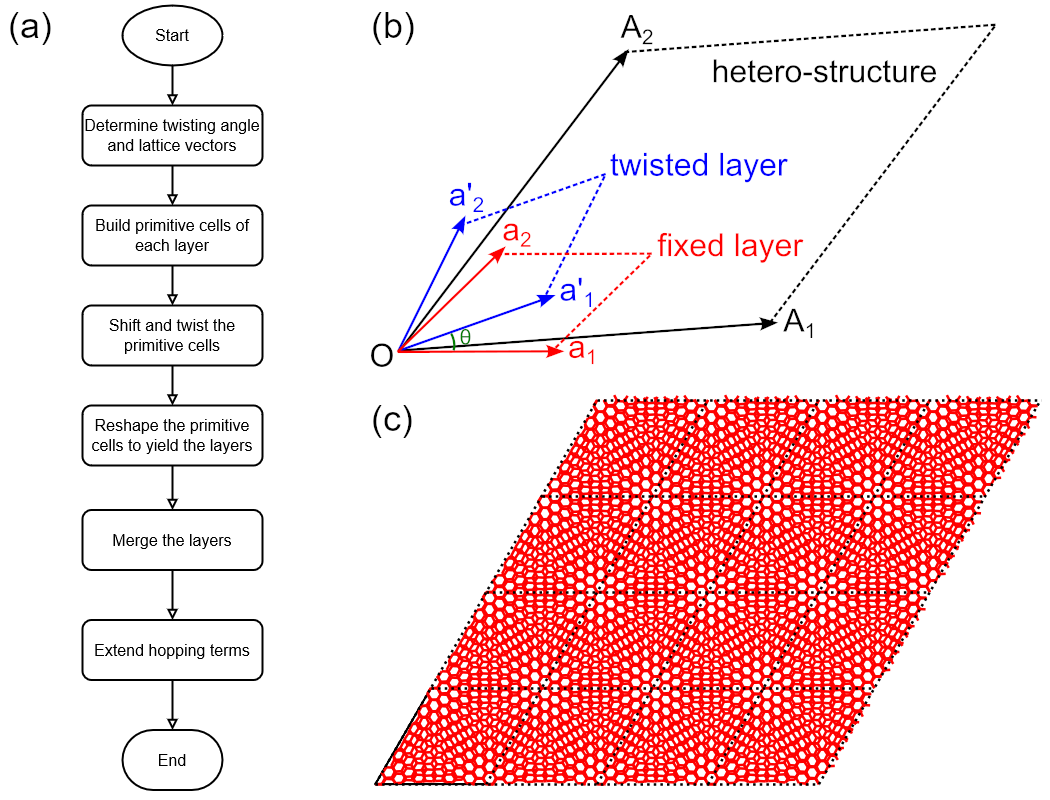}
	\caption{(a) Workflow of constructing hetero-structure. (b) Schematic plot of lattice vectors of fixed ($\mathbf a_1$, $\mathbf a_2$) and twisted  ($\mathbf a_1^\prime$, $\mathbf a_2^\prime$) primitive cells and the hetero-structure ($\mathbf A_1$, $\mathbf A_2$), as well as the twisting angle $\theta$. (c) Twisted bilayer graphene sample with $4\times4\times1$ merged cells of $i=5$.}
	\label{fig:model_tbg}
\end{figure}

The workflow of constructing hetero structures is shown in Fig. \ref{fig:model_tbg}(a). First of all, we determine the twisting angle and lattice vectors of the hetero-structure. Then we build the primitive cells of each layer, shift the twisted layer along $z$-axis by the interlayer distance and rotate it by the twisting angle. After that, we reshape the primitive cells to the lattice vectors of the hetero-structure to yield the layers, as depicted in Fig. \ref{fig:model_tbg}(b). When all the layers are ready, we merge them into one cell and add the intralayer and interlayer hopping terms up to a given cutoff distance. For the visualization of Moir\'{e} pattern, we also need to build a sample from the merged cell.

Before constructing the model, we need to import the required packages and define some necessary functions. The packages are imported by
\begin{lstlisting}
import math
import numpy as np
from numpy.linalg import norm
import tbplas as tb
\end{lstlisting}
The twisting angle and lattice vectors are determined following the formulation in Ref. \cite{dos2007graphene}
\begin{align}
    \theta_i &= \arccos \frac{3i^2+3i+1/2}{3i^2+3i+1}, \\
    \mathbf{A}_1 &= i\cdot\mathbf{a}_1 + (i+1)\cdot\mathbf{a}_2, \\
    \mathbf{A}_2 &= -(i+1)\cdot\mathbf{a}_1 + (2i+1)\cdot\mathbf{a}_2,
\end{align}
where $\mathbf{a}_1$ and $\mathbf{a}_2$ are the lattice vectors of the primitive cell of fixed layer and $i$ is the index of hetero-structure. We define the following functions accordingly
\begin{lstlisting}
def calc_twist_angle(i):
    cos_ang = (3 * i**2 + 3 * i + 0.5) / (3 * i**2 + 3 * i + 1)
    return math.acos(cos_ang)


def calc_hetero_lattice(i, prim_cell_fixed):
    hetero_lattice = np.array([[i, i + 1, 0],
                               [-(i + 1), 2 * i + 1, 0],
                               [0, 0, 1]])
    hetero_lattice = tb.frac2cart(prim_cell_fixed.lat_vec, hetero_lattice)
    return hetero_lattice
\end{lstlisting}
\api{calc\_twist\_angle} returns the twisting angle in radians, while \api{calc\_hetero\_lattice} returns the Cartesian coordinates of lattce vectors in nm. After merging the layers, we need to add the interlayer hopping terms. Meanwhile, the intralayer hoppings terms should also be extended in the same approach. We define the \api{extend\_hop} function to achieve these goals
\begin{lstlisting}
def extend_hop(prim_cell, max_distance=0.75):
    neighbors = tb.find_neighbors(prim_cell, a_max=1, b_max=1,
                                  max_distance=max_distance)
    for term in neighbors:
        i, j = term.pair
        prim_cell.add_hopping(term.rn, i, j, calc_hop(term.rij))
\end{lstlisting}
\change{Here in line 2 we call the \api{find\_neighbors} function to get the neighboring orbital pairs up to the cutoff distance \api{max\_distance}. Then the hopping terms are evaluated according to the displacement vector \api{rij} with the \api{calc\_hop} function and added to the primitive cell.} The \api{calc\_hop} function is defined according to the formulation in Ref. \cite{Laissardiere2012}
\begin{lstlisting}
def calc_hop(rij):
    a0 = 0.1418
    a1 = 0.3349
    r_c = 0.6140
    l_c = 0.0265
    gamma0 = 2.7
    gamma1 = 0.48
    decay = 22.18
    q_pi = decay * a0
    q_sigma = decay * a1
    dr = norm(rij).item()
    n = rij.item(2) / dr
    v_pp_pi = - gamma0 * math.exp(q_pi * (1 - dr / a0))
    v_pp_sigma = gamma1 * math.exp(q_sigma * (1 - dr / a1))
    fc = 1 / (1 + math.exp((dr - r_c) / l_c))
    hop = (n**2 * v_pp_sigma + (1 - n**2) * v_pp_pi) * fc
    return hop
\end{lstlisting}

With all the functions ready, we proceed to build the hetero-structure. In line 2-4 we evaluate the twisting angle of bilayer graphene for $i=5$. Then we construct the primitive cells of the fixed and twisted layers with the \api{make\_graphene\_diamond} function. The fixed primitive cell is located at $z=0$ and does not need rotation or shifting. On the other hand, the twisted primitive cell needs to be rotated counter-clockwise by the twisting angle and shifted towards $+z$ by 0.3349 nm, which is done with the \api{spiral\_prim\_cell} function. After that, we reshape the primitive cells to the lattice vectors of hetero-structure with the \api{make\_hetero\_layer} function, which is a wrapper to coordinate conversion and \api{reshape\_prim\_cell}. Then the layers are merged with \api{merge\_prim\_cell} and the hopping terms are extended with \api{extend\_hop} using a cutoff distance of 0.75 nm. Finally, a sample with $4\times4\times1$ merged cells is created and plotted, with the hopping terms below 0.3 eV hidden for clarity. The output is shown in Fig. \ref{fig:model_tbg} (c), where the Moir\'{e} pattern can be clearly observed.
\begin{lstlisting}
def main():
    # Evaluate twisting angle
    i = 5
    angle = calc_twist_angle(i)

    # Prepare primitive cells of fixed and twisted layer
    prim_cell_fixed = tb.make_graphene_diamond()
    prim_cell_twisted = tb.make_graphene_diamond()

    # Shift and rotate the twisted layer
    tb.spiral_prim_cell(prim_cell_twisted, angle=angle, shift=0.3349)

    # Reshape primitive cells to the lattice vectors of hetero-structure
    hetero_lattice = calc_hetero_lattice(i, prim_cell_fixed)
    layer_fixed = tb.make_hetero_layer(prim_cell_fixed, hetero_lattice)
    layer_twisted = tb.make_hetero_layer(prim_cell_twisted, hetero_lattice)

    # Merge layers
    merged_cell = tb.merge_prim_cell(layer_fixed, layer_twisted)

    # Extend hopping terms
    extend_hop(merged_cell, max_distance=0.75)

    # Visualize Moire pattern
    sample = tb.Sample(tb.SuperCell(merged_cell, dim=(4, 4, 1), pbc=(True, True, False)))
    sample.plot(with_orbitals=False, hop_as_arrows=False, hop_eng_cutoff=0.3)


if __name__ == "__main__":
    main()
\end{lstlisting}

\subsubsection{Quasicrystal}
\label{quasi_crystal}

\begin{figure}[h]
	\centering
	\includegraphics[width=0.5\linewidth]{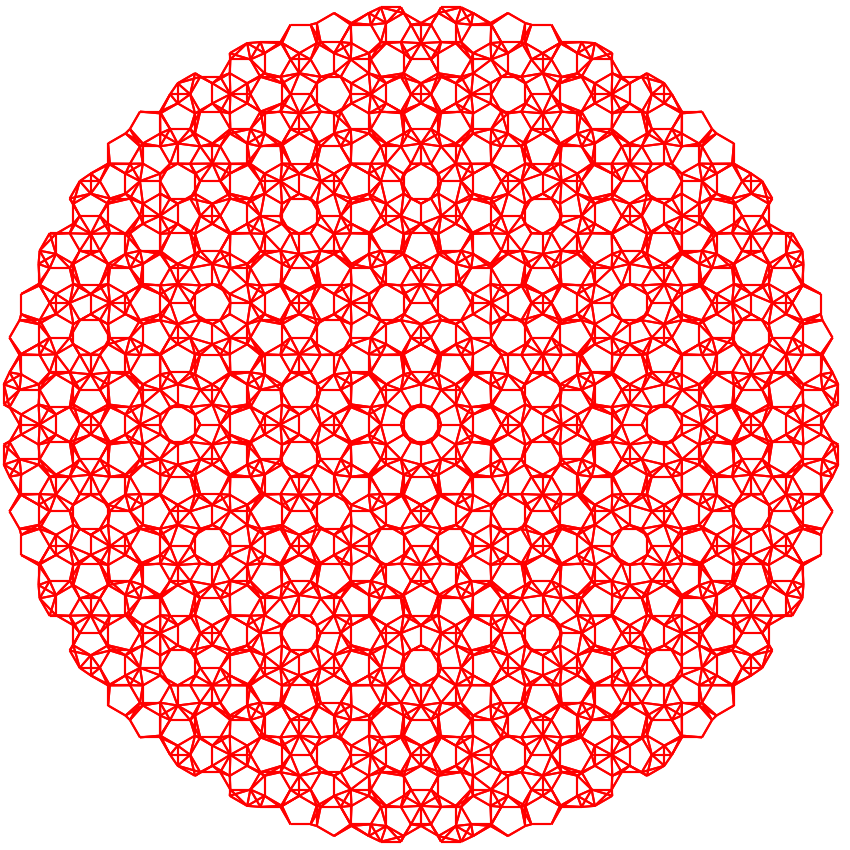}
	\caption{Plot of the quasicrystal formed from the incommensurate $30^\circ$ twisted bilayer graphene with a radius of 3 nm.}
	\label{fig:quasi_crystal}
\end{figure}

Here we consider the construction of hetero structure-based quasicrystal, in which we also need to shift, twist, reshape and merge the cells. Taking bilayer graphene quasicrystal as an example, a quasicrystal with 12-fold symmtery is formed by twisting one layer by $30^\circ$ with respect to the center of $\mathbf{c} = \frac{2}{3}\mathbf{a}_1 + \frac{2}{3}\mathbf{a}_2$, where $\mathbf{a}_1$ and $\mathbf{a}_2$ are the lattice vectors of the primitive cell of fixed layer. We begin with defining the geometric parameters
\begin{lstlisting}
angle = 30 / 180 * math.pi
center = (2./3, 2./3, 0)
radius = 3.0
shift = 0.3349
dim = (33, 33, 1)
\end{lstlisting}
Here \api{angle} is the twisting angle and \api{center} is the fractional coordinate of twisting center. The radius of the quasicrystal is controlled by \api{radius}, while \api{shift} specifies the interlayer distance. We need a large cell to hold the quasicrystal, whose dimension is given in \api{dim}. After introducing the parameters, we build the fixed and twisted layers by
\begin{lstlisting}
prim_cell = tb.make_graphene_diamond()
layer_fixed = tb.extend_prim_cell(prim_cell, dim=dim)
layer_twisted = tb.extend_prim_cell(prim_cell, dim=dim)
\end{lstlisting}
Then we shift and rotate the twisted layer with respect to the center and reshape it to the lattice vectors of fixed layer
\begin{lstlisting}
# Get the Cartesian coordinate of twisting center
center = np.array([dim[0]//2, dim[1]//2, 0]) + center
center = np.matmul(center, prim_cell.lat_vec)

# Twist, shift and reshape top layer
tb.spiral_prim_cell(layer_twisted, angle=angle, center=center, shift=shift)
conv_mat = np.matmul(layer_fixed.lat_vec, np.linalg.inv(layer_twisted.lat_vec))
layer_twisted = tb.reshape_prim_cell(layer_twisted, conv_mat)
\end{lstlisting}
Since we have extended the primitive cell by $33\times33\times1$ times, and we want the quasicrystal to be located in the center of the cell, we need to convert the coordinate of twisting center in line 2-3. The twisting operation is done by the \api{spiral\_prim\_cell} function, where the Cartesian coordinate of the center is given in the \api{center} argument. The fixed and twisted layers have the same lattice vectors after reshaping, so we can merge them safely
\begin{lstlisting}
# Merge bottom and top layers
final_cell = tb.merge_prim_cell(layer_twisted, layer_fixed)
\end{lstlisting}
Then we remove unnecessary orbitals to produce a round quasicrystal with finite radius. This is done by a loop over orbital positions to collect the indices of unnecessary orbitals, and function calls to \api{remove\_orbitals} and \api{trim} functions
\begin{lstlisting}
# Remove unnecessary orbitals
idx_remove = []
orb_pos = final_cell.orb_pos_nm
for i, pos in enumerate(orb_pos):
    if np.linalg.norm(pos[:2] - center[:2]) > radius:
        idx_remove.append(i)
final_cell.remove_orbitals(idx_remove)

# Remove dangling orbitals
final_cell.trim()
\end{lstlisting}
Finally, we extend the hoppings and visualize the quasicrystal
\begin{lstlisting}
extend_hop(final_cell)
final_cell.plot(with_cells=False, with_orbitals=False, hop_as_arrows=False, hop_eng_cutoff=0.3)
\end{lstlisting}
The output is shown in Fig. \ref{fig:quasi_crystal}.

\subsubsection{Fractal}
\label{fractal}

\begin{figure}[h]
	\centering
	\includegraphics[width=1.0\linewidth]{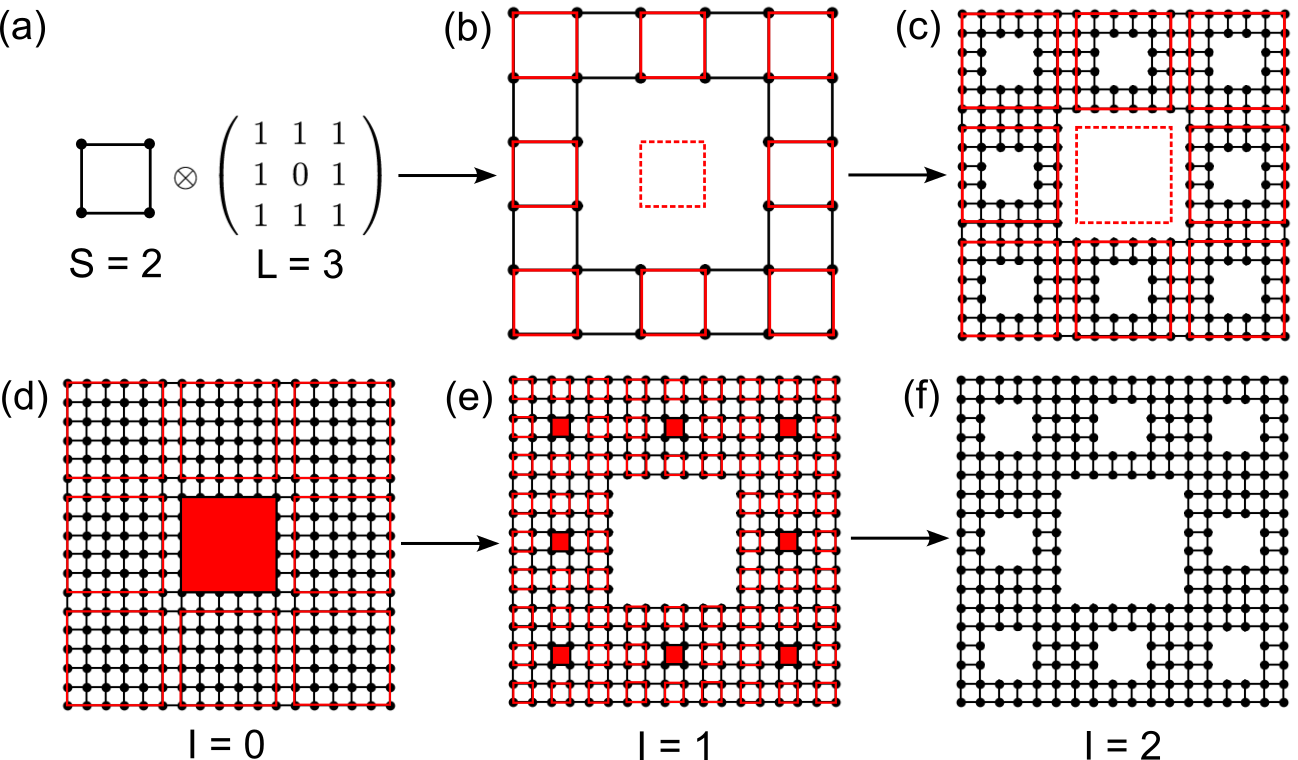}
	\caption{Schematic plot of constructing Sierpi\'nski carpet with $S=2$, $L=3$ and $I=2$ in (a)-(c) bottom-up and (d)-(f) top-down approaches. The dashed squares in (a)-(c) and filled squares in (d)-(f) indicate the void areas in the fractal.}
	\label{fig:fractal_schematics}
\end{figure}

Generally, fractals can be constructed in two approaches, namely \textit{bottom-up} and \textit{top-down}, as demonstrated in Fig. \ref{fig:fractal_schematics}. The bottom-up approach builds the fractal by iteratively replicating the fractal of low iteration number following some specific pattern. On the contrary, the top-down approach builds a large model at first, then recursively removes unnecessary orbitals and hopping terms following the pattern. Both approaches can be implemented with \api{TBPLaS}, while the top-down approach is faster.

In this section, we will take the Sierpi\'nski carpet as an example and built it in the top-down approach. We begin with defining the following auxiliary classes
\begin{lstlisting}
class Box:
    def __init__(self, i0, j0, i1, j1, void=False):
        self.i0 = i0
        self.j0 = j0
        self.i1 = i1
        self.j1 = j1
        self.void = void

class Mask:
    def __init__(self, starting_box, num_grid, num_iter=0):
        self.boxes = [starting_box]
        self.num_grid = num_grid
        for i in range(num_iter):
            new_boxes = []
            for box in self.boxes:
                new_boxes.extend(self.partition_box(box))
            self.boxes = new_boxes

    def partition_box(self, box):
        if box.void:
            sub_boxes = [box]
        else:
            sub_boxes = []
            di = (box.i1 - box.i0 + 1) // self.num_grid
            dj = (box.j1 - box.j0 + 1) // self.num_grid
            for ii in range(self.num_grid):
                i0 = box.i0 + ii * di
                i1 = i0 + di
                for jj in range(self.num_grid):
                    j0 = box.j0 + jj * dj
                    j1 = j0 + dj
                    if (1 <= ii < self.num_grid - 1 and
                        1 <= jj < self.num_grid - 1):
                        void = True
                    else:
                        void = False
                    sub_boxes.append(Box(i0, j0, i1, j1, void))
        return sub_boxes

    def etch_prim_cell(self, prim_cell, width):
        prim_cell.sync_array()
        masked_id_pc = []
        for box in self.boxes:
            if box.void:
                id_pc = [(ia, ib)
                         for ia in range(box.i0, box.i1)
                         for ib in range(box.j0, box.j1)]
                masked_id_pc.extend(id_pc)
        masked_id_pc = [i[0]*width + i[1] for i in masked_id_pc]
        prim_cell.remove_orbitals(masked_id_pc)
        prim_cell.sync_array()
\end{lstlisting}
Here the \api{Box} represents a rectangular area spanning from $[i_0, j_0]$ to $(i_1, j_1)$. If the box is marked as void, then the orbitals inside it will be removed. The \api{Mask} class is a collection of boxes, which recursively partitions them into smaller boxes and marks the central boxes as void. It offers the \api{etch\_prim\_cell} function to produce the fractal by removing orbitals falling into void boxes.

\begin{figure}[h]
	\centering
	\includegraphics[width=0.6\linewidth]{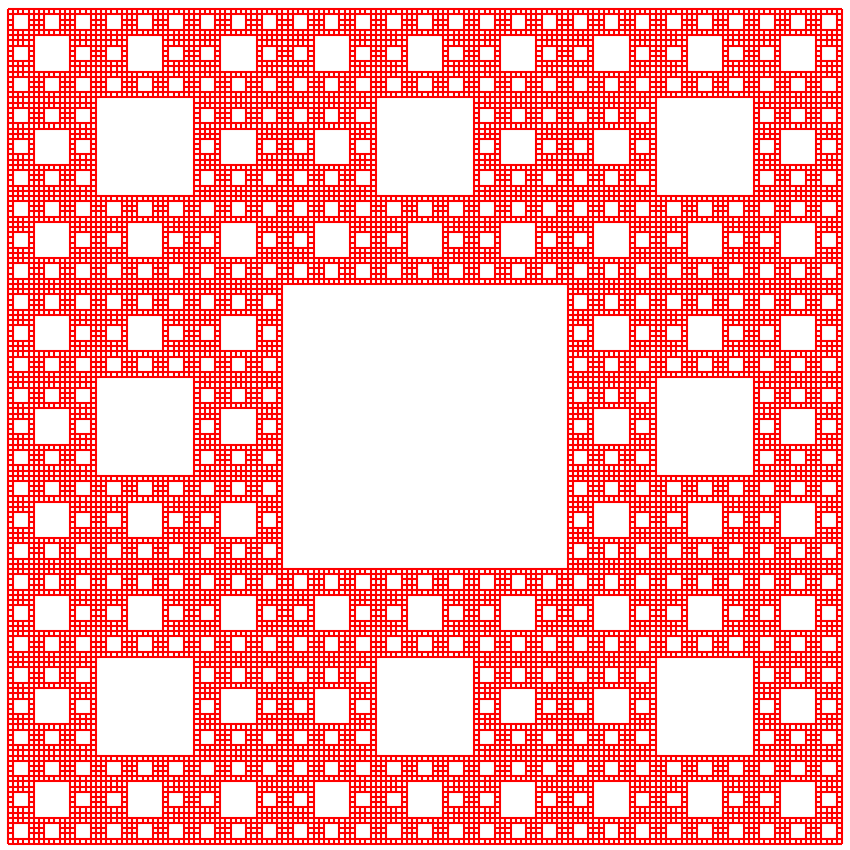}
	\caption{Sierpi\'nski carpet with $S=2$, $L=3$ and $I=4$.}
	\label{fig:sierpinski}
\end{figure}

To demonstrate the usage of the auxiliary classes, we define the geometric parameters and create a square primitive cell
\begin{lstlisting}
# Geometric parameters
start_width = 2
extension = 3
iteration = 4

# Create a square primitive cell
lattice = np.eye(3, dtype=np.float64)
prim_cell = tb.PrimitiveCell(lattice)
prim_cell.add_orbital((0, 0))
prim_cell.add_hopping((1, 0), 0, 0, 1.0)
prim_cell.add_hopping((0, 1), 0, 0, 1.0)
\end{lstlisting}
The Sierpi\'nski carpet is characterized by 3 parameters: the starting width $S$, the extension $L$ which controls the pattern, and the iteration number $I$, as shown in Fig. \ref{fig:fractal_schematics}. We extend the square primitive cell to the final width of the carpet, which is determined as $D = S \cdot L^I$
\begin{lstlisting}
# Create the extended cell
final_width = start_width * extension**iteration
extended_cell = tb.extend_prim_cell(prim_cell, dim=(final_width, final_width, 1))
extended_cell.apply_pbc((False, False, False))
\end{lstlisting}
Then we create a box covering the whole extended cell and a mask from the box. The bottom-left corner of the box is located at $[0, 0]$, while the top-right corner is at $(D-1, D-1)$
\begin{lstlisting}
# Create the mask
start_box = Box(0, 0, final_width-1, final_width-1)
mask = Mask(start_box, num_grid=extension, num_iter=iteration)
\end{lstlisting}
Then we call the \api{etch\_prim\_cell} function to remove the orbitals falling into void boxes of the mask
\begin{lstlisting}
# Remove orbitals
mask.etch_prim_cell(extended_cell, final_width)
\end{lstlisting}
Finally, we visualize the fractal
\begin{lstlisting}
# Plot the fractal
extended_cell.plot(with_orbitals=False, with_cells=False, with_conj=False, hop_as_arrows=False)
\end{lstlisting}
The output is shown in Fig. \ref{fig:sierpinski}.

\subsection{Strain and external fields}
\label{strain_field}

In this section, we introduce the common procedure of applying strain and external fields on the model. It is difficult to design common \textit{out-of-the-box} user APIs that offer such functionalities since they are strongly case-dependent. Generally, the user should implement these perturbations by modifying model attributes such as orbital positions, on-site energies and hopping integrals. For the primitive cell, it is straightforward to achieve this goal with the \api{set\_orbital} and \api{add\_hopping} functions, as mentioned in section \ref{prim_cell}. The \api{Sample} class, on the contrary, does not offer such functions. Instead, the user should work with the attributes directly. In the \api{Sample} class, orbital positions and on-site energies are stored in the \api{orb\_pos} and \api{orb\_eng} attributes. Hopping terms are represented with 3 attributes: \api{hop\_i} and \api{hop\_j} for orbital indices, and \api{hop\_v} for hopping integrals. There is also an auxiliary attribute \api{dr} which holds the hopping vectors. All the attributes are NumPy arrays. The on-site energies and hopping terms can be modified directly, while the orbital positions should be changed via a modifier function. The hopping vectors are updated from the orbital positions and hopping terms automatically, thus no need of explicit modification. 

As the example, we will investigate the propagation of wave function in a graphene sample. We begin with defining the functions for adding strain and external fields, then calculate and plot the time-dependent wave function to check their effects on the propagation. The impact of magnetic field on electronic structure will also be discussed.

\subsubsection{Functions for strain}
\label{func_strain}

Strain will introduce deformation into the model, changing both orbital positions and hopping integrals. It is a rule that orbital positions should not be modified directly, but through a modifier function. We consider a Gaussian bump deformation, and define the following function to generate the modifier
\begin{lstlisting}
def make_deform(center, sigma=0.5, extent=(1.0, 1.0), scale=(0.5, 0.5)):
    def _deform(orb_pos):
        x, y, z = orb_pos[:, 0], orb_pos[:, 1], orb_pos[:, 2]
        dx = (x - center[0]) * extent[0]
        dy = (y - center[1]) * extent[1]
        amp = np.exp(-(dx**2 + dy**2) / (2 * sigma**2))
        x += amp * dx * scale[0]
        y += amp * dy * scale[0]
        z += amp * scale[1]
    return _deform
\end{lstlisting}
Here \api{center}, \api{sigma} and \api{extent} control the location, width and extent of the bump. For example, if \api{extent} is set to $(1.0, 0.0)$, the bump will become one-dimensional which varies along $x$-direction while remains constant along $y$-direction. \api{scale} specifies the scaling factors for in-plane and out-of-plane displacements. The \api{make\_deform} function returns another function as the modifier, which updates the orbital positions \textit{in place} according to the following expression
\begin{align}
	\mathbf r_i &\rightarrow \mathbf r_i + \Delta_i, \\
	\Delta_i^{\parallel} &= A_i \cdot (\mathbf r_i^{\parallel} - \mathbf c_0^{\parallel}) \cdot s^{\parallel}, \\
	\Delta_i^{\perp} &= A_i \cdot s^{\perp}, \\
	A_i &= \exp \left[-\frac{1}{2\sigma^2}\sum_{j=1}^{2} (\mathbf r_i^j - \mathbf c_0^j)^2 \cdot \eta^j \right],
\end{align}
where $\mathbf r_i$ is the position of $i$-th orbital, $\Delta_i$ is the displacement, $s$ is the scaling factor, $\parallel$ and $\perp$ are the in-plane and out-of-plane components. The location, width and extent of the bump are denoted as $\mathbf c_0$, $\sigma$ and $\eta$, respectively.

In addition to the orbital position modifier, we also need to update hopping integrals
\begin{lstlisting}
def update_hop(sample):
    sample.init_hop()
    sample.init_dr()
    for i, rij in enumerate(sample.dr):
        sample.hop_v[i] = calc_hop(rij)
\end{lstlisting}
As we will make use of the hopping terms and vectors, we should call the \api{init\_hop} and \api{init\_dr} functions to initialize the attributes. Similar rule holds for the on-site energies and orbital positions, as discussed in section \ref{super_cell}. Then we loop over the hopping terms to update the integrals in \api{hop\_v} according to the vectors in \api{dr} with the \api{calc\_hop} function, which is defined in section \ref{model_tbg}.

\subsubsection{Functions for external fields}
\label{func_field}

The effects of external electric field can be modeled by adding position-dependent potential to the on-site energies. We consider a Gaussian-type scattering potential described by
\begin{equation}
    V_i = V_0 \cdot A_i
\end{equation}
and define the following function to add the potential to the sample
\begin{lstlisting}
def add_efield(sample, center, sigma=0.5, extent=(1.0, 1.0), v_pot=1.0):
    sample.init_orb_pos()
    sample.init_orb_eng()
    orb_pos = sample.orb_pos
    orb_eng = sample.orb_eng
    for i, pos in enumerate(orb_pos):
        dx = (pos.item(0) - center[0]) * extent[0]
        dy = (pos.item(1) - center[1]) * extent[1]
        orb_eng[i] += v_pot * math.exp(-(dx**2 + dy**2) / (2 * sigma**2))
\end{lstlisting}
The arguments \api{center}, \api{sigma} and \api{extent} are similar to that of the \api{make\_deform} function, while \api{v\_pot} specifies $V_0$. Similar to \api{update\_hop}, we need to call \api{init\_orb\_pos} and \api{init\_orb\_eng} to initialize orbital positions and on-site energies before accessing them. Then the position-dependent scattering potential is added to the on-site energies.

The effects of magnetic field can be modeled with Peierls substitution, as discussed in section \ref{method}. For homogeneous magnetic field perpendicular to the $xOy$-plane along $-z$ direction,  the \api{Sample} class offers an API \api{set\_magnetic\_field}, which follows the Landau gauge of vector potential $\mathbf A = (By, 0, 0)$ and updates the hopping terms as
\begin{equation}
    t_{ij} \rightarrow t_{ij} \cdot \exp \left[\mathrm i\frac{eB}{2\hbar c} \cdot (\mathbf r_j^x - \mathbf r_i^x) \cdot (\mathbf r_j^y + \mathbf r_i^y) \right]
\end{equation}
where $B$ is the intensity of magnetic field, $\mathbf r_i$ and $\mathbf r_j$ are the positions of $i$-th and $j$-th orbitals, respectively.

\subsubsection{Initial wave functions}
\label{init_wfc}

The initial wave function we consider here as an example for the propagation is a Gaussian wave-packet, which is defined by
\begin{lstlisting}
def init_wfc_gaussian(sample, center, sigma=0.5, extent=(1.0, 1.0)):
    sample.init_orb_pos()
    orb_pos = sample.orb_pos
    wfc = np.zeros(orb_pos.shape[0], dtype=np.complex128)
    for i, pos in enumerate(orb_pos):
        dx = (pos.item(0) - center[0]) * extent[0]
        dy = (pos.item(1) - center[1]) * extent[1]
        wfc[i] = math.exp(-(dx**2 + dy**2) / (2 * sigma**2))
    wfc /= np.linalg.norm(wfc)
    return wfc
\end{lstlisting}
Note that the wave function should be a complex vector whose length must be equal to the number of orbitals. Also, it should be normalized before being returned.

\subsubsection{Propagation of wave function}
\label{prop_wfc}

\begin{figure}[h]
	\centering
	\includegraphics[width=1.0\linewidth]{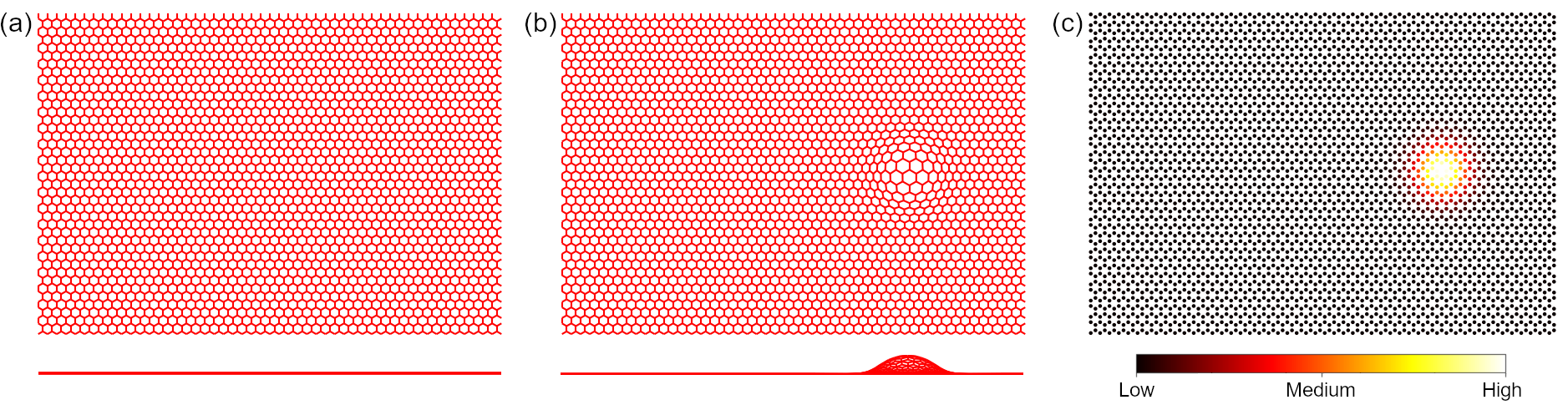}
	\caption{Top and side views of (a) pristine graphene sample and (b) sample with deformation. (c) Plot of on-site energies of graphene sample with scattering potential.}
	\label{fig:sample_struct}
\end{figure}

We consider a rectangular graphene sample with $50\times20\times1$ primitive cells, as shown in Fig. \ref{fig:sample_struct}(a). We begin with importing the necessary packages and defining some geometric parameters
\begin{lstlisting}
import math
import numpy as np
from numpy.linalg import norm
import tbplas as tb

prim_cell = tb.make_graphene_rect()
dim = (50, 20, 1)
pbc = (True, True, False)
x_max = prim_cell.lat_vec[0, 0] * dim[0]
y_max = prim_cell.lat_vec[1, 1] * dim[1]
wfc_center = (x_max * 0.5, y_max * 0.5)
deform_center = (x_max * 0.75, y_max * 0.5)
\end{lstlisting}
Here \api{dim} and \api{pbc} define the dimension and boundary condition. \api{x\_max} and \api{y\_max} are the lengths of the sample along $x$ and $y$ directions. The initial wave function will be a Gaussian wave-packet located at the center of the sample given by \api{wfc\_center}. The deformation and scattering potential will be located at the center of right half of the sample, as specified by \api{deform\_center} and shown in Fig. \ref{fig:sample_struct} (b)-(c).

\begin{figure}[h]
	\centering
	\includegraphics[width=1.0\linewidth]{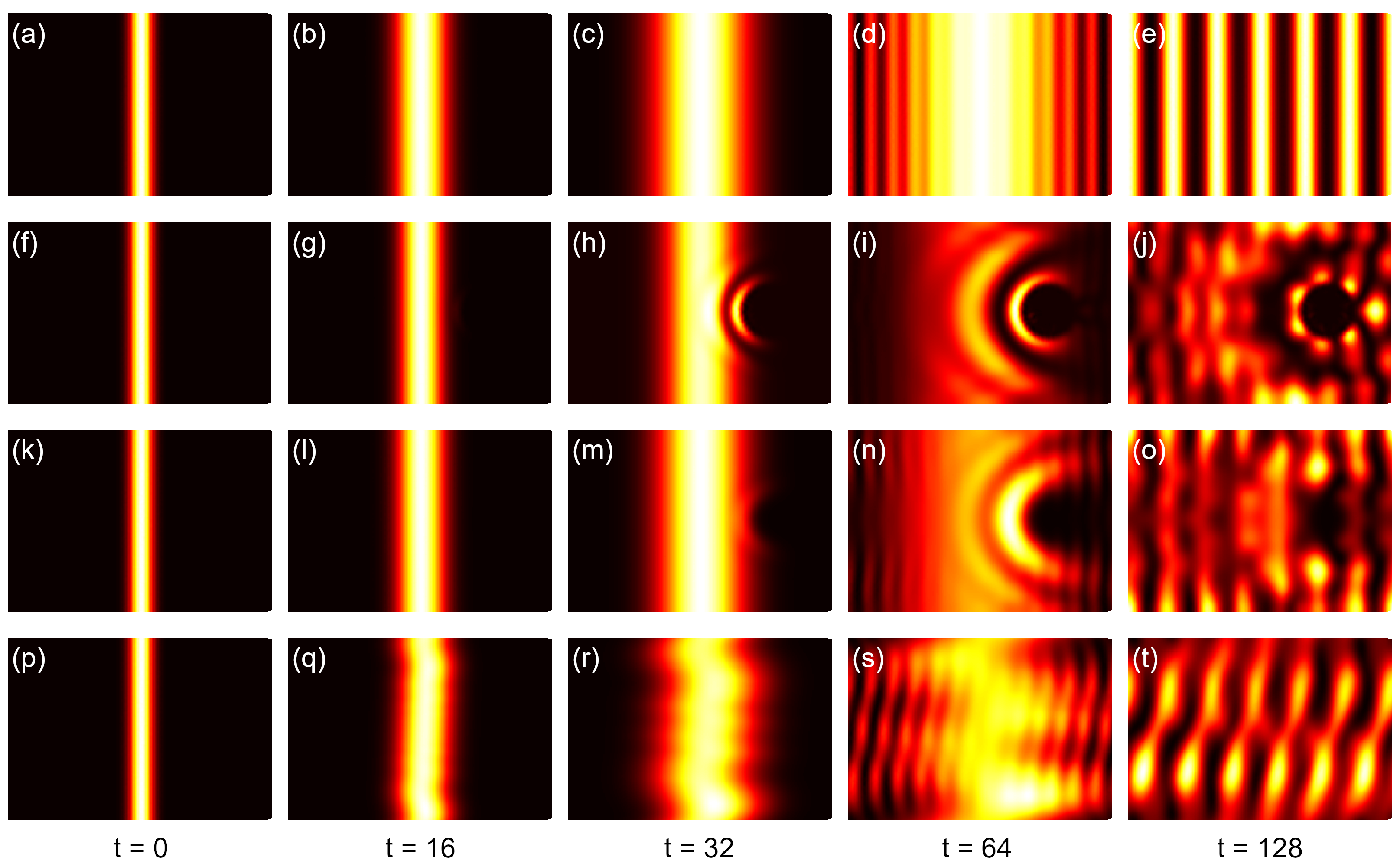}
	\caption{(a)-(e) Propagation of one-dimensional Gaussian wave-packet in pristine graphene sample. (f)-(j) Propagation in graphene sample with deformation, (k)-(o) with scattering potential and (p)-(t) with magnetic field of 50 Tesla.}
	\label{fig:sample_wfc}
\end{figure}

We firstly investigate the propagation of a one-dimensional Gaussian wave-packet in pristine sample, which is given by
\begin{lstlisting}
# Prepare the sample and inital wave function
sample = tb.Sample(tb.SuperCell(prim_cell, dim, pbc))
psi0 = init_wfc_gaussian(sample, center=wfc_center, extent=(1.0, 0.0))

# Propagate the wave function
config = tb.Config()
config.generic["nr_time_steps"] = 128
time_log = np.array([0, 16, 32, 64, 128])
sample.rescale_ham()
solver = tb.Solver(sample, config)
psi_t = solver.calc_psi_t(psi0, time_log)

# Visualize the time-dependent wave function
vis = tb.Visualizer()
for i in range(len(time_log)):
    vis.plot_wfc(sample, np.abs(psi_t[i])**2, cmap="hot", scatter=False)
\end{lstlisting}
As the propagation is performed with the \api{calc\_psi\_t} function of \api{Solver} class, it follows the common procedure of TBPM calculation. We propagate the wave function by 128 steps, and save the snapshots in \api{psi\_t} at the time steps specified in \api{time\_log}. The snapshots are then visualized by the \api{plot\_wfc} function of \api{Visualizer} class, as shown in Fig. \ref{fig:sample_wfc}(a)-(e), where the wave-packet diffuses freely, hits the boundary and forms interference pattern.

We then add the bump deformation to the sample, by assigning the modifier function to the supercell and calling \api{update\_hop} to update the hopping terms
\begin{lstlisting}
deform = make_deform(center=deform_center)
sample = tb.Sample(tb.SuperCell(prim_cell, dim, pbc, orb_pos_modifier=deform))
update_hop(sample)
\end{lstlisting}
The propagation of wave-packet in deformed graphene sample is shown in Fig. \ref{fig:sample_wfc}(f)-(j). Obviously, the wave function gets scattered by the bump. Although similar interference pattern is formed, the propagation in the right part of the sample is significantly hindered, due to the increased inter-atomic distances and reduced hopping integrals at the bump.

Similar phenomena are observed when the scattering potential is added to the sample by
\begin{lstlisting}
add_efield(sample, center=deform_center)
\end{lstlisting}
The time-dependent wave function is shown in Fig. \ref{fig:sample_wfc}(k)-(o). Due to the higher on-site energies, the probability of emergence of electron is suppressed near the scattering center.

As for the effects of magnetic field, it is well known that Landau levels will emerge in the DOS, as shown in Fig. \ref{fig:sample_dos_ac}(d). The analytical solution to Schr\"{o}dinger's equation for free electron in homogeneous magnetic field with $\mathbf A=(By, 0, 0)$ shows that the wave function will propagate freely along $x$ and $z$-directions while oscillates along $y$-direction. To simulate this process, we apply the magnetic field to the sample by
\begin{lstlisting}
sample.set_magnetic_field(50)
\end{lstlisting}
The snapshots of time-dependent wave function are shown in Fig. \ref{fig:sample_wfc}(p)-(t). The interference pattern is similar to the case without magnetic field, as the wave function propagates freely along $x$ direction. However, due to the oscillation along $y$-direction, the interference pattern gets distorted during the propagation. These phenomena agree well with the analytical results.

\subsection{Miscellaneous}
\label{misc}

\subsubsection{\change{Wannier90 interface, Slater-Koster formula and parameter fitting}}
\label{misc_wan_sk_fit}

\change{In this section, we demonstrate the usage of Wannier90 interface \api{wan2pc}, Slater-Koster table \api{SK} and parameter fitting tool \api{ParamFit}, by reducing an 8-band graphene primitive cell imported from the output of Wannier90. We achieve this by truncating the hopping terms to the second nearest neighbor, and refitting the on-site energies and Slater-Koster parameters to minimize the residual between the reference and fitted band data, i.e.,
\begin{equation}
    \min_{\mathbf{x}} \sum_{i,\mathbf{k}}\omega_i\left|\bar{E}_{i,\mathbf{k}} - E_{i,\mathbf{k}}(\mathbf{x})\right|^2
\end{equation}
where $\mathbf{x}$ are the fitting parameters, $\omega$ are the fitting weights, $\bar{E}$ and $E$ are the reference and fitted band data from the original and reduced cells, $i$ and $\mathbf{k}$ are band and $\mathbf{k}$-point indices, respectively.}

We begin with importing the necessary packages
\begin{lstlisting}
import numpy as np
import matplotlib.pyplot as plt
import tbplas as tb
\end{lstlisting}
and define the following function to build the primitive cell from the Slater-Koster parameters
\begin{lstlisting}
def make_cell(sk_params):
    # Lattice constants and orbital info.
    lat_vec = np.array([
        [2.458075766398899, 0.000000000000000, 0.000000000000000],
        [-1.229037883199450, 2.128755065595607, 0.000000000000000],
        [0.000000000000000, 0.000000000000000, 15.000014072326660],
    ])
    orb_pos = np.array([
        [0.000000000, 0.000000000, 0.000000000],
        [0.666666667, 0.333333333, 0.000000000],
    ])
    orb_label = ("s", "px", "py", "pz")

    # Create the cell and add orbitals
    e_s, e_p = sk_params[0], sk_params[1]
    cell = tb.PrimitiveCell(lat_vec, unit=tb.ANG)
    for pos in orb_pos:
        for label in orb_label:
            if label == "s":
                cell.add_orbital(pos, energy=e_s, label=label)
            else:
                cell.add_orbital(pos, energy=e_p, label=label)

    # Add Hopping terms
    neighbors = tb.find_neighbors(cell, a_max=5, b_max=5,
                                  max_distance=0.25)
    sk = tb.SK()
    for term in neighbors:
        i, j = term.pair
        label_i = cell.get_orbital(i).label
        label_j = cell.get_orbital(j).label
        hop = calc_hop(sk, term.rij, term.distance, label_i, label_j,
                       sk_params)
        cell.add_hopping(term.rn, i, j, hop)
    return cell
\end{lstlisting}
In line 3-12 we define the lattice vectors, orbital positions and labels. The \api{SK} class will utilize the orbital labels to evaluate the hopping integrals, so they \textit{must} be chosen from a set of predefined strings, namely \api{s} for $s$ orbitals, \api{px}/\api{py}/\api{pz} for $p$ orbitals, and \api{dxy}/\api{dx2-y2}/\api{dyz}/\api{dzx}/\api{dz2} for $d$ orbitals, respectively. Then in line 15-22 we add the orbitals with on-site energies taken from the first 2 elements of \api{sk\_params} and the predefined labels. In line 25 we call \api{find\_neighbors} to find all the orbital pairs within the cutoff distance of 0.25 nm, where the arguments \api{a\_max} and \api{b\_max} specify the searching range. After that, we create a Slater-Koster table from the \api{SK} class, and loop over the orbital pairs to add the hopping terms, which are evaluated by the \api{calc\_hop} function depending on the displacement vector \api{rij}, the distance \api{distance}, orbital labels \api{label\_i} and \api{label\_j}, and Slater-Koster parameters \api{sk\_params}. The \api{calc\_hop} function is defined as
\begin{lstlisting}
def calc_hop(sk, rij, distance, label_i, label_j, sk_params):
    # 1st and 2nd hopping distances in nm
    d1 = 0.1419170044439990
    d2 = 0.2458074906840380
    if abs(distance - d1) < 1.0e-5:
        v_sss, v_sps, v_pps, v_ppp = sk_params[2:6]
    elif abs(distance - d2) < 1.0e-5:
        v_sss, v_sps, v_pps, v_ppp = sk_params[6:10]
    else:
        raise ValueError(f"Too large distance {distance}")
    return sk.eval(r=rij, label_i=label_i, label_j=label_j,
                   v_sss=v_sss, v_sps=v_sps,
                   v_pps=v_pps, v_ppp=v_ppp)
\end{lstlisting}
where we extract the first and second-nearest Slater-Koster parameters from \api{sk\_params}, and call the \api{eval} function of \api{SK} class to evaluate the hopping integral, taking the displacement vector, orbital labels and SK parameters as input.

The fitting tool \api{ParamFit} is an abstract class. The users should derive their own fitting class from it, and implement the \api{calc\_bands\_ref} and \api{calc\_bands\_fit} functions, which return the reference and fitted band data, respectively. We define a \api{MyFit} class as
\begin{lstlisting}
class MyFit(tb.ParamFit):
    def calc_bands_ref(self):
        cell = tb.wan2pc("graphene")
        k_len, bands = cell.calc_bands(self.k_points)
        return bands

    def calc_bands_fit(self, sk_params):
        cell = make_cell(sk_params)
        k_len, bands = cell.calc_bands(self.k_points, echo_details=False)
        return bands
\end{lstlisting}
In \api{calc\_bands\_ref}, we import the primitive cell with the Wannier90 interface \api{wan2pc}, then calculate and return the band data. The \api{calc\_bands\_fit} function does a similar job, with the only difference that the primitive cell is constructed from Slater-Koster parameters with the \api{make\_cell} function we have just created.

\begin{figure}[h]
	\centering
	\includegraphics[width=1.0\linewidth]{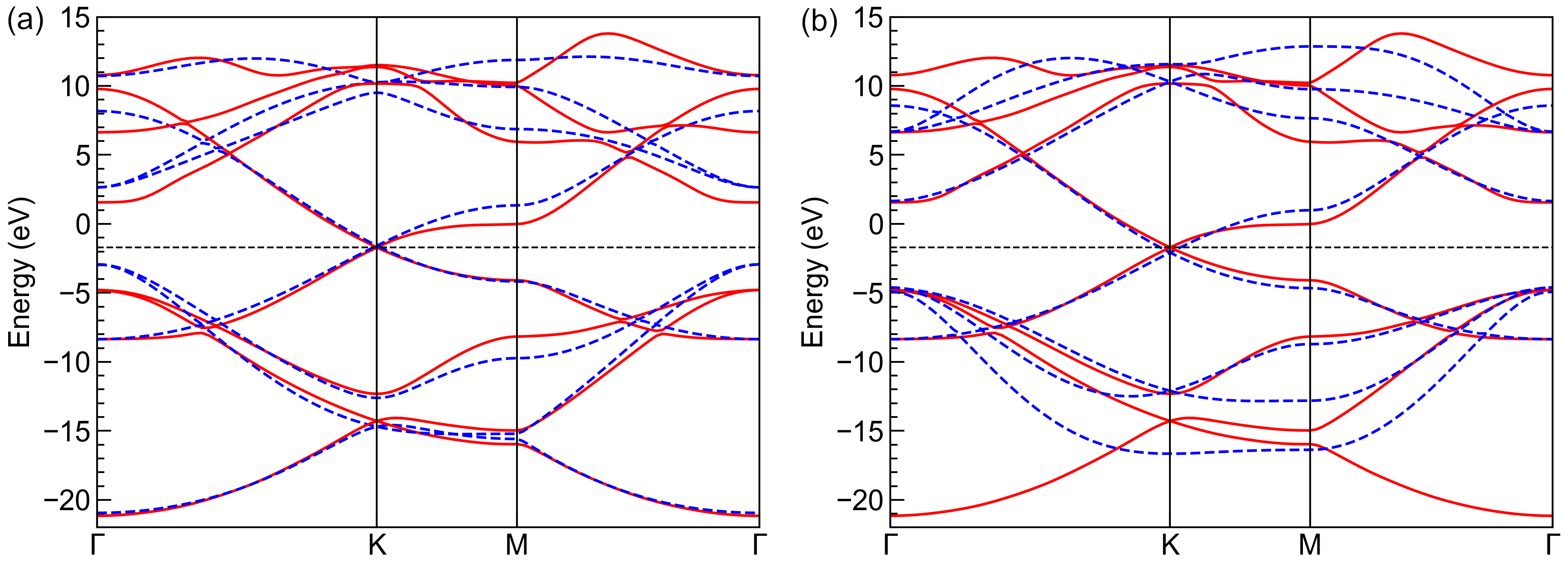}
	\caption{\change{Band structures from reference (solid red lines) and fitted (dashed blue lines) primitive cells with (a) equal weights for all bands and (b) lower weights for bands 1-2 and 7-8. The horizontal dashed black lines indicate the Fermi level.}}
	\label{fig:wan_sk_fit}
\end{figure}

The application of \api{MyFit} class is as following
\begin{lstlisting}
def main():
    # Fit the sk parameters
    k_points = tb.gen_kmesh((120, 120, 1))
    weights = np.array([0.1, 0.1, 1.0, 1.0, 1.0, 1.0, 0.1, 0.1])
    fit = MyFit(k_points, weights)
    sk0 = np.array([-8.370, 0.0,
                    -5.729, 5.618, 6.050, -3.070,
                    0.102, -0.171, -0.377, 0.070])
    sk1 = fit.fit(sk0)
    print("SK parameters after fitting:")
    print(sk1[:2])
    print(sk1[2:6])
    print(sk1[6:10])

    # Plot fitted band structure
    cell_ref = tb.wan2pc("graphene")
    cell_fit = make_cell(sk1)
    k_points = np.array([
        [0.0, 0.0, 0.0],
        [1./3, 1./3, 0.0],
        [1./2, 0.0, 0.0],
        [0.0, 0.0, 0.0],
    ])
    k_path, k_idx = tb.gen_kpath(k_points, [40, 40, 40])
    k_len, bands_ref = cell_ref.calc_bands(k_path)
    k_len, bands_fit = cell_fit.calc_bands(k_path)
    num_bands = bands_ref.shape[1]
    for i in range(num_bands):
        plt.plot(k_len, bands_ref[:, i], color="red", linewidth=1.0)
        plt.plot(k_len, bands_fit[:, i], color="blue", linewidth=1.0)
    plt.show()


if __name__ == "__main__":
    main()
\end{lstlisting}
To create a \api{ParamFit} instance, we need to specify the $\mathbf{k}$-points and fitting weights, as shown in line 3-4. For the $\mathbf{k}$-points, we are going to use a $\mathbf{k}$-grid of $120\times120\times1$. The length of weights should be equal to the number of orbitals of the primitive cell, which is 8 in our case. We assume all the bands to have the same weights, and set them to 1. Then we create the \api{ParamFit} instance, define the initial guess of parameters from Ref. \cite{Konschuh2010}, and get the fitted results with the \api{fit} function. The output should look like
\begin{lstlisting}
SK parameters after fitting:
[-3.63102899 -1.08477167]
[-5.27742318  5.87219052  4.61650991 -2.75652966]
[-0.24734558  0.17599166  0.14798703  0.16545428]
\end{lstlisting}
The first two numbers are the on-site energies for $s$ and $p$ orbitals, while the following numbers are the Slater-Koster paramets $V_{ss\sigma}$, $V_{sp\sigma}$, $V_{pp\sigma}$ and $V_{pp\pi}$ at first and second nearest hopping distances, respectively. We can also plot and compare the band structures from the reference and fitted primitive cells, as shown in Fig. \ref{fig:wan_sk_fit}(a). It is clear that the fitted band structure agrees well with the reference data near the Fermi level (-1.7 eV) and at deep (-20 eV) or high energies (10 eV). However, the derivation from reference data of intermediate bands (-5 eV and 5 eV) is non-negligible. To improve this, we lower the weights of band 1-2 and 7-8 by
\begin{lstlisting}
weights = np.array([0.1, 0.1, 1.0, 1.0, 1.0, 1.0, 0.1, 0.1])
\end{lstlisting}
and refitting the parameters. The results are shown in Fig. \ref{fig:wan_sk_fit}(b), where the fitted and reference band structures agree well from -5 to 5 eV.

\subsubsection{\change{$\mathbb{Z}_2$ topological invariant and spin-orbital coupling}}
\label{z2_soc}

\change{In this section, we demonstrate the usage of \api{Z2} and \api{SOC} classes by calculating the topological invariant of bilayer bismuth \cite{Murakami2006} and check the effects of SOC. We consider the intra-atom SOC term
\begin{equation}
    H^{\mathrm{soc}} = \lambda \mathbf{L}\cdot\mathbf{S}
\end{equation}
and evaluate its matrix elements in the direct product basis of $\ket{l}\otimes\ket{s}$, where $\ket{l}$ are the $s$/$p$/$d$ orbitals and $\ket{s}$ are the eigenstates of Pauli matrix $\sigma_z$. We prefer this basis set because it does not require the evaluation of Clebsch-Gordan coefficients, thus much easier to implement. In this basis, the matrix element of SOC becomes
\begin{equation}
\label{soc_mtxel}
H_{ij}^{\mathrm{soc}} = \lambda\langle \mathbf{L}\cdot\mathbf{S} \rangle_{ij} = \lambda\langle l_i,s_i\left|\mathbf{L}\cdot\mathbf{S}\right|l_j,s_j\rangle
\end{equation}
The \api{eval} function of \api{SOC} class calculates $\langle \mathbf{L}\cdot\mathbf{S} \rangle_{ij}$ taking the orbital and spin labels as input. The orbital labels should follow the notations in \ref{misc_wan_sk_fit}, while the spin labels should be either \api{up} or \api{down}. In actual calculations, we firstly double the orbitals and hopping terms in the primitive cell to yield the product basis, then add SOC as hopping terms between basis functions following Eq. (\ref{soc_mtxel}).}

We begin with importing the necessary packages
\begin{lstlisting}
from math import sqrt, pi
import numpy as np
from numpy.linalg import norm
import tbplas as tb
import tbplas.builder.exceptions as exc
\end{lstlisting}
Then we define the function to build the primitive cell without SOC
\begin{lstlisting}
def make_cell():
    # Lattice constants
    a = 4.5332
    c = 11.7967
    mu = 0.2341

    # Lattice vectors of bulk
    a1 = np.array([-0.5*a, -sqrt(3)/6*a, c/3])
    a2 = np.array([0.5*a, -sqrt(3)/6*a, c/3])
    a3 = np.array([0, sqrt(3)/3*a, c/3])

    # Lattice vectors and atomic positions of bilayer
    a1_2d = a2 - a1
    a2_2d = a3 - a1
    a3_2d = np.array([0, 0, c])
    lat_vec = np.array([a1_2d, a2_2d, a3_2d])
    atom_position = np.array([[0, 0, 0], [1/3, 1/3, 2*mu-1/3]])

    # Create cell and add orbitals
    cell = tb.PrimitiveCell(lat_vec, unit=tb.ANG)
    atom_label = ("Bi1", "Bi2")
    e_s, e_p = -10.906, -0.486
    orbital_energy = {"s": e_s, "px": e_p, "py": e_p, "pz": e_p}
    for i, pos in enumerate(atom_position):
        for orbital, energy in orbital_energy.items():
            label = f"{atom_label[i]}:{orbital}"
            cell.add_orbital(pos, label=label, energy=energy)

    # Add hopping terms
    neighbors = tb.find_neighbors(cell, a_max=5, b_max=5, max_distance=0.454)
    sk = tb.SK()
    for term in neighbors:
        i, j = term.pair
        label_i = cell.get_orbital(i).label
        label_j = cell.get_orbital(j).label
        hop = calc_hop(sk, term.rij, label_i, label_j)
        cell.add_hopping(term.rn, i, j, hop)
    return cell
\end{lstlisting}
The \api{make\_cell} function is much similar to that of section \ref{misc_wan_sk_fit}, where we firstly define the lattice vectors and orbital positions according to Ref. \cite{Murakami2006,Liu1995}, then add the orbitals and hopping terms using Slater-Koster formulation. Note that we have included atom labels in the orbital labels, namely \api{Bi1} and \api{Bi2}, in order to distinguish the intra-atom terms when adding SOC afterwards. The hopping terms are evaluated by the \api{calc\_hop} function, which is also similar to that of section \ref{misc_wan_sk_fit}
\begin{lstlisting}
def calc_hop(sk, rij, label_i, label_j):
    dict1 = {"v_sss": -0.608, "v_sps": 1.320, "v_pps": 1.854, "v_ppp": -0.600}
    dict2 = {"v_sss": -0.384, "v_sps": 0.433, "v_pps": 1.396, "v_ppp": -0.344}
    dict3 = {"v_sss": 0.0, "v_sps": 0.0, "v_pps": 0.156, "v_ppp": 0.0}
    r_norm = norm(rij)
    if abs(r_norm - 0.30628728) < 1.0e-5:
        data = dict1
    elif abs(r_norm - 0.35116131) < 1.0e-5:
        data = dict2
    else:
        data = dict3
    lm_i = label_i.split(":")[1]
    lm_j = label_j.split(":")[1]
    return sk.eval(r=rij, label_i=lm_i, label_j=lm_j,
                   v_sss=data["v_sss"], v_sps=data["v_sps"],
                   v_pps=data["v_pps"], v_ppp=data["v_ppp"])
\end{lstlisting}

The intra-atom SOC is implemented in the \api{add\_soc} function, which is defined as
\begin{lstlisting}
def add_soc(cell):
    # Double the orbitals and hopping terms
    cell = tb.merge_prim_cell(cell, cell)

    # Add spin notations to the orbitals
    num_orb_half = cell.num_orb // 2
    num_orb_total = cell.num_orb
    for i in range(num_orb_half):
        label = cell.get_orbital(i).label
        cell.set_orbital(i, label=f"{label}:up")
    for i in range(num_orb_half, num_orb_total):
        label = cell.get_orbital(i).label
        cell.set_orbital(i, label=f"{label}:down")

    # Add SOC terms
    soc_lambda = 1.5
    soc = tb.SOC()
    for i in range(num_orb_total):
        label_i = cell.get_orbital(i).label.split(":")
        atom_i, lm_i, spin_i = label_i

        for j in range(i+1, num_orb_total):
            label_j = cell.get_orbital(j).label.split(":")
            atom_j, lm_j, spin_j = label_j

            if atom_j == atom_i:
                soc_intensity = soc.eval(label_i=lm_i, spin_i=spin_i,
                                         label_j=lm_j, spin_j=spin_j)
                soc_intensity *= soc_lambda
                if abs(soc_intensity) >= 1.0e-15:
                    try:
                        energy = cell.get_hopping((0, 0, 0), i, j)
                    except exc.PCHopNotFoundError:
                        energy = 0.0
                    energy += soc_intensity
                    cell.add_hopping((0, 0, 0), i, j, soc_intensity)
    return cell
\end{lstlisting}
In line 3-13, we double the orbitals and hopping terms and add spin labels to the orbitals. Then we define the spin-orbital coupling intensity $\lambda$ and create an \api{SOC} instance in 16-17. Afterwards, we loop over the upper-triangular orbital pairs to add the SOC terms, while the conjugate terms are handled automatically. In line 26 we check if the two orbitals reside on the same atom, while in line 27 we call the \api{eval} function to calcualte the matrix element $\langle\mathbf{L}\cdot\mathbf{S}\rangle_{ij}$. If the corresponding hopping term already exists, the SOC term will be added to it. Otherwise, a new hopping term will be created.

\begin{figure}[h]
	\centering
	\includegraphics[width=1.0\linewidth]{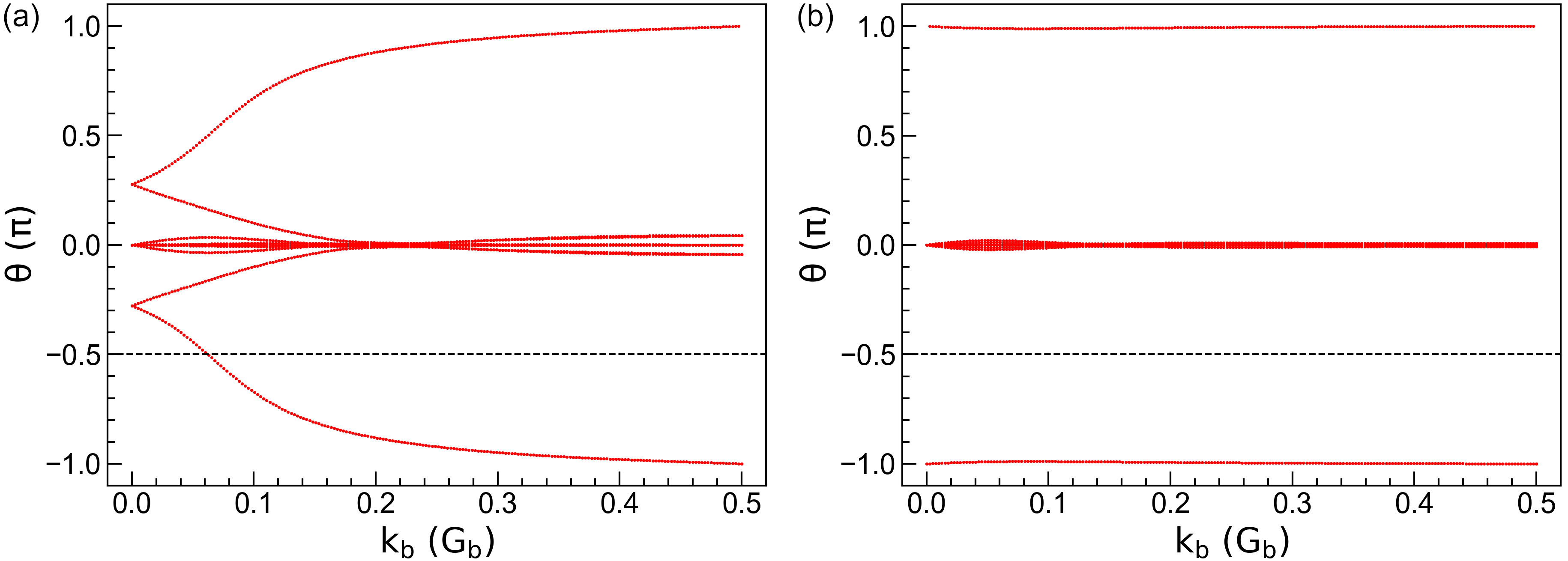}
	\caption{\change{Topological phases $\theta_m^D$ of bilayer bismuth under SOC intensity of (a) $\lambda$ = 1.5 eV and (b) $\lambda$ = 0.15 eV. The horizontal dashed lines indicates the reference lines with which the crossing number is determined.}}
	\label{fig:z2_soc}
\end{figure}

With all the auxiliary functions ready, we now proceed to calculate the $\mathbb{Z}_2$ invariant number of bilayer bismuth
\begin{lstlisting}
def main():
    # Create cell and add soc
    cell = make_cell()
    cell = add_soc(cell)

    # Evaluate Z2
    ka_array = np.linspace(-0.5, 0.5, 200)
    kb_array = np.linspace(0.0, 0.5, 200)
    kc = 0.0
    z2 = tb.Z2(cell, num_occ=10)
    kb_array, phases = z2.calc_phases(ka_array, kb_array, kc)

    # Plot phases
    vis = tb.Visualizer()
    vis.plot_phases(kb_array, phases / pi)


if __name__ == "__main__":
    main()
\end{lstlisting}
To calculate $\mathbb{Z}_2$ we need to sample the $\mathbf{k}_a$ from $-\frac{1}{2}\mathbf{G}_a$ to $\frac{1}{2}\mathbf{G}_a$, and $\mathbf{k}_b$ from $\mathbf{0}$ to $\frac{1}{2}\mathbf{G}_b$. Then we create a \api{Z2} instance and its \api{calc\_phases} function to get the topological phases $\theta_m^D$ defined in Eq. (\ref{theta_D}). After that, we plot $\theta_m^D$ as the function of $\mathbf{k}_b$ in Fig. \ref{fig:z2_soc}(a). It is clear that the crossing number of phases against the reference line is 1, indicating that bilayer bismuth is a topological insulator. We then decrease the SOC intensity $\lambda$ to 0.15 eV and re-calculate the phases. The results are shown in Fig. \ref{fig:z2_soc}(b), where the crossing number is 0, indicating that bilayer bismuth becomes a normal insulator under weak SOC, similar to the case of bilayer Sb \cite{yu2011equivalent}.

\subsection{Parallelization}
\label{para_use}

In this section, we give the general guidelines to set up the parallelization environment and show how to run calculations in parallel mode within \api{TBPLaS}. It should be noted that the determination of optimal parallelization configuration is a non-trivial task and strongly case-dependent. So, the guidelines provided here serve only as a \textit{starting point}, whereas intensive tests and benchmarks are required before production runs.

\subsubsection{General guidelines}
\label{para_use_gen}

The technical details of parallelism of \api{TBPLaS} have been discussed in section \ref{parallel}. Up to now, hybrid MPI+OpenMP parallelization has been implemented for the evaluation of band structure and DOS from exact-diagonalization, response properties from Lindhard function, \change{topological invariant $\mathbb{Z}_2$} and TBPM calculations. Both MPI and OpenMP can be switched on/off separately on demand, while pure OpenMP mode is enabled by default.

The number of OpenMP threads is controlled by the \api{OMP\_NUM\_THREADS} environment variable. If \api{TBPLaS} has been compiled with MKL support, then the \api{MKL\_NUM\_THREADS} environment variable will also take effect. If none of the environment variables has been set, OpenMP will make use of all the CPU cores on the computing node. To switch off OpenMP, set the environment variables to 1. On the contrary, MPI-based parallelization is disabled by default, but can be easily enabled with a single option. The \api{calc\_bands} and \api{calc\_dos} functions of \api{PrimitiveCell} and \api{Sample} classes, the initialization functions of \api{Lindhard}, \change{\api{Z2},} \api{Solver} and \api{Analyzer} classes all accept an argument named \api{enable\_mpi} whose default value is taken to be False. If set to True, MPI-based parallelization is turned on, provided that the MPI4PY package has been installed. Hybrid MPI+OpenMP parallelization is achieved by enabling MPI and OpenMP simultaneously. The number of processes is controlled by the MPI launcher, which receives arguments from the command line, environment variables or configuration file. The user is recommended to check the manual of job queuing system on the computer for properly setting the environment variables and invoking the MPI launcher. For computers without a queuing system, e.g., laptops, desktops and standalone workstations, the MPI launcher should be \api{mpirun} or \api{mpiexec}, while the number of processes is controlled by the \api{-np} command line option.

The optimal parallelization configuration, i.e., the numbers of MPI processes and OpenMP threads, depend on the hardware, the model size and the type of calculation. Generally speaking, matrix diagonalization for a single $\mathbf k$-point is poorly parallelized over threads. But the diagonalization for multiple $\mathbf k$-points can be efficiently parallelized over processes. Therefore, for band structure and DOS calculations, as well as response properties from Lindhard function \change{and topological invariant from \api{Z2},} it is recommended to run in pure MPI-mode by setting the number of MPI processes to the total number of allocated CPU cores and the number of OpenMP threads to 1. However, MPI-based parallelization uses more RAM since every process has to keep a copy of the wave functions and energies. So, if the available RAM imposes a limit, try to use less processes and more threads. Anyway, the product of the numbers of processes and threads should be equal to the number of allocated CPU cores. For example, if you have allocated 16 cores, then you can try 16 processes $\times$ 1 thread, 8 processes $\times$ 2 threads, 4 processes $\times$ 4 threads, etc. For TBPM calculations, the number of random initial wave functions should be divisible by the number of processes. For example, if you are going to consider 16 initial wave functions, then the number of processes should be 1, 2, 4, 8, or 16. The number of threads should be set according to the number of processes. Again, if the RAM size is a problem, try to decrease the number of processes and increase the number of threads.

If MPI-based parallelization is enabled, either in pure MPI or hybrid MPI+OpenMP mode, special care should be taken to output and plotting part of the job script. These operations should be performed on the master process only, otherwise the output will mess up or files get corrupted, since all the processes will try to modify the same file or plotting the same data. This situation is avoided by checking the rank of the process before action. The \api{Lindhard}, \change{\api{Z2},} \api{Solver}, \api{Analyzer} and \api{Visualizer} classes all offer an \api{is\_master} attribute to detect the master process, whose usage will be demonstrated in the following sections.

Last but not least, we have to mention that all the calculations in previous sections can be run in either interactive or batch mode. You can input the script line-by-line in the terminal, or save it to a file and pass the file to the Python interpreter. However, MPI-based parallelization supports only the batch mode, since there is no possibility to input anything in the terminal for multiple processes in one time. In the following sections, we assume the script file to be \api{test\_mpi.py}. A common head block of the script is given in \ref{para_use_bands} and will not be explicitly repeated in subsequent sections.

\subsubsection{Band structure and DOS}
\label{para_use_bands}

We demonstrate the usage of \api{calc\_bands} and \api{calc\_dos} in parallel mode by calculating the band structure and DOS of a $12\times12\times1$ graphene sample. Procedure shown here is also valid for the primitive cell. To enable MPI-based parallelization, we need to save the script to a file, for instance, \api{test\_mpi.py}. The head block of this file should be
\begin{lstlisting}[escapechar=^]
^\#!^ /usr/bin/env python

import numpy as np
import tbplas as tb


timer = tb.Timer()
vis = tb.Visualizer(enable_mpi=True)
\end{lstlisting}
where the first line is a magic line declaring that the script should be interpreted by the Python program. In the following lines we import the necessary packages. To record and report the time usage, we need to create a timer from the \api{Timer} class. We also need a visualizer for plotting the results, where the \api{enable\_mpi} argument is set to \api{True} during initialization. This head block also is essential for other examples in subsequent sections.

For convenience, we will not build the primitive cell from scratch, but import it from the material repository with the \api{make\_graphene\_diamond} function
\begin{lstlisting}
cell = tb.make_graphene_diamond()
\end{lstlisting}
Then we build the sample by
\begin{lstlisting}
sample = tb.Sample(tb.SuperCell(cell, dim=(12, 12, 1), pbc=(True, True, False)))
\end{lstlisting}
The evaluation of band structure in parallel mode is similar to the serial mode, which also involves generating the $\mathbf k$-path and calling \api{calc\_bands}. The only difference is that we need to set the \api{enable\_mpi} argument to \api{True} when calling \api{calc\_bands}
\begin{lstlisting}
k_points = np.array([
    [0.0, 0.0, 0.0],
    [2./3, 1./3, 0.0],
    [1./2, 0.0, 0.0],
    [0.0, 0.0, 0.0],
])
k_path, k_idx = tb.gen_kpath(k_points, [40, 40, 40])
timer.tic("band")
k_len, bands = sample.calc_bands(k_path, enable_mpi=True)
timer.toc("band")
vis.plot_bands(k_len, bands, k_idx, k_label)
if vis.is_master:
    timer.report_total_time()
\end{lstlisting}
The \api{tic} and \api{toc} functions begin and end the recording of time usage, which receive a string as the argument for tagging the record. The visualizer is aware of the parallel environment, so no special treatment is needed when plotting the results. Finally, the time usage is reported with the \api{report\_total\_time} function on the master process only, by checking the \api{is\_master} attribute of the visualizer.

We run \api{test\_mpi.py} by
\begin{lstlisting}
$ export OMP_NUM_THREADS=1
$ mpirun -np 1 ./test_mpi.py
\end{lstlisting}
With the environment variable \api{OMP\_NUM\_THREADS} set to 1, the script will run in pure MPI-mode. We invoke 1 MPI process by the \api{-np} option of the MPI launcher (\api{mpirun}). The output should look like
\begin{lstlisting}
    band :      11.03s
\end{lstlisting}
So, the evaluation of bands takes 11.03 seconds on 1 process. We try with more processes
\begin{lstlisting}
$ mpirun -np 2 ./test_mpi.py
    band :       5.71s
$ mpirun -np 4 ./test_mpi.py
    band :       2.93s
\end{lstlisting}
Obviously, the time usage scales reversely with the number of processes. Detailed discussion on the time usage and speedup under different parallelization configurations will be discussed in section \ref{para_use_benchmark}.

Evaluation of DOS can be parallelized in the same approach, by setting the \api{enable\_mpi} argument to \api{True}
\begin{lstlisting}
k_mesh = tb.gen_kmesh((20, 20, 1))
timer.tic("dos")
energies, dos = sample.calc_dos(k_mesh, enable_mpi=True)
timer.toc("dos")
vis.plot_dos(energies, dos)
if vis.is_master:
    timer.report_total_time()
\end{lstlisting}
The script can be run in the same approach as evaluating the band structure.

\subsubsection{Response properties from Lindhard function}
\label{para_use_lindhard}

To evaluate response properties in parallel mode, simply set the \api{enable\_mpi} argument to \api{True} when creating the Lindhard calculator
\begin{lstlisting}
lind = tb.Lindhard(cell=cell, energy_max=10.0, energy_step=2048, kmesh_size=(600, 600, 1), mu=0.0, temperature=300.0, g_s=2, back_epsilon=1.0, dimension=2, enable_mpi=True)
\end{lstlisting}
Subsequent calls to the functions of \api{Lindhard} class does not need further special treatment. For example, the optical conductivity can be evaluated in the same approach as in serial mode
\begin{lstlisting}
timer.tic("ac_cond")
omegas, ac_cond = lind.calc_ac_cond(component="xx")
timer.toc("ac_cond")
vis.plot_xy(omegas, ac_cond)
if vis.is_master:
    timer.report_total_time()
\end{lstlisting}

\subsubsection{\change{Topological invariant from \api{Z2}}}
\label{para_z2}
\change{The evaluation of phases $\theta_m^D$ can be paralleled in the same approach as response functions}
\begin{lstlisting}
z2 = tb.Z2(cell, num_occ=10, enable_mpi=True)
timer.tic("z2")
kb_array, phases = z2.calc_phases(ka_array, kb_array, kc)
timer.toc("z2")
vis.plot_phases(kb_array, phases / pi)
if vis.is_master:
    timer.report_total_time()
\end{lstlisting}
\change{where we only need to set \api{enable\_mpi} argument to \api{True} when creating the \api{Z2} instance.}

\subsubsection{Properties from TBPM}
\label{para_use_tbpm}

TBPM calculations in parallel mode are similar to the evaluation of response functions. The user only needs to set the \api{enable\_mpi} argument to \api{True}. To make the time usage noticeable, we build a larger sample first
\begin{lstlisting}
sample = tb.Sample(tb.SuperCell(cell, dim=(240, 240, 1), pbc=(True, True, False)))
\end{lstlisting}
Then we create the configuration, solver and analyzer, with the argument \api{enable\_mpi=True}
\begin{lstlisting}
sample.rescale_ham(9.0)
config = tb.Config()
config.generic["nr_random_samples"] = 4
config.generic["nr_time_steps"] = 256
solver = tb.Solver(sample, config, enable_mpi=True)
analyzer = tb.Analyzer(sample, config, enable_mpi=True)
\end{lstlisting}
Correlation function can be obtained and analyzed in the same way as in serial mode
\begin{lstlisting}
timer.tic("corr_dos")
corr_dos = solver.calc_corr_dos()
timer.toc("corr_dos")
energies, dos = analyzer.calc_dos(corr_dos)
vis.plot_dos(energies, dos)
if vis.is_master:
    timer.report_total_time()
\end{lstlisting}

\subsubsection{Benchmarks}
\label{para_use_benchmark}
The time usages and speedups of different types of calculations are summarized in Table \ref{tab:para_benchmark}. The benchmarks have been performed on an Intel$^\circledR$ Xeon$^\circledR$ Gold 6248 CPU, with 16 cores allocated at most. It is obvious that for the evaluation of band structure and DOS, increasing the number of MPI processes significantly boosts the calculation. However, the efficiency enhancement of increasing OpenMP threads is much lower. The average speedup drops significantly when OpenMP is enbaled, indicating a poor scaling versus the number of CPU cores. This is due to the fact that matrix diagonalization cannot be efficiently parallelized over threads. On the contrary, pure MPI-based parallelization has the best efficiency, with an almost linear scaling (average speedup $\approx 1$).

The evaluation of optical conductivity has an additional post-processing step after diagonalization, which is suitable for both MPI and OpenMP-based parallelization. So, the speedup and scaling versus the number of threads improve slightly. \change{$\mathbb{Z}_2$ topological invariant shows a similar scaling behavior as band structure and DOS, i.e., pure MPI parallelization has the best efficiency.} For TBPM calculations, the speedups and efficiencies of multi-threading and multi-processing are almost equal, since sparse matrix-vector multiplication can be efficiently parallelized over threads. Although pure MPI-mode still has the best efficiency, the number of processes is limited by the number of random initial wave functions and available RAM size, as discussed in section \ref{para_use_gen}. So, pure OpenMP or hybrid MPI+OpenMP paralelization is recommended for TBPM calculations, with the optimal numbers of processes and and threads determined from benchmarks. 

\begin{table}
	\begin{center}
		\caption{Time usages and speedups of benchmarks for various calculation types with respect to the numbers of MPI processes ($n_p$) and OpenMP threads ($n_t$) per process. The standard ($t_0$) of each type is defined as the time usage on 1 process $\times$ 1 thread, while the speedup is defined as $t_0/t_{n_p n_t}$. Numbers in the brackets are the average speedups to each CPU core defined as $t_0/(t_{n_p n_t} \times n_p \times n_t$).}
		\label{tab:para_benchmark}
		\begin{tabular}{cccrrr}
			\hline\hline
			\multirow{2}{*}{Type} & \multirow{2}{*}{$t_0/s$} & \multirow{2}{*}{$n_p$} & \multicolumn{3}{c}{$n_t$} \\
			\cline{4-6} \multicolumn{3}{c}{} & \multicolumn{1}{c}{1} & \multicolumn{1}{c}{2} & \multicolumn{1}{c}{4} \\
			\hline
			\multirow{3}{*}{Band structure} & \multirow{3}{*}{2.56} & 1 & 1.00 (1.00) & 1.19 (0.60) & 1.45 (0.36) \\
                                            &                       & 2 & 1.92 (0.96) & 1.61 (0.40) & 2.03 (0.25) \\
                                            &                       & 4 & 4.00 (1.00) & 3.05 (0.38) & 4.06 (0.25) \\
			\hline
			\multirow{3}{*}{DOS} & \multirow{3}{*}{10.62} & 1 & 1.00 (1.00) & 1.17 (0.58) & 1.33 (0.33) \\
			                     &                        & 2 & 1.84 (0.92) & 1.74 (0.44) & 2.00 (0.25) \\
			                     &                        & 4 & 3.74 (0.93) & 3.23 (0.40) & 3.88 (0.24) \\
			\hline
			\multirow{3}{*}{Optical conductivity} & \multirow{3}{*}{24.45} & 1 & 1.00 (1.00) & 1.58 (0.79) & 2.25 (0.56) \\
			                                 &                        & 2 & 1.76 (0.88) & 2.61 (0.65) & 3.49 (0.44) \\
			                                 &                        & 4 & 3.30 (0.83) & 4.57 (0.57) & 5.93 (0.37) \\
			\hline
			\multirow{3}{*}{\change{$\mathbb{Z}_2$ invariant}} & \multirow{3}{*}{\change{34.37}} & \change{1} & \change{1.00 (1.00)} & \change{0.99 (0.50)} & \change{1.00 (0.25)} \\
			                                       &                           & \change{2} & \change{1.67 (0.84)} & \change{1.72 (0.43)} & \change{1.71 (0.21)} \\
			                                       &                           & \change{4} & \change{3.32 (0.83)} & \change{3.34 (0.42)} & \change{3.38 (0.21)} \\
			\hline
			\multirow{3}{*}{TBPM} & \multirow{3}{*}{24.71} & 1 & 1.00 (1.00) & 1.91 (0.96) & 3.48 (0.87) \\
			                      &                        & 2 & 1.96 (0.98) & 3.80 (0.95) & 6.84 (0.86) \\
			                      &                        & 4 & 3.55 (0.89) & 6.68 (0.83) & 12.80 (0.80) \\
			\hline\hline
		\end{tabular}
	\end{center}
\end{table}
\section{Examples}
\label{examples}

\begin{figure}[h]
	\centering
	\includegraphics[width=0.8\linewidth]{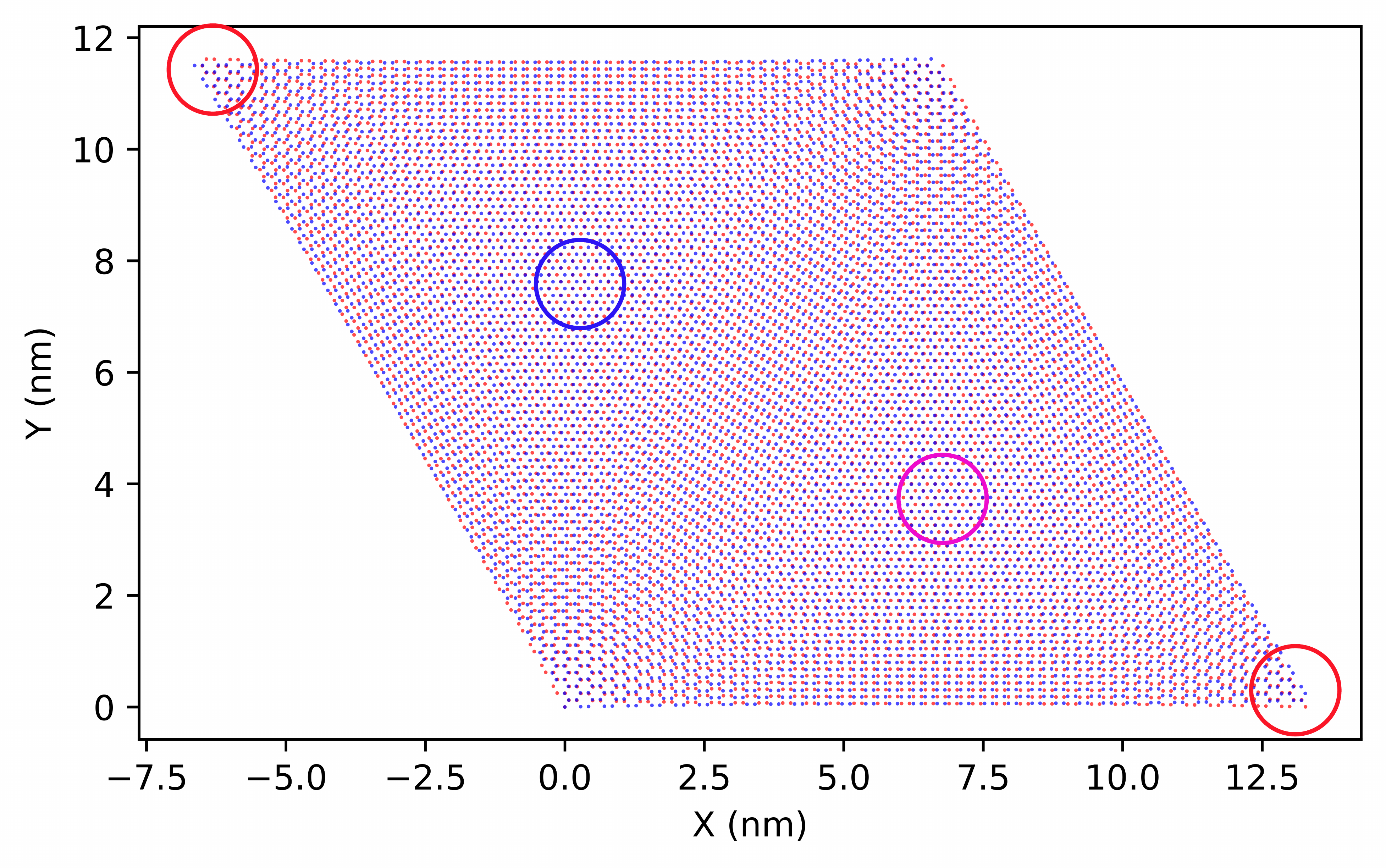}
	\caption{Atomic structure of TBG with twist angle $\theta=1.05^\circ$. Highly-symmetric stacking regions of AA, AB and BA are marked by red, blue and magenta circles, respectively. Carbon atoms in the top and bottom layers are depicted in blue and red, respectively.}
	\label{fig:TBG_struc}
\end{figure}

As mentioned in previous sections, \api{TBPLaS} is capable of tackling complex systems with tens of billions of atoms. In this section, we present an example utilizing \api{TBPLaS} to calculate the properties of TBG with magic angle $\theta=1.05^\circ$. For TBG with the magic angle, flat bands appear near the Fermi level, which provide a platform to explore strongly correlated phases and superconductivity \cite{cao2018unconventional,cao2018correlated,bistritzer2011moire}. The moir\'{e} supercell of twisted bilayer graphene is constructed by identifying a common periodicity between the two layers. We start with a AA stacking bilayer graphene ($\theta=0^\circ$), and choose the rotation origin (O) at an atom site. Then, we rotate one layer relatively to the other one by the angle $\theta$. Fig. \ref{fig:TBG_struc} shows the atomic structure of the magic angle TBG. The moir\'{e} superlattice contains three types of high-symmetry staking patterns, namely AA, AB and BA stacking. For TBG with twist angles smaller that $1.2^\circ$, the system suffers significant lattice reconstruction due to the interplay between the interlayer van der Waals interaction and the in-plane strain field \cite{gargiulo2017structural}. The lattice relaxation (both the out-of-plane and in-plane) of TBG is performed with the LAMMPS package \cite{plimpton1995fast}. The intralyer and interlayer interactions of TBG are simulated with the long-range carbon bond-order potential \cite{los2005improved} and Kolmogorov-Crespi potential \cite{kolmogorov2005registry}, respectively.

The properties of both rigid (without lattice relaxation) and relaxed (with lattice relaxation) TBG with magic angle are calculated with a full tight-binding model based on $p_z$ orbitals \cite{Laissardiere2012}. The on-site energies $\epsilon_i$ are set to zero, and the hopping parameters between sites $i$ and $j$ are described by a distance-dependent function as
\begin{equation}
    t_{ij}=n^2V_{pp\sigma}(r_{ij})+(1-n^2)V_{pp\pi}(r_{ij})
\end{equation}
where $r_{ij}=|\mathbf{r}_{ij}|$ is the distance between two sites located at $\mathbf{r}_i$ and $\mathbf{r}_j$, $n$ is the direction cosine of $\mathbf{r}_{ij}$ along the direction that is perpendicular to the graphene layer. The Slater-Koster parameters $V_{pp\sigma}$ and $V_{pp\pi}$ are 
\begin{align}
     V_{pp\pi}(r_{ij}) &= -t_0\mathrm{e}^{q_\pi(1-r_{ij}/d)}F_c(r_{ij})\\
     V_{pp\sigma}(r_{ij}) &= t_1\mathrm{e}^{q_\sigma (1-r_{ij}/h)}F_c(r_{ij})
\end{align}
where $d=1.42$ \AA \; and $h=3.349$ \AA \; are the nearest in-plane and out-of-plane carbon-carbon distances, respectively. $t_0=2.8$ eV and $t_1=0.44$ eV are re-optimized to obtain the magic angle at rotation angle $\theta=1.05^\circ$ \cite{kuang2021collective}. The parameters $q_\sigma$ and $q_\pi$ satisfy $q_\sigma/h=q_\pi/d=2.218$ \AA$^{-1}$, and the smooth function is defined as $F_c(r)=(1+\mathrm{e}^{(r-r_c)/l_c})^{-1}$ with $l_c=0.265$ \AA and $r_c=5.0$ \AA.

\begin{figure}[h!]
	\centering
	\includegraphics[width=1\linewidth]{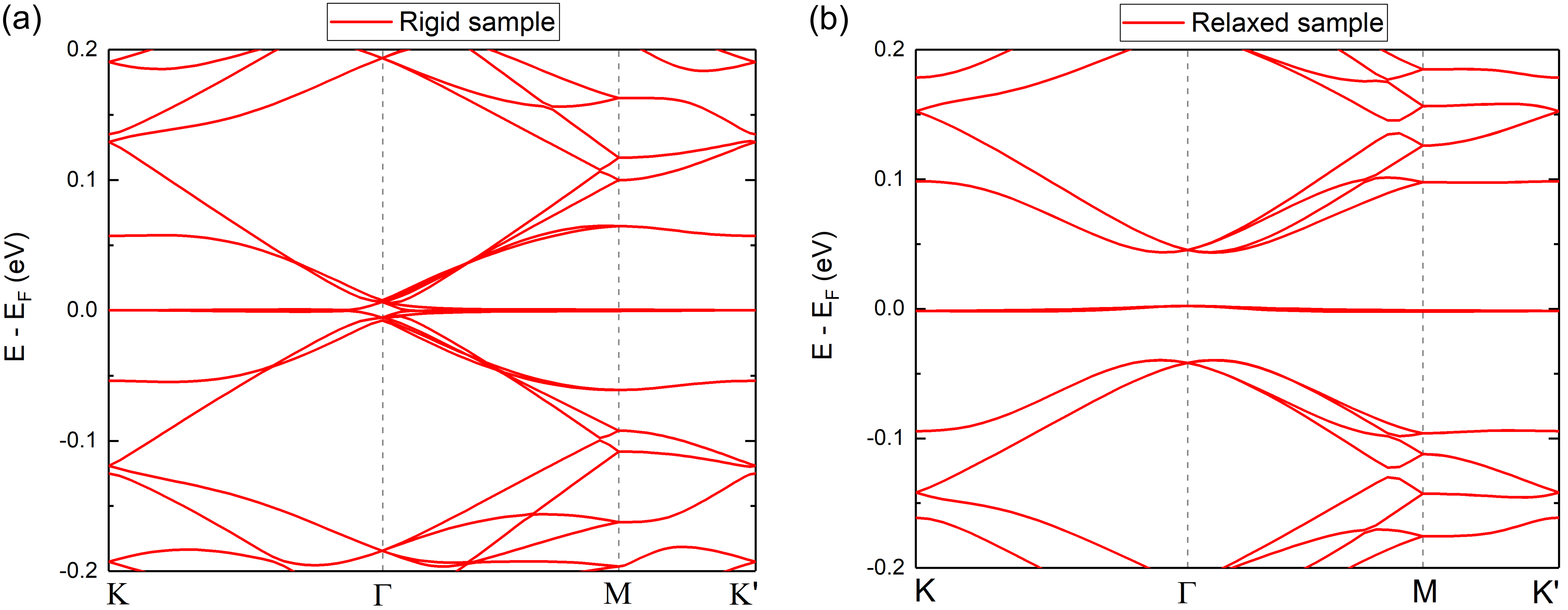}
	\caption{Band structures of (a) rigid and (b) relaxed TBG with $\theta=1.05^\circ$.}
	\label{fig:TBG_bs}
\end{figure}

\begin{figure}[h]
	\centering
	\includegraphics[width=1\linewidth]{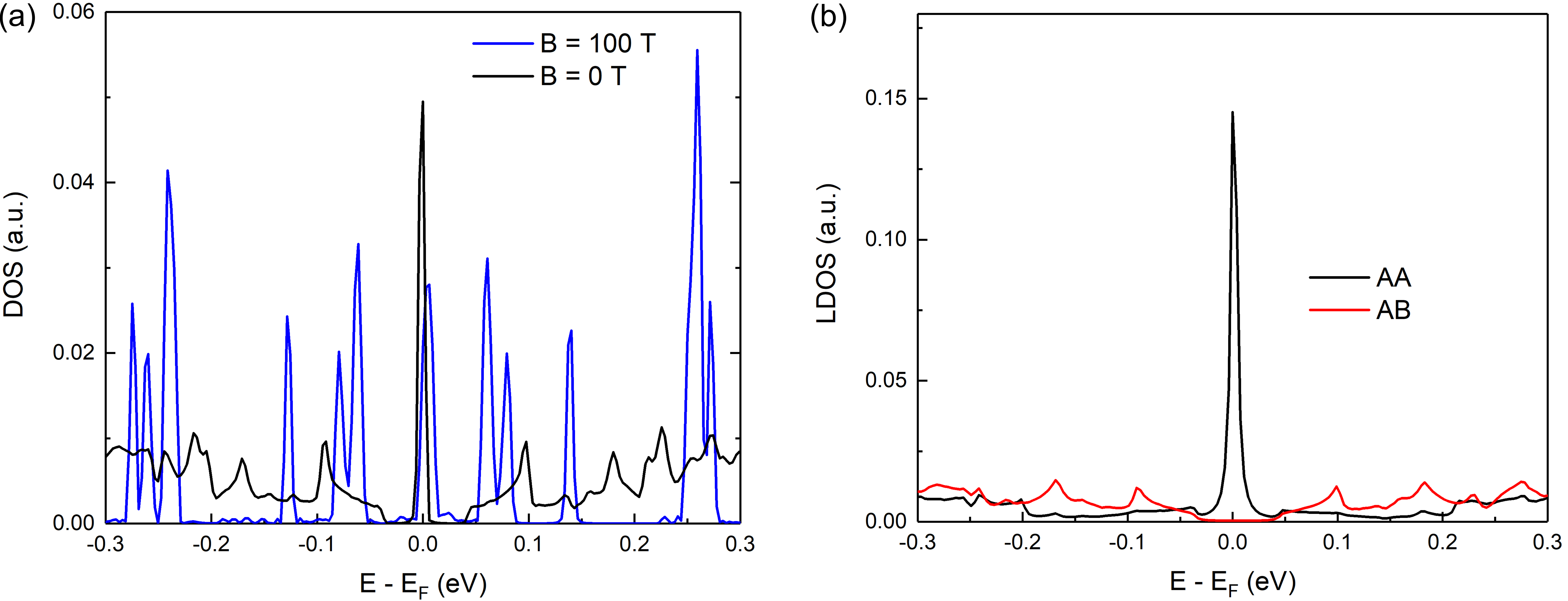}
	\caption{(a) Density of states of relaxed magic angle TBG with (blue line) and without (black line) magnetic field. (b) Local density of states of the highly-symmetric stacking regions of AA (black line) and AB (red line) in relaxed magic angle TBG.}
	\label{fig:TBG_dos}
\end{figure}

Fig. \ref{fig:TBG_bs} shows the band structure of rigid and relaxed TBG with twist angle $\theta=1.05^\circ$, which are obtained by exact diagonalization. In TBG without lattice relaxation (rigid sample), ultraflat bands appear in the charge neutrality. The bandwidth (energy difference between the K and $\Gamma$ points of the Brillouin zone) of the flat band is 7 meV, and the bandgap (energy difference between the flat band and the remote bands at the $\Gamma$ point) is zero. In relaxed sample (with lattice relaxation), the bandwidth and bandgap are 4 meV and 43 meV, respectively. Obviously, the lattice relaxation has a significant effect on the electronic structure of magic angle TBG. The black line in Fig. \ref{fig:TBG_dos}(a) is the density of states of relaxed TBG with magic angle, which is calculated via the TBPM in Eq. (\ref{dos_tbpm}). In the calculations, the accuracy of the DOS can be guaranteed by utilizing a large enough model with more than ten million atoms. The number of time integration steps is 4096, which gives an energy resolution of 3.7 meV. In DOS a sharp peak appears in the charge neutrality, which corresponds to the flat bands. When a perpendicular magnetic field is applied, Landau levels appear in the DOS. The splitting of the peak around the energy $E=68$ meV is the lifting of the twofold degeneracy due to the Dirac point splitting in twisted bilayer graphene \cite{lee2011quantum}.

\begin{figure}[h]
	\centering
	\includegraphics[width=1\linewidth]{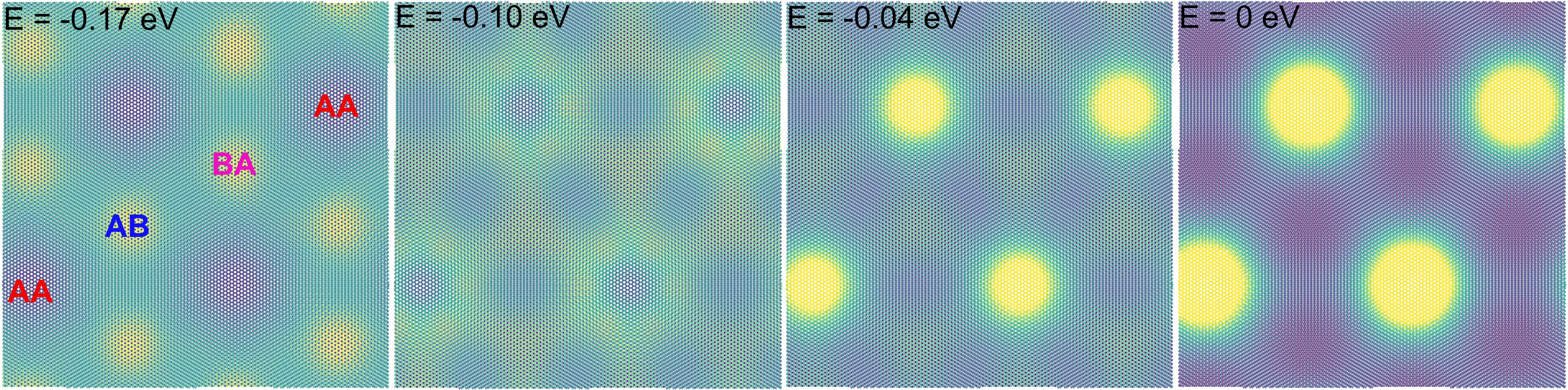}
	\caption{Local density of states mappings of magic angle TBG (with lattice relaxation) at energies $E=-0.17$ eV, $-0.10$ eV, $-0.04$ eV and $0$ eV.}
	\label{fig:TBG_quasi}
\end{figure}

The LDOS is an important quantity to describe the local properties of a system, which can be utilized to simulate the $\mathrm{d}I/\mathrm{d}V$ spectra obtained with the STM in experiments. \api{TBPLaS} provides three approaches to evaluate the LDOS, i.e. exact-diagonalization, TBPM and the recursion method. Both TBPM and the recursion method are capable of dealing with very large models. The LDOS of different stacking regions in magic angle TBG obtained with TBPM are shown in Fig. \ref{fig:TBG_dos}. It is clear that the LDOS of the AA and AB regions have obvious different features. Only the LDOS of the AA region has a sharp peak at energy $E=0$, which means that the states of the flat bands are mainly localized in the AA region. The LDOS of the AB region has some peaks located at high energies. Such strong LDOS modulation shows spatially localized electronic states with specific energies, which can be justified by calculating the LDOS mapping (quasieigenstates) via Eq. (\ref{quasi}). The LDOS mappings at different energies are shown in Fig. \ref{fig:TBG_quasi}. At energy $E=0$, states are mainly localized in the AA regions. At the energy $E=-0.17$ eV, states are mainly localized in the AB and BA regions. Such periodic variation of the local electronic structure is a consequence of different interlayer couplings in TBG. The LDOS mapping is equivalent to the $\mathrm{d}I/\mathrm{d}V$ mapping observed experimentally with STM. 

\begin{figure}[h]
	\centering
	\includegraphics[width=0.7\linewidth]{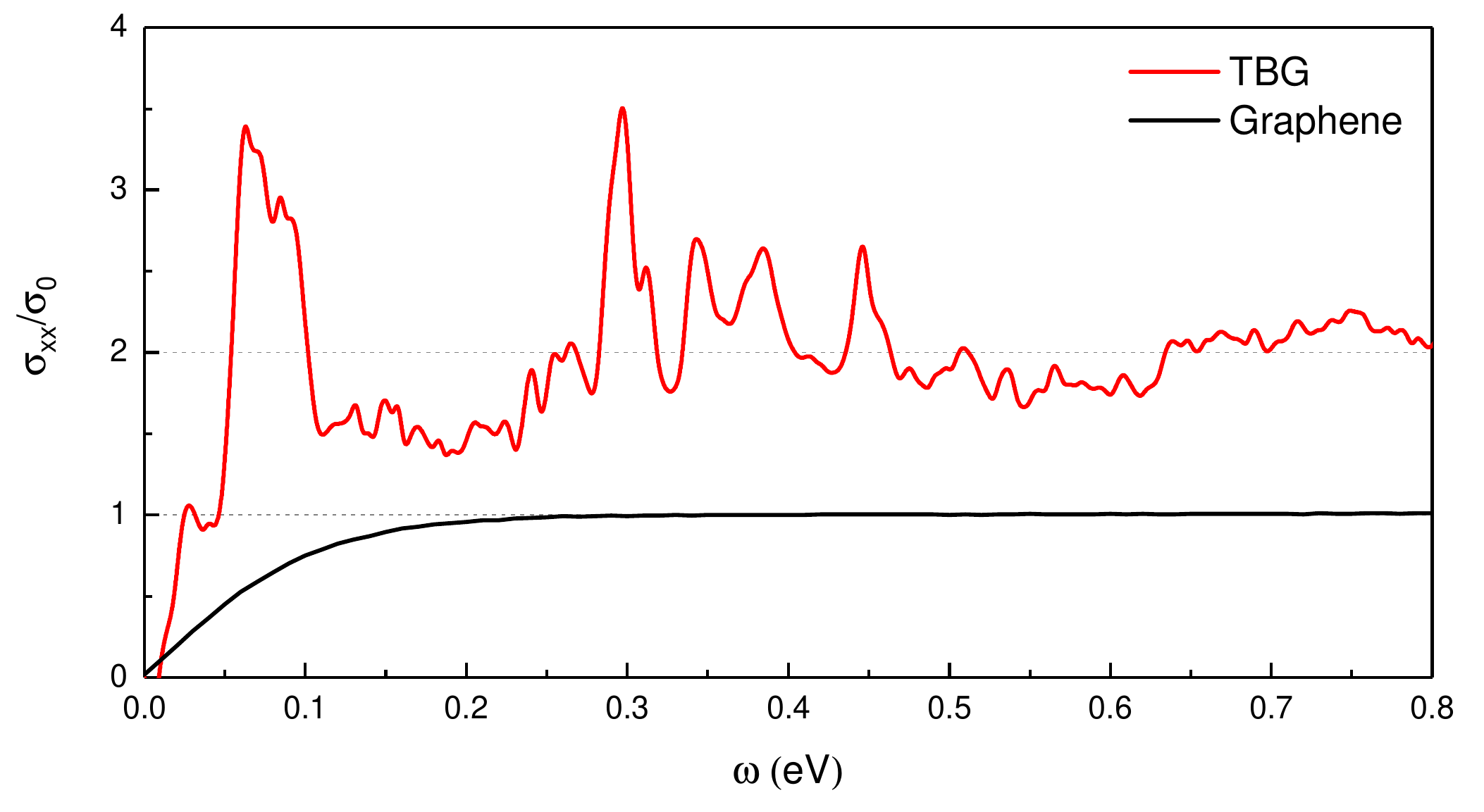}
	\caption{Optical conductivity of relaxed TBG with twist angle $\theta=1.05^\circ$ and monolayer graphene. The temperature is $T=300$ K and the chemical potential is $\mu=0$ eV.}
	\label{fig:TBG_optical}
\end{figure}

In \api{TBPLaS}, we can also investigate the optical conductivity via the Kubo formula or the Lindhard function. The Lindhard function is more suitable for small models since it requires a diagonalization process. On the contrary, by combing the Kubo formula and TBPM, we can tackle large models that contain tens of millions of orbitals. In Fig. \ref{fig:TBG_optical}, the optical conductivity of the magic angle TBG and monolayer graphene is calculated with TBPM. Note that we omit the Drude weight part in the calculation. For TBG the peak with energy around $E=0.1$ eV is due to the transition between the flat bands and their adjacent bands. A dip-peak feature around $E=0.1$ eV is due to the electron-hole asymmetry \cite{moon2013optical}. 

\begin{figure}[h]
	\centering
	\includegraphics[width=1\linewidth]{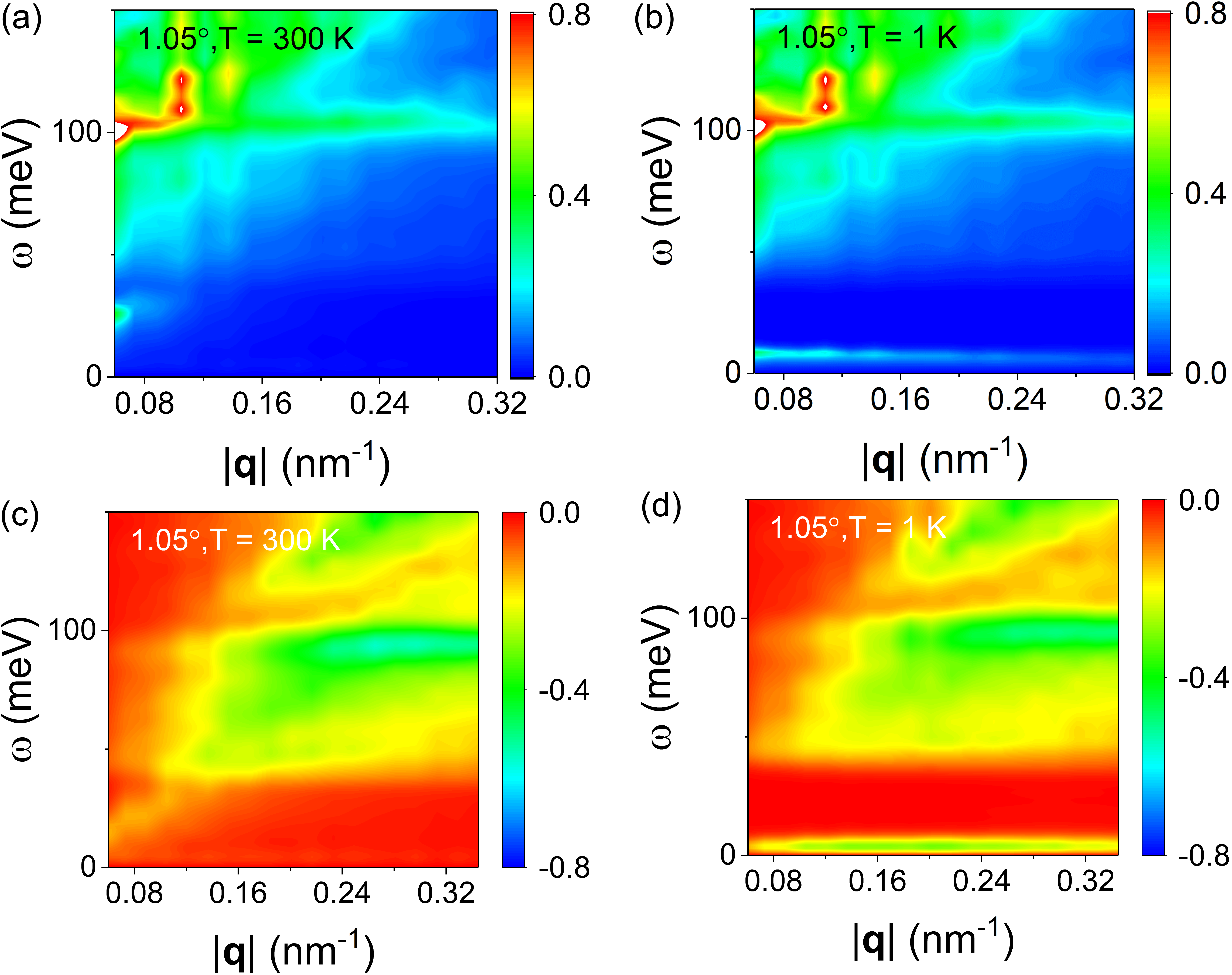}
	\caption{Plot of the loss function $-\mathrm{Im}\ \left[1/\epsilon(\mathbf{q}, \omega)\right]$ as function of frequency $\omega$ and wave vector $\mathbf{q}$ for relaxed TBG with twist angel $\theta=1.05^\circ$ at temperatures (a) $T=300$ K and (b) $T = 1$ K \cite{kuang2021collective}. Plot of the particle-hole continuum $-\mathrm{Im}\  \Pi(\mathbf{q},\omega)$ with respect to the frequency $\omega$ and wave vector $\mathbf{q}$ at (c) $T=300$ K and (d) $T = 1$ K. The chemical potential is $\mu=0$ and the background dielectric constant $\kappa=3.03$ of hBN substrate.}
	\label{fig:TBG_plasmon}
\end{figure}

In addition to the optical conductivity, many other response properties can also be obtained with \api{TBPLaS}. Fig. \ref{fig:TBG_plasmon} shows the electron energy loss function of the magic angle TBG. Firstly, we calculate the dynamical polarization by using the Kubo formula in Eq. (\ref{kubo_dp}). Then the dielectric function and energy loss function are obtained within the random phase approximation with Eqs. (\ref{eps_rpa}) and (\ref{EEL}), respectively. The plasmon mode can be detected by many experimental techniques, e.g. the scattering-type scanning near-field optical microscope (s-SNOM) and electron energy loss spectroscopy. In experiments, when a plasmon mode with frequency $\omega_p$ exists with low damping, the energy loss spectra possess a sharp peak at $\omega=\omega_p$. For the magic angle TBG, interband plasmon modes close to 100 meV appear at both temperature $T=300$ K and 1 K, which are attributed to the interband transitions from the flat bands to bands located at energy of 100 meV. These modes originate from the collective oscillations of electrons in the AA region \cite{hesp2021observation}. The $\omega_p=100$ meV plasmon mode disperses within the particle-hole continuum in Figs. \ref{fig:TBG_plasmon}(c) and \ref{fig:TBG_plasmon}(d) with fast damping into electron-hole pairs. It becomes clear with a fine and flat shape with momentum larger than $0.2\;\mathrm{nm^{-1}}$. Single-particle transitions are almost forbidden in flat bands below 40 meV, corresponding to the value of band gap between the flat bands and the excited bands at $\Gamma$ point, from which the continuum spectrum rises to non-zero zone in Fig. \ref{fig:TBG_plasmon} (c). When the temperature declines to the critical temperature 1 K at which the superconductivity can be detected in the magic-angle system \cite{cao2018unconventional}, a thin plasmon mode with energy 9 meV emerges and stretches to large q in Fig. \ref{fig:TBG_plasmon}(b), which is contributed to the collective excitations among flat bands, i.e. flat-band plasmon. Meanwhile, underneath the collective flat-band plasmon mode, the particle-hole transitions arise with occupying a tiny energy region ranging from 0 meV to 8 meV in Fig. \ref{fig:TBG_plasmon}(d). As a result, this plasmon mode extends above the edge of this tiny energy zone and is free from the Laudau damping.

\section{Summary}
\label{summary}

In summary, we have introduced the \api{TBPLaS} package, an open-source software suite for accurate electronic structure, optical properties, plasmon and transport calculations in real-space based on the tight-binding theory. It has an intuitive Python API for convenient simulation set-up, and Cython/Fortran cores for efficient performance. The main advantage of \api{TBPLaS} is that the numerical calculations is based on the TBPM without diagonalization. Both the memory and CPU costs have a linear scaling with the system size. So we can tackle models contain tens of millions of atoms or even billions of atoms if necessary. In addition to TBPM, exact diagonalization-based methods are also implemented. Moreover, crystalline defects, vacancies, adsorbates, charge impurity centres, strains and external perturbations can be easily and intuitively set up in \api{TPLaS}, which allows us to simulate large and complex models. With a wide range of pre-defined functions, the numerical calculations can be performed only with a few lines of code.

In the first release, \api{TBPLaS} already features a large variety of functionalities, e.g. the band structure, DOS, LDOS, wave functions,  plasmon, optical conductivity, electric conductivity, Hall conductivity, quasi-eigenstate, real-space electron density and wave packet propagation. Moreover, thanks to its extensible and modular nature, it is easy to implemented other algorithms involving the tight-binding Hamiltonian. Further developments and extensions of \api{TBPLaS}, for instance, the real-space self-consistent Hartree and Hubbard methods for large systems \cite{guinea2018electrostatic,goodwin2020hartree} and support for GPU acceleration, will be implemented in the future.

~\\
{\textbf{Acknowledgments}}                                 
~\\

We thank Edo van Veen, Guus Slotman, Kaixiang Huang, and Yalei Zhang for their contributions to the earlier version of the code. This work is supported by the National Natural Science Foundation of China (Grant No.12174291) and the National Key R\&D Program of China (Grant No. 2018YFA0305800). Numerical calculations presented in this paper have been performed in the Supercomputing Center of Wuhan University. 

~\\
{\textbf{Declaration of competing interest}}                                 
~\\

The authors declares no competing interests.

\bibliographystyle{unsrt}
\bibliography{bibliography.bib}

\end{document}